\documentstyle[11pt]{article}
\let\la=\label

\def\be{\begin{equation}}
\def\ee{\end{equation}}
\def\bea{\begin{eqnarray}}
\def\eea{\end{eqnarray}}
\def\ba{\begin{array}}
\def\ea{\end{array}}
\def\endb{\end{thebibliography}}
\def\ed{\end{document}}

\newcommand{\eq}[1]{(\ref{#1})}
\newcommand{\w}[1]{\\[0.#1cm]}
\def\eqs#1#2{(\ref{#1}-\ref{#2})}
\def\nn{\nonumber}

\let\bm=\bibitem

\def\ft#1#2{{\textstyle{{\scriptstyle #1} \over {\scriptstyle #2}}}}
\def\fft#1#2{{#1 \over #2}}
\newcommand{\mx}[4]{\left#1\begin{array}{#2}#3\end{array}\right#4}
\def\mi#1#2#3{#1_{#2}\cdots #1_{#3}}

\def\bd{{\dot \beta}}
\def\ad{{\dot \alpha}}
\def\cd{{\dot \gamma}}
\def\dd{{\dot \delta}}
\def\yb{{\bar y}}
\def\zb{{\bar z}}
\def\se{\;\;=\;\;}
\def\ub{\bar{u}}
\def\vb{\bar{v}}
\def\kb{\overline{\k}}

\def\bp{\bar{\pi}}

\def\a{\alpha}
\def\b{\beta}
\def\c{\gamma}
\def\C{\Gamma}

\def\d{\delta}

\def\e{\epsilon}
\def\vare{\varepsilon}
\def\f{\phi}

\def\vf{\varphi}

\def\k{\kappa}
\def\l{\lambda}
\def\L{\Lambda}
\def\m{\mu}
\def\n{\nu}
\def\p{\pi}

\def\r{\rho}
\def\s{\sigma}
\def\sb{\bar{\s}}

\def\t{\tau}
\def\th{\theta}

\def\x{\xi}
\def\o{\omega}
\def\O{\Omega}
\def\y{\eta}
\def\z{\zeta}
\def\ybar{\bar{\y}}

\def\hty{\hat{y}}
\def\htyb{\hat{\bar{y}}}
\def\htt{\hat{\th}}
\def\hta{\hat{a}}
\def\htad{\hat{a}^{\dagger}}
\def\htp{\hat{\psi}}
\def\htpd{\hat{\psi}^{\dagger}}
\def\hvare{\hat{\vare}}

\def\una{\underline{\a}}
\def\unb{\underline{\b}}

\def\del{\partial}

\newcommand{\ket}[1]{\left|\left. #1 \right\rangle\right.}
\newcommand{\pd}[2]{\fft{\del}{\del#1^{#2}}}
\newcommand{\zeval}[1]{\left.#1\right|_{Z=0}}

\def\smpl{ +}
\def\smm{ -}
\def\ns{\normalsize}

\def\alg{$shs^E(8|4)$}
\def\aext{$\widehat{\cal A}$}
\def\alge{$\widehat{shs}^E(8|4)$}

\def\pl#1#2#3{Phys.~Lett.~{\bf {#1}B} (19{#2}) #3}

\def\np#1#2#3{Nucl.~Phys.~{\bf B{#1}} (19{#2}) #3}

\def\cites#1#2{\cite{#1}-\cite{#2}}

\newcommand{\hoch}[1]{$\, ^{#1}$}

\def\g{\u g}

\begin{document}

\pagestyle{empty}

\rightline{CTP TAMU-19/98}
\rightline{hep-th/9805125}

\vspace{1.5truecm}
\centerline{\Large \bf Higher Spin $N=8$ Supergravity}

\vspace{1truecm}
\centerline{E. Sezgin\hoch1 and P. Sundell\hoch2}
\bigskip
\centerline{\it Center for Theoretical Physics, Texas A\&M University,}
\centerline{\it College Station, Texas 77843, USA}
\vspace{1.3truecm}
\centerline{ABSTRACT}
\vspace{.5truecm}

The product of two $N=8$ supersingletons yields an infinite tower of
massless states of higher spin in four dimensional anti de Sitter space.
All the states with spin $s\geq 1$ correspond to generators of
Vasiliev's super higher spin algebra $shs^E(8|4)$ which contains the
$D=4, N=8$ anti de Sitter superalgebra $OSp(8|4)$. Gauging the higher
spin algebra and introducing a matter multiplet in a quasi-adjoint
representation leads to a consistent and fully nonlinear equations of
motion as shown sometime ago by Vasiliev. We show the embedding of the
$N=8$ AdS supergravity equations of motion in the full system at the
linearized level and discuss the implications for the embedding of the
interacting theory. We furthermore speculate that the boundary $N=8$
singleton field theory yields the dynamics of the $N=8$ AdS supergravity
in the bulk, including all higher spin massless fields, in an unbroken
phase of M-theory.

{\vfill\leftline{}\vfill
\vskip  10pt
\footnoterule
{\footnotesize

\hoch{1} Research supported in part by NSF Grant PHY-9722090
\\
\hoch{2} Research supported by The Natural Science Research Council of
Sweden (NFR) \vskip -12pt }

\vfill\eject

\pagestyle{plain}

%%%%%%%%%%%%%%%%%%%%%%%%%%%%%%%%%%%%%%%%%%%%%%%%%%%%%%%%%%%%%%%%%%%%%%%%%%%

\section{Introduction}

%%%%%%%%%%%%%%%%%%%%%%%%%%%%%%%%%%%%%%%%%%%%%%%%%%%%%%%%%%%%%%%%%%%%%%%%%%%

Sometime ago, the singleton representation of the super AdS group
$OSp(4|8)$ was encountered in the spectrum analysis of $D=11$
supergravity compactified on $AdS_4 \times S^7$ \cite{es1}. This is the
ultra short representation of the AdS supergroup $OSp(4|8)$ in four
dimensions, consisting of 8 bosonic and 8 fermionic states \cite{g1,ns},
which cannot be described in terms of local fields in the bulk of AdS,
as was discovered long ago by Dirac \cite{dirac} in the context of the
singleton representation of the $AdS$ group $SO(3,2)$. Indeed, it has
been shown \cite{es2} that the $OSp(4|8)$ singleton multiplet occurring
in the Kaluza-Klein spectrum is pure gauge. It has also been shown that,
though gauge modes, the singletons are needed to fill the
representations of the spectrum generating algebra $SO(8,1)$ \cite{g6}.

After the eleven dimensional supermembrane was discovered \cite{bst1},
it was speculated in \cite{duff1} that singletons may play a role in its
description. Soon after, it was conjectured in \cite{bd1} and \cite{nst}
that a whole class of AdS compactifications of supergravity theories may
be closely related to various singleton/doubleton field theories. The
field theories for them were constructed on the boundary of AdS as free
superconformal field theories, but the main thrust of the conjecture
remained to be tested.

Notwithstanding this state of affairs, in \cite{bsst1,bsst2,bst0} the
$N=8$ supersingleton field theory was assumed to be a quantum consistent
theory of the supermembrane, and its quantization was studied. In
particular, the spectrum of massless states was considered. In that
context, a well known and a remarkable property of the singletons was
used, namely the fact that the symmetric product of two supersingletons
yields an infinite tower of massless higher spin states \cite{ff1}. The
detailed form of this result for $OSp(8|4)$ will be spelled out in
section 3. It was argued in \cite{bsst1,bsst2,bst0} that all of these
states should arise in the quantum supermembrane theory!
Furthermore,
the occurrence of the infinitely many massless higher spin fields
implies the existence of infinitely many (local) gauge symmetries
analogous to the Yang-Mills, general coordinate and local
supersymmetries associated with spin 1, 2 and 3/2, respectively.

In a later development, in the course studying the $D=11$ supermembranes in
$AdS_4 \times S^7$, an attempt was made to obtain the singleton field theory
from the $D=11$ supermembrane action by expanding around $AdS_4 \times S^7$
\cite{mc1,mc2,dps}. However, for reasons explained in \cite{mc2} the $N=8$
singleton field theory on $S^2 \times S^1$ boundary of $AdS_4$ did not emerge
fully as the critical mass term for the boson was lacking. This problem has been
recently circumvented by embedding the membrane worldvolume in target space in
such a way that the resulting worldvolume field theory is a conformal field
theory on a three dimensional Minkowski space that serves as the boundary of
$AdS_4$ \cite{ktv}.

Interestingly enough, in a related development Fradkin and Vasiliev
\cite{fv0,fv1} were in the course of developing a higher spin gauge
theory in its own right (see \cite{fv0} for references to earlier
work). These authors succeeded in constructing interacting field
theories for higher spin fields. It was observed that the previous
difficulties in constructing higher spin theories can be bypassed by
formulating the theory in AdS space and to consider an infinite tower
of gauge fields controlled by various higher spin algebras based on
certain infinite dimensional extensions of super AdS algebras. In
particular, the AdS radius could not be taken to infinity since its
positive powers occurred in the higher spin interactions and therefore
one could not take a naive Poincar\'e limit.

In a series of papers Vasiliev \cites{v1}{v11} pursued the program of
constructing the $AdS$ higher spin gauge theory and simplified the
construction considerably. In \cite{v4,v6} the spin 0 and 1/2 fields
were introduced to the system within the framework of free differential
algebras. The theory was furthermore cast into an elegant geometrical
form in \cite{v7,v8,v9,v10} by extending the higher spin algebra to
include new auxiliary spinorial variables.

Remarkably, applying the formalism of Vasiliev to a suitable higher spin
algebra that contains the maximally extended super AdS algebra
$OSp(8|4)$, the resulting spectrum of gauge fields and spin $s\leq \ft12$ fields
coincide with the massless states resulting from the symmetric product
of two $OSp(8|4$ supersingletons! \cite{bsst1}

With recent observation of Maldacena \cite{jm} that there is a
correspondence between the physics in AdS space and at its boundary,
which has been successfully tested affirmatively in a number of examples
with very interesting results, the issue of whether the singletons of
$AdS_4$ could indeed play a role in the description of bulk physics has
again arisen. As far as the singleton field theory is concerned, as in
the past, we still expect all the massless higher spin states to arise
as two singleton states and the massive states to arise in the product
of three and more singletons. An interesting question to which we have
no answer at this time is how such states might arise in M-theory.
However, reviving the conjecture of \cite{bsst1,bsst2,bst0} we expect
that they will arise as a new phase of M-theory that is yet to be
uncovered.

The purpose of this paper is to address a more modest aspect of this
problem by examining the Vasiliev theory of higher spin fields (which is
applicable to a wide class of higher spin superalgebras) and determine
the precise manner in which the $N=8$ de Wit-Nicolai gauged supergravity
\cite{dn1,dn2}, which is the gauged version of the Cremmer-Julia $N=8$
supergravity in four dimensions \cite{cj1,cj2}, can be described within
this framework. We will indeed show how this embedding works in the
higher spin AdS supergravity based on the higher spin superalgebra known
as \alg algebra \cite{kv1,kv2}. (The details of this algebra will be
discussed in the next section). In our opinion, this constitutes a
positive step towards the understanding of the M-theoretic origin of the
massless higher spin gauge theory. After presenting our results, we will
comment on further aspects of this issue in the concluding section.

Since the techniques for constructing higher spin theories that we
shall employ in this paper are not all too well-known we have chosen to
make this paper as self-contained as possible by including some basic
features of the construction of higher spin theories. However, we would
like to stress that the main purpose of the paper is to argue that
higher spin supergravity plays a physical role within the context of
M-theory and in particular that AdS bulk/boundary could be instrumental
for bringing the subject of higher spin theories to a point where more
nontrivial issues are addressed than previously. Therefore, while the
formulation of higher spin theories allows for quite general extended
algebras without any obvious truncation to supergravity, we believe
that the extended higher spin theories that allow truncation to
supergravity - which we shall refer to as higher spin supergravities -
are of particular interest.

The organization of the paper is as follows: In section 2 we define the
gauge algebra \alg\ and discuss its infinite dimensional UIR's based on
the $N=8$ supersingleton. In section 3 and 4 we give the field content
of the theory and the form of the nonlinear higher spin field
equations. The perturbative treatment of the theory around the anti de
Sitter vacuum with $N=8$ supersymmetry is explained in section 5, where
we also derive the linearized equations of motion. In section 6 we
analyze the linear dynamics in more detail and in section 7 we treat
the particular case of the $N=8$ AdS supergravity. Section 8 is devoted
to a further discussion of our results and some speculations. Our
conventions and notation are collected in Appendices A and B. Some
details of the \alg algebra are given in Appendices C and some useful
lemmas are collected in Appendix D.

%%%%%%%%%%%%%%%%%%%%%%%%%%%%%%%%%%%%%%%%%%%%%%%%%%%%%%%%%%%%%%%%%%%%%%%

\section{The Higher Spin Superalgebra \alg}

%%%%%%%%%%%%%%%%%%%%%%%%%%%%%%%%%%%%%%%%%%%%%%%%%%%%%%%%%%%%%%%%%%%%%%%

The higher spin gauge algebra \alg\ is a Lie superalgebra which we shall
define as a subspace of an associative algebra ${\cal A}$ with Lie
bracket given by the ordinary commutator based on the associative
product of ${\cal A}$. ${\cal A}$ is the space of fully symmetrized
functions of the two-component, complex, Grassmann even spinor elements
$y^\a$, its complex conjugate $\yb^\ad$ (which together form a
four-component Majorana spinor $Y_{\una}=(y_{\a},\yb^{\ad})$ of
the AdS group $SO(3,2)$; our spinor conventions are given in
Appendix A) and $8$ real, Grassmann odd elements $\th^i$ forming
an $SO(8)$ vector\footnote{
%
%%%%%%%%%%footnote
%
In an equivalent formulation, the Grassmann odd $\th^i$ can be replaced
by Grassmann even $\c$-matrices $\C^i$ of the $SO(8)$ Clifford
algebra.}. These generating elements therefore obey the ``classical"
commutation relations

\bea
{}[ y_{\a},y_{\b}]&=&[\yb_{\ad},y_{\b}]\se
{}[\yb_{\ad},\yb_{\bd}]\se 0\ ,\qquad \a,\ad=1,2\nn\w2
{}[y_{\a},\th^{i}]&=&[\yb_{\ad},\th^{i}]\se0\ ,\nn\w2
\{\th^i,\th^j\} &=&0\ ,\qquad\qquad\qquad
\qquad\qquad\qquad\quad i=1,...,8\ , \qquad\la{ca}
\eea

and the reality conditions

\be
(y_{\a})^{\dagger}\se \yb_{\ad}\ ,\qquad
(\th^{i})^{\dagger}\se \th^{i}\ .\la{app7}
\ee

A general element of ${\cal A}$ can thus be written as a formal power
series

\be
F(y,\yb,\th\,) \se\sum_{\ba{l} m,n\geq 0 \\ k=0,...,8 \ea }
{1\over k! } ~F_{i_{1}\cdots i_{k}}(m,n)\,\th^{i_{1}\cdots i_{k}}\ ,
\la{F}
\ee

where we have used the notation

\bea
F_{i_{1}\cdots i_{k}}(m,n)& := & {1\over m!~n!}\, F_{\a_{1}\cdots \a_m
\bd_1\cdots \bd_n i_1 \cdots i_k}~ y^{\a_1\cdots \a_m}~\yb^{\bd_1\cdots
\bd_n}\ ,\nn\w2 y^{\a_1\cdots \a_m} & := & y^{\a_1} \cdots y^{\a_m}\
, \qquad
\yb^{\bd_1\cdots \bd_n}\;\;  := \;\; \yb^{\bd_1} \cdots \yb^{\bd_n}\ ,
\nn\w2
\th^{i_1\cdots i_k}& := & \th^{i_1}\cdots \th^{i_k}\ ,
\la{fmn}
\eea

where the coefficients $F_{\a_1\cdots \a_m \bd_1\cdots \bd_n i_1 \cdots
i_k}$ are Grassmann numbers carrying fully symmetrized spinor indices
and antisymmetrized $SO(8)$ vector indices. It is important to note that
the $y$ and $\yb$ dependence of \eq{F} is not restricted by $SO(3,2)$
invariance\label{foot1}\footnote{
%
%%%%%%%%%%%%% footnote %%%%%%%%%%%%
$SO(3,2)$ invariance requires a reality condition, such as
$(F(y,\yb,\th))^{\dagger}=F(y,\yb,\th)$, in which case $F(y,\yb,\th)$
can be expanded in terms of $SO(3,2)$ invariant monomials
$F_{\mi{\una}1m}(\th)Y^{\mi{\una}1m}$ where $F_{\mi{\una}1m}(\th)$ are
real functions of $\th$.}. For convenience we shall use the shorter
notation

\be
F(Y,\th\,)\;\; := \;\; F(y,\yb,\th\,)\la{conv}
\ee

for arbitrary elements in ${\cal A}$. The argument of $F$ will be
suppressed when there is no ambiguity in notation.

The associative algebra product $\star$ of ${\cal A}$ is defined by the
following $SO(3,2)\times SO(8)$ invariant "contraction rules"\footnote{
%
%%%%%%%%%%%footnote
%
The $\star$ product of functions $F(y)$ is isomorphic to the algebra of
fully symmetrized functions $F(\hat{y})$ of the Heisenberg algebra
$[\hat{y}_{\a},\hat{y}_{\b}]=2i\e_{\a\b}$, obtained from $F(y)$ by
replacing $y\rightarrow \hat{y}$; see Appendix C.}:

\bea
&& y_{\a_{1}\cdots\a_{m}}~\star ~y_{\b_{1}\cdots\b_{n}} \se y_{a_{1}\cdots
\a_{m}\b_{1}\cdots \b_{n}}+i\,m\,n\, ~y_{\a_{1}\cdots\a_{m-1}\b_{1}\cdots
\b_{n-1}}~\e_{\a_{m}\b_{n}}
\nn\w2
&& +i^2\,{m\,(m-1)n\,(n-1)\over 2!}~y_{\a_{1}\cdots\a_{m-2}\b_{1}\cdots
\b_{n-2}}~\e_{\a_{m-1}\b_{n-1}}~\e_{\a_{m}\b_{n}}+\cdots
\nn\w2
&&+i^{\,k}\,k!\,{m \choose k}{n \choose  k}
~y_{\a_{1}\cdots\a_{m-k}\b_{1}\cdots
\b_{n-k}}~\e_{\a_{m-k+1}\b_{n-k+1}}
\cdots \e_{\a_{m}\b_{n}}+\cdots\ ,
\la{yc}\w3
&& \left(y_{\mi\a1m}\yb_{\mi{\ad}1n}\right)\star
\left(y_{\mi\b1p}\yb_{\mi{\bd}1q}\right)\se
\left(y_{\mi\a1m}\star y_{\mi\b1p}\right)
\left(\yb_{\mi{\ad}1n}\star\yb_{\mi{\bd}1q}\right)\ ,
\la{yybstar}\w3
&& \th^{i_{1}\cdots i_{m}}\star\th_{j_{1}\cdots j_{n}}\se
\th^{i_{1}\cdots i_{m}}{}_{j_{1}\cdots j_{n}} +m\,n~
\th^{i_{1}\cdots i_{m-1}}{}_{j_{2}\cdots j_{n}}~\d^{i_{m}}_{j_{1}}
\nn\w2
&&+\fft{m\,(m-1)n\,(n-1)}{2!}~\th^{i_{1}\cdots i_{m-2}}{}_{j_{3}\cdots j_{n}}
~\d^{i_{m-1}i_{m}}_{j_{2}j_{1}}+\cdots
\nn\w2
&& +k!\,{m \choose k}{n \choose k}
~\th^{i_{1}\cdots i_{m-k}}{}_{j_{k+1}\cdots j_{n}}~\d^{i_{m-k+1}\cdots i_{m}}
_{j_{k}\cdots j_{1}}+\cdots\quad\ ,
\la{thc}
\eea

where we use the following {\it (anti-)symmetrization convention}: all
$SO(3,1)$ spinor indices of the  same type are automatically
subject to unit strength symmetrization, and all $SO(8)$ vector
indices are automatically subject to unit strength
anti-symmetrization. The $\star$ product rules \eqs{yc}{yybstar}
for commuting spinor elements can be summarized more concisely by
the following manifestly $SO(3,2)$ invariant formula

\be
F(Y\,)\star G(Y\,)\se  \int d^4U\,d^4V\,
F(Y+U\,)~G(Y+V\,)~
\exp i \left(u_{\a}v^{\,\a}+\ub_{\ad}\vb^{\,\ad}\right)
\ ,\la{star}\ee

where the normalization of the integration measure is such that
$1\star F=F$.

The $\star$ product rule \eq{thc} for the anti-commuting elements
$\th^{i}$ is equivalent to the decomposition rule for generalized Dirac
matrices of the (associative) $SO(8)$ Clifford algebra

\be
\th^{i}\star \th^{j}\se  \d^{ij}+\th^{i}\th^{j}\ .\la{tt}
\ee

It is also straight forward to verify the associativity of \eq{star}. We
also notice that as special cases of \eq{yc} we have

\be
y_{\a}\star y_{\b}\se \y_{\a}y_{\b}+i\,\e_{\a\b}\ ,\qquad
\yb^{\ad}\star \yb^{\bd}\se \yb^{\ad}\yb^{bd}+i\,\e^{\ad\bd}\ ,\la{spec}
\ee

which can be written in a manifestly $SO(3,2)$ covariant form

\be
Y_{\una}\star Y_{\unb}\se Y_{\una}Y_{\unb}+ i C_{\underline{\a\b}}\ .
\la{mso32}
\ee

The hermitian conjugation acts as an anti-linear, anti-involution of
the $\star$ algebra

\be
(F\star G)^{\dagger}\se (G)^{\dagger}\star (F)^{\dagger}\ ,\la{hc}
\ee

provided that we define $(\x\y)^{\dagger}=\y^{\dagger}\x^{\dagger}$ for
Grassmann odd quantities $\x$ and $\y$. In particular

\be
\left(\th^{i_{1}\cdots i_{k}}\right)^{\dagger}\se
(-1)^{\ft{k(k-1)}{2}} \th^{i_{1}\cdots i_{k}}\ .
\la{app8}
\ee

We next introduce the map

\be
\t(y_{\a})\se i~y_{\a}\ ,\quad \t(\yb_{\ad})\se i~\yb_{\ad}\ ,\quad
\t(\th^{i})\se i~\th^{i}\ .\la{deft}
\ee

The action of $\t$ on a general element in ${\cal A}$ is then
defined by declaring $\t$ to be an involution of the classical
product \eq{ca}. From the definition of the $\star$ algebra in
\eq{thc} and \eq{star} it follows that $\t$ is a graded
anti-involution of ${\cal A}$:

\be
\t(F\star G)\se (-1)^{FG}\,\t(G)\star \t(F)\ ,\qquad
F,\ G\,\in{\cal A}\ ,\la{tfg}
\ee

where the indices in the exponent indicate Grassmann parities.

We are now ready to define \alg\ as the Lie superalgebra given by the
subspace of ${\cal A}$ spanned by Grassmann {\it even} elements
$P$ in ${\cal A}$ obeying the conditions \cite{kv2}

\be
\t(P)\se -P\ ,\quad\quad  P^{\dagger}\se-P\ ,\la{pdef}
\ee

and with Lie bracket defined by

\be
[\,P\,,\,Q\,]_{\star}\se P\star Q - Q\star P\ .\la{lie}
\ee

From \eq{hc} and \eq{tfg} it follows that \alg\ is closed under the
bracket. The general solution to \eq{pdef} is

\bea
P(Y,\th) \se \fft{1}{2i}\sum_{k=0}^{\infty} &{}
&\!\!\!\!\!\!\!\!\!\!\!\!\left(
\quad\;\sum_{m+n=4k}\;\; \left(\ft1{2!} P_{ij}(m,n)\th^{ij}
+\ft1{6!}P_{i_{1}\cdots i_{6}}(m,n)\th^{i_{1}\cdots i_{6}} \right)
\right. \nn\w2
&+&\sum_{m+n=4k+1}\left( P_{i}(m,n)\th^{i}
+\ft1{5!}P_{i_{1}\cdots i_{5}}(m,n)\th^{i_{1}\cdots i_{5}}\right)\nn\w2
&+&\sum_{m+n=4k+2}\left( P(m,n)
+\ft1{4!}P_{i_{1}\cdots i_{4}}(m,n)\th^{i_{1}\cdots i_{4}}
+\ft1{8!}P_{i_{1}\cdots i_{8}}(m,n)\th^{i_{1}\cdots i_{8}}\right)\nn\w2
&+&\left. \sum_{m+n=4k+3}\left( \ft1{3!}P_{ijk}(m,n)\th^{ijk}
+\ft1{7!}P_{i_{1}\cdots i_{7}}(m,n)\th^{i_{1}\cdots i_{7}}\right)\right)
\ ,\la{p}\w4
\left(P_{\mi\a1m\mi{\bd}1n\mi i1k}\right)^{\dagger} &=&
(-1)^{\ft{k\,(k-1)}{2}}~P_{\mi\b1n\mi{\ad}1m\mi i1k}\ ,
\la{realpr}
\eea

where we have used the notation defined in \eq{fmn}. The Grassmann
parity of $P_{\mi\a1m\mi{\bd}1n\mi i1k}$ is $(m+n)$ mod $2$. Hence the
bosonic (fermionic) fields carry an even (odd) number of spinor indices
and even-rank (odd-rank) antisymmetric tensor representations of
$SO(8)$, that is, the $1$, $28$ and $70=35_++35_-$ representations
(the $8$ and $56$ representations)
\footnote{
%%%%%%%%%%%%%%%%%%%%%%%% footnote  %%%%%%%%%%%%%%%%%%%%%%%%%%%%%%%%%%
The reality condition in \eq{pdef} implies that the expansion \eq{p} can
be expressed in terms of $Y_{\underline{\a}}$ (see footnote on p.
\pageref{foot1}). For example $\sum_{m+n=4k+2}P(m,n)\rightarrow
\sum_{k}P(4k+2)$ where $P(4k+2)$ is a $(4k+2)$'th order homogeneous
polynomial in $Y_{\underline{\a}}$. Also note the $SO(3,2)$ invariance
of the map $\t$, which multiplies $Y_{\una}$ by a factor of $i$.}.
%%%%%%%%%%%%%%%%%%%%%%%%%%%%%%%%%%%%%%%%%%%%%%%%%%%%%%%%%%%%%%%%%%%%%

The expansion \eq{p} is a sum of homogeneous polynomials
$P_{4k+2}(Y,\th)$ of degree $4k+2$, that is

\be
P_{4k+2}(\l Y,\l\th)\se \l^{4k+2}P_{4k+2}( Y,\th)
\ ,\qquad k=0,1,2,...\ ,\la{hompoly}
\ee

where $\l$ is a complex number. Hence

\be
shs^E(8|4)\se \bigoplus _{k=0}^{\infty} L_{k}\ ,\la{lk}
\ee

where the $k$'th ``level" $L_{k}$ is the vector space spanned by the polynomials
$P_{4k+2}$. The $\star$ commutator has the following schematical
structure:

\be
{}[\,L_{k}\,,\,L_{l}\,]\se \bigoplus_{m=|k-l|}^{k+l} L_{m}\ .
\la{pkl}
\ee

Notice that since $P$ is Grassmann even, the $\star$ commutator of two elements in
\alg\ receives contributions only from odd numbers of contractions
between the generators.

Eq.\,\eq{pkl} shows that $L_{0}$ form a closed subalgebra of \alg. In
fact $L_{0}\simeq OSp(8|4)$ is the maximal finite dimensional
subalgebra of \alg\ containing $SO(3,2)\simeq Sp(4,R)$ (see Appendix B
for conventions and normalizations). The $SO(8)$ subalgebra of $L_0$ is
the diagonal subalgebra of the $SO(8)_{+}\times SO(8)_{-}$ subalgebra
of \alg\ with generators $\th^{ij}\star
\ft12(1\pm\C)$, where $\C$ is the hermitian and $\t$-invariant $SO(8)$
chirality operator

\be
\C\se \th^1\cdots \th^8\ ,\qquad \{\,\C\,,\,\th^i\,\}_{\star}\se0\ ,\qquad
\C^2\se 1\ ,\qquad (\C)^{\dagger}\se
\t(\C)\se \C\ .\la{chop}
\ee

Thus the higher spin gauge theory based on \alg\ contains the gauge
fields of $SO(8)_{+}\times SO(8)_{-}$ while the truncation of the
theory to the gauged $N=8$ supergravity theory contains the gauge
fields of the diagonal $SO(8)$.

For $l=0$ and arbitrary $k$, \eq{pkl} shows that $L_{k}$ forms an
$OSp(8|4)$ irreducible representation spanned schematically by the
polynomials
\be
Y^{\,4k+2}\, ,\ Y^{\,4k+1}\th\, ,\ ...\, ,\ Y^{\,4k-6}\th^{\,8}
\ ,\la{yyy}
\ee

where $Y^{m}\th^{n}$ denotes a hermitian monomial of $(y,\yb)$ and
$\th^{i}$ of homogeneous degree $m$ and $n$ respectively. The space
spanned by $Y^{m}\th^{n}$ forms an irreducible representations of
$SO(3,2)\times SO(8)$ with $SO(3,2)$ spin $s=\ft{m}2$ and real dimension
${m+3\choose 3}{8\choose n}$. (The spin is the eigenvalue of the
$SO(3,2)$ generator $M_{12}$). Hence, for $k\geq 1$ the real dimension
of the subspace of $L_{k}$ spanned by bosonic generators is given by

\be
N^{b}_k\se \sum_{n=0}^{4}{4k-3+2n\choose 3}{8\choose 2n}\ ,\la{nb}
\ee

and the real dimension of the subspace of $L_{k}$ spanned by fermionic
generators is given by

\be
N^{f}_k\se \sum_{n=0}^3{4k-2+2n\choose 3}{8\choose 2n+1}
\ .\la{nf}
\ee

As expected

\be
\ft12 \mbox{dim}_{\mbox{{\small{\bf R}}}}
(L_{k})\se N^b_k\se N^f_k\se
\ft{256}{3}k(5+16k^2)\ , \qquad k=1,2,...\ .\la{nbnf}
\ee

The dimension grows rapidly; for example $N^b_1=N^f_1=1,792$ and
$N^b_2=N^f_2=11,776$.

In gauging \alg\ the physical states coming from the gauge fields (see
section 6 for a detailed spectral analysis) are given in the $s\geq 1$
entries of Table 1 \cite{bsst1,bsst2,kv1,kv2}. At any level $k\geq 2$,
the $512$ physical states described by the gauge fields form
$N=8$ supermultiplets (consisting of two irreducible $128+128$ sub-multiplets
related by CPT conjugation such that the full multiplet is CPT
invariant). At level $k=0,1$, however, spin $ s \leq 1/2 $ states are
needed to form supermultiplets. The $k=1$ supermultiplet also has a
total of
$516$ physical states, while the $k=0$ multiplet, which is
CPT-conjugate, contains a total of $256$ physical states.

{\footnotesize
\tabcolsep=1mm
\begin{table}[t]
\begin{center}
\begin{tabular}{|c|cc|cccccccccccccc|}\hline
& & & & & & & & & & & & & & & &\\
{\large${}_{k}\backslash s$} & $0$ & \ns{$\ft12$} & $1$
& \ns{$\ft32$} & $2$ &
\ns{$\ft52$} &
$3$ & \ns{$\ft72$} & $4$ & $\cdots$ & $2s$ & $2s+$\ns{$\ft12$} & $2s+1$ &
$2s+$\ns{$\ft32$}
& $2s+2$ & $\cdots$ \\
& & & & & & & & & & & & & & & &\\ \hline
& & & & & & & & & & & & & & & &\\
$0$ & $35_{\smpl}\!+\!35_{\smm}$ & $56$ & $28$ & $8$ & $1$ &
& & & & & & & & & &\\
$1$ & $1\!+\!1$ & $8$ & $28$ & $56$ & $35_{\smpl}\!+\!35_{\smm}$
& $56$ & $28$ & $8$ & $1$ &
& & & & & & \\
$2$ & & & & & $1$ & $8$ & $28$ & $56$ & $35_{\smpl}\!+\!35_{\smm}$
& $\cdots$
& & & & & & \\
$\vdots$ & & & & & & & & & & & & & & & & \\
$s-1$ & & & & & & & & & & $\cdots$ & $1$ & & & & & \\
$s$ & & & & & & & & & & $\cdots$ & $35_{\smpl}\!+\!35_{\smm}$
& $56$ & $28$ & $8$ & $1$
& $\cdots$\\
$s+1$ & & & & & & & & & & & $1$ & $8$ & $28$ & $56$
& $35_{\smpl}\!+\!35_{\smm}$ & $\cdots$\\
$\vdots$ & & & & & & & & & & & & & & & & \\ \hline
\end{tabular}
\end{center}
\caption{{\small The $SO(3,2)\times SO(8)$ content of the symmetric tensor
product of two $N=8$ singletons. This product is a unitary irreducible
representation (UIR) of \alg\ which decomposes into infinitely many
$OSp(8|4)$ supermultiplets labeled by the level number $k$ defined in
\eq{lk}. The states with $s \geq 1$ span the spectrum of the \alg\ gauge
fields. }}
\la{table}
\end{table}
}

Physical consistency of a gauge theory built on \alg\ actually requires
that the complete particle spectrum forms a unitary representation of
the full, infinite dimensional algebra \alg\ \cite{kv1}; see end of
section 5.1. Not all higher spin algebras are admissible in this sense.
The admissibility of \alg\ is based on the fact that the particle
spectrum of the \alg\ gauge field (that will be analyzed in detail in
section 6) fits into the symmetric tensor product of two
$OSp(8|4)$ singleton supermultiplets \cite{bsst1,kv1,kv2}. Each
singleton supermultiplet consists of a singleton $Rac$ in the $8_{s}$
representation of $SO(8)$ and a singleton $Di$ in the $8_{c}$
representation:

\be
Rac \oplus Di\se \left[ D(\ft12,0)\otimes 8_{s}\right] \oplus
\left[ D(1,\ft12)\otimes 8_{c}\right]\ ,\la{dirac}
\ee

where $D(E_{0},s)$ denotes an UIR of $SO(3,2)$ for which $E_{0}$ is the
minimal energy eigenvalue of the energy operator $M_{04}$, and $s$ is
the maximum eigenvalue of the spin operator $M_{12}$ in the lowest
energy sector. From the oscillator representation of \alg\ given in
Appendix C it follows that the space \eq{dirac} actually forms a UIR of
\alg. Therefore also the symmetric and the anti-symmetric tensor
products of two copies of the space \eq{dirac} form UIR's of \alg. In
particular, the decomposition of the symmetric tensor product

\bea
&& \left[ \left(D(\ft12,0)\otimes 8_s\right)
\otimes \left(D(\ft12,0)\otimes 8_s\right)\right]_S =
\nn\w2
&&
\sum_{s=0,2,4,...}\left[D(s+1,s)\otimes 1 + D(s+1,s)\otimes 35_+\right]
+ \sum_{s=1,3,5,...}D(s+1,s)\otimes 28
\nn\w4
&& \left[ \left(D(1,\ft12)\otimes 8_c\right)
\otimes \left(D(1,\ft12)\otimes 8_c\right)\right]_S =
\nn\w2
&&
\sum_{s=0,2,4,...}\left[D(s+1,s)\otimes 1 + D(s+1,s)\otimes 35_-\right]
+\sum_{s=1,3,5,...}D(s+1,s)\otimes 28
\nn\w4
&& \left[ \left(D(\ft12,0)\otimes 8_s\right)
\otimes \left(D(1,\ft12)\otimes 8_c\right)\right]_S =
\nn\w2
&&
\sum_{s=\ft12,\ft32,\ft52,...}\left( D(s+1,s)\otimes 8_v
+ D(s+1,s)\otimes 56_v\right)
\la{ssd}
\eea

under the $OSp(8|4)$ subalgebra of \alg\ leads to the UIR's of
$OSp(8|4)$ given in Table 1. There are a number of ways to derive this
result. See, for example, \cite{bsst1}. Another derivation is given in
Appendix C (see \eq{diractp} and \eq{so8tp}).

We thus expect \alg\ to be a suitable algebra for a supergauge theory
of higher spin $AdS$ supergravity. However, the determination of the
particle content of the \alg\ gauge theory requires a lot more
analysis; there are auxiliary gauge fields which must be eliminated
algebraically in favor of the true dynamical gauge fields and Table 1
makes it clear that one has to couple the gauge fields to a finite set
of fields with spin $s \leq \ft12$.

%%%%%%%%%%%%%%%%%%%%%%%%%%%%%%%%%%%%%%%%%%%%%%%%%%%%%%%%%%%%%%%%%%%%%%

\section{The Field Content}

%%%%%%%%%%%%%%%%%%%%%%%%%%%%%%%%%%%%%%%%%%%%%%%%%%%%%%%%%%%%%%%%%%%%%%

\subsection{The \alg\ Connection}

%%%%%%%%%%%%%%%%%%%%%%%%%%%%%%%%%%%%%%%%%%%%%%%%%%%%%%%%%%%%%%%%%%%%%

The one-form master gauge connection

\be
\o\,(Y,\th)\se dx^{\m}\o_{\m}(Y,\th)\la{gf}
\ee

is by definition Grassmann even and takes its values in the algebra
\alg:

\be
\t(\o)\se -\o\ ,\quad\quad (\o)^{\dagger}\se -\o\ .\la{w}
\ee

The curvature of $\o$ is the \alg-valued $2$-form is defined
by\footnote{
%
%%%%%%%%%%%% footnote %%%%%%%%
We shall temporarily set the gauge coupling equal to one. We shall
discuss the gauge coupling and the gravitational coupling in
section 7. Also note that in the spin $s=2$ sector one must not confuse
the $SO(3,1)$ valued curvature $R_{\m\n,\a\b}$ with the Riemann tensor
$r_{\m\n,\a\b}$. The latter is by definition the curvature of the
$SO(3,1)$ gauge field $\o_{\m,\a\b}$, while
$R_{\m\n,\a\b}$ contains contributions from
$\o\star\o$ that are bilinear in gauge fields
corresponding to generators whose commutators contain $SO(3,1)$ generators.}

\bea
R&=& d\,\o-\o\star\o\ ,\qquad R\;\; := \;\;
 \ft12 dx^{\m}\wedge dx^{\n} R_{\m\n}\ ,\nn\w2
dR&=& \o\star R- R\star \o\ ,
\la{r}
\eea

where $d=dx^{\m}\del_{\m}$ is the exterior derivative and the wedge
products are suppressed. The curvature transforms covariantly under
\alg\-valued gauge transformations

\be
\d_{\vare}\,\o \se d\,\vare -[\,\o\,,\,\vare\,]_{\star}\ ,\qquad
\d_{\vare}\,R\se [\,\vare\,,\,R\,]_{\star}\ ,\la{dw}
\ee

where the gauge parameter $\vare$ is an arbitrary \alg-valued function

From \eq{p} it follows that $\o$ has the expansion

\bea
\o(Y,\th) \se \fft{1}{2i}\sum_{k=0}^{\infty} &{}
&\!\!\!\!\!\!\!\!\!\!\!\!\left(
\quad\;\sum_{m+n=4k}\;\; \left(\ft1{2!} \o_{ij}(m,n)\th^{ij}
+\ft1{6!}\o_{i_{1}\cdots i_{6}}(m,n)\th^{i_{1}\cdots i_{6}} \right)
\right. \nn\w2
&+&\sum_{m+n=4k+1}\left( \o_{i}(m,n)\th^{i}
+\ft1{5!}\o_{i_{1}\cdots i_{5}}(m,n)\th^{i_{1}\cdots i_{5}}\right)\nn\w2
&+&\sum_{m+n=4k+2}\left( \o(m,n)
+\ft1{4!}\o_{i_{1}\cdots i_{4}}(m,n)\th^{i_{1}\cdots i_{4}}
+\ft1{8!}\o_{i_{1}\cdots i_{8}}(m,n)\th^{i_{1}\cdots i_{8}}\right)\nn\w2
&+&\left. \sum_{m+n=4k+3}\left( \ft1{3!}\o_{ijk}(m,n)\th^{ijk}
+\ft1{7!}\o_{i_{1}\cdots i_{7}}(m,n)\th^{i_{1}\cdots i_{7}}\right)\right)
\ . \la{wexp}
\eea

Note that the bosonic gauge fields are always in the $1$, $28$ and
$35_++35_-$ representations of $SO(8)$, while the fermionic fields are
always in $8$ and $56$ representations.

By defining $\s^{a}(1,1) := (\s^a)_{\a\ad} y^\a\yb^{\ad}$,
$\s^{ab}(2,0) :=  \ft12(\s^{ab})_{\a\b}y^{\a\b}$ and
$\sb^{ab}(0,2) :=  \ft12(\sb^{ab})_{\ad\bd}\yb^{\ad\bd}$, we can decompose
the gauge field $\o_{\m,\mi{i}1k}(m,n)$ and its curvature
$R_{\m\n,\mi{i}1k}(m,n)$ into irreducible Lorentz tensors $\y_{\mi
i1k}(p,q)$ of spin $s=\ft{p+q}{2}$ by expanding the following $\star$
products

\bea
\s^a (1,1) \star \o_{a,\,\mi i1k}(m,n)& := &
\z_{\mi i1k}(m+1,n+1)+\z_{\mi i1k}(m+1,n-1)
\nn\w2
&&+\z_{\mi i1k}(m-1,n+1)+\z_{\mi i1k}(m-1,n-1)\ ,
\nn\w2
\s^{ab}(2,0)\star R_{ab,\,\mi i1k}(m,n)& := &
\x_{\mi i1k}(m+2,n)+\x_{\mi i1k}(m,n)+\x_{\mi i1k}(m-2,n)\ ,
\nn\w2
\sb^{ab}(0,2)\star R_{ab,\,\mi i1k}(m,n)& := &
\y_{\mi i1k}(m,n+2)+\y_{\mi i1k}(m,n)+\y_{\mi i1k}(m,n-2)\ ,
\la{decomp}
\eea

where the curved index $\m$ has been converted into a flat index $a$ by
means of the vierbein $\o_{\m,a}$ (which is given in terms of the gauge
field $\o_{\m,\a\ad}$ by \eq{appb}). If $m\neq n$ then these irreps are
complex. If $m=n$ then $\z_{\mi i1k}(m+1,m+1)$ and $\z_{\mi
i1k}(m-1,m-1)$ are real or purely imaginary (depending on $k$), while
$\z_{\mi i1k}(m-1,m+1)$ and $\z_{\mi i1k}(m+1,m-1)$ and $\x_{\mi i1k}(p,m)$
and $\y_{\mi i1k}(m,p)\ (p=m,m\pm 2)$ are related by hermitian
conjugation. The chiral components $\x_{\mi i1k}(2s,0)$ and
$\y_{\mi i1k}(0,2s)$ ($s=1,\ft32,2,...$) are called the {\it
generalized Weyl tensors}.

\bigskip\bigskip
%%%%%%%%%%%%%%%%%%%%%%%%%%%%%%%%%%%%%%%%%%%%%%%%%%%%%%%%%%%%%%%%%%%%%%%%%%%
\centerline{{\it Auxiliary and Dynamical Gauge Fields and Gauge
Symmetries}}
%%%%%%%%%%%%%%%%%%%%%%%%%%%%%%%%%%%%%%%%%%%%%%%%%%%%%%%%%%%%%%%%%%%%%%%%%%
\bigskip

It is important to notice that the number of algebraically independent
components of the level $L_k$ gauge fields, as given by \eq{nbnf} for
$k\geq 1$, grows rapidly with $k$, while the corresponding
supermultiplets in Table 1 each contain $256+256$ states. In fact,
gauge symmetry alone cannot remove all the unphysical states. One also
has to impose some (curvature) constraints that serve to determine
algebraically a certain subset of the gauge fields, known as {\it
auxiliary gauge fields}, in terms of the remaining, {\it dynamical
gauge fields}, as well as to yield dynamical field equations for the
latter. (The curvature constraints also incorporate the spacetime
diffeomorphisms into the gauge group). The constraints actually involve
all curvature components except the generalized Weyl tensors
$\x_{\mi i1k}(2s,0)$ and $\y_{\mi i1k}(0,2s)$ ($s=1,\ft32,2,...$)
defined in \eq{decomp}. As a result only the gauge fields $\o(m,n,\th)$
with $|m-n|\leq 1$ are dynamical, as will be shown in section 6.1; see
also Figure \ref{wfig}.
\bigskip
%%%%%%%%%%%%% figure: field content of \o

\def\bw#1#2{\put(#1,#2){\circle*{3}}}
\def\fw#1#2{\put(#1,#2){\circle{3}}}
\def\lc#1#2{\put(#1,#2){$\diamondsuit$}}
\def\aux#1#2{\put(#1,#2){{\large $\times$}}}
\def\cdo#1#2{\put(#1,#2){{\huge $......$}}}
\def\vdo#1#2{\put(#1,#2){{\Huge $\vdots$}}}

\begin{figure}[!h]
\begin{center}
\unitlength=.6mm
\begin{picture}(120,120)(0,-10)
\put(-2,-1.5){{\Large $\star$}}
\bw{10}{10}\bw{20}{20}\bw{30}{30}\bw{40}{40}\bw{50}{50}
\bw{60}{60}\bw{70}{70}\bw{80}{80}
\fw{0}{10}\fw{10}{20}\fw{20}{30}\fw{30}{40}\fw{40}{50}
\fw{50}{60}\fw{60}{70}\fw{70}{80}
\fw{10}{0}\fw{20}{10}\fw{30}{20}\fw{40}{30}\fw{50}{40}
\fw{60}{50}\fw{70}{60}\fw{80}{70}
\lc{18}{-2}\lc{28}{8}\lc{38}{18}\lc{48}{28}\lc{58}{38}
\lc{68}{48}\lc{78}{58}
\lc{-2}{18}\lc{8}{28}\lc{18}{38}\lc{28}{48}\lc{38}{58}
\lc{48}{68}\lc{58}{78}
\aux{-2.5}{28.5}\aux{-2.5}{38.5}\aux{-2.5}{48.5}\aux{-2.5}{58.5}
\aux{-2.5}{68.5}\aux{-2.5}{78.5}
\aux{7.5}{38.5}\aux{7.5}{48.5}\aux{7.5}{58.5}\aux{7.5}{68.5}\aux{7.5}{78.5}
\aux{17.5}{48.5}\aux{17.5}{58.5}\aux{17.5}{68.5}\aux{17.5}{78.5}
\aux{27.5}{58.5}\aux{27.5}{68.5}\aux{27.5}{78.5}
\aux{37.5}{68.5}\aux{37.5}{78.5}
\aux{47.5}{78.5}
\aux{27.5}{-1.5}\aux{37.5}{-1.5}\aux{47.5}{-1.5}\aux{57.5}{-1.5}
\aux{67.5}{-1.5}\aux{77.5}{-1.5}
\aux{37.5}{8.5}\aux{47.5}{8.5}\aux{57.5}{8.5}\aux{67.5}{8.5}\aux{77.5}{8.5}
\aux{47.5}{18.5}\aux{57.5}{18.5}\aux{67.5}{18.5}\aux{77.5}{18.5}
\aux{57.5}{28.5}\aux{67.5}{28.5}\aux{77.5}{28.5}
\aux{67.5}{38.5}\aux{77.5}{38.5}
\aux{77.5}{48.5}
\vdo{-.5}{90}\vdo{-.5}{94.4}
\vdo{9.5}{90}\vdo{9.5}{94.4}
\vdo{19.5}{90}\vdo{19.5}{94.4}
\vdo{29.5}{90}\vdo{29.5}{94.4}
\vdo{39.5}{90}\vdo{39.5}{94.4}
\vdo{49.5}{90}\vdo{49.5}{94.4}
\vdo{59.5}{90}\vdo{59.5}{94.4}
\vdo{69.5}{90}\vdo{69.5}{94.4}
\vdo{79.5}{90}\vdo{79.5}{94.4}
\cdo{90}{80}\cdo{90}{70}\cdo{90}{60}\cdo{90}{50}\cdo{90}{40}
\cdo{90}{30}\cdo{90}{20}\cdo{90}{10}\cdo{90}{0}
\put(-7,110){$n$}
\put(115,-8){$m$}
\put(0,103){\vector(0,1){20}}
\put(110,0){\vector(1,0){20}}
\end{picture}
\end{center}
\caption{{\small Each entry of the integer grid, $m,n=0,1,2,...$, represents a
component $\o(m,n;\th\,)$ of the \alg\ valued connection one form $\o$
according to the following rule: the $\star$ denote the spin $s=1$
component; a {\Large $\bullet$} denotes a generalized vierbein; a
{\Large $\circ$} denotes a generalized gravitino; a $\diamondsuit$
denotes an auxiliary generalized Lorentz connection and the $\times$'s
denote the remaining auxiliary connections. }}
\la{wfig}
\end{figure}
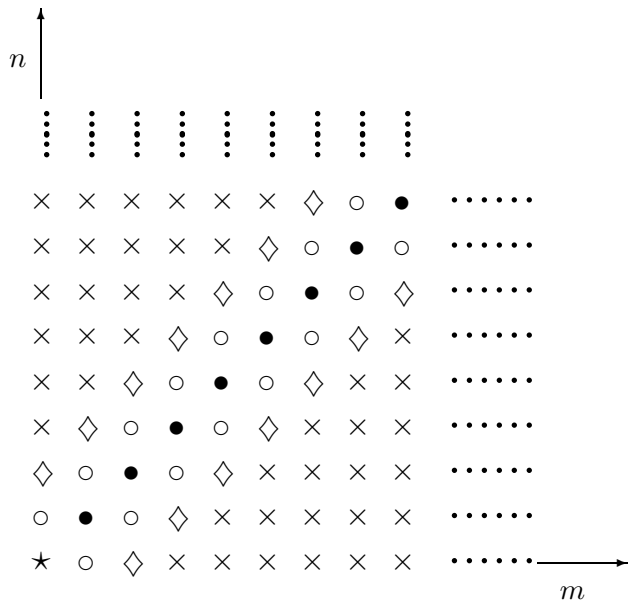
\bigskip

The auxiliary gauge fields (denoted by $\times$'s and
$\diamondsuit$'s in Figure \ref{wfig}) are

\be
\o_{a}(m,n,\th\,)\ ,\qquad |m-n|\geq 2\la{aux1}\ ,
\ee

and their hermitian conjugates. Notice that there are no auxiliary
gauge fields with spin $s\leq \ft32$. In particular we shall refer to
the auxiliary gauge fields $\o_{a}(s-2,s,\th\,)$ ($s=2,3,...$) and
their hermitian conjugates as the {\it generalized Lorentz connections}
(and these are denoted by $\diamondsuit$'s in Figure \ref{wfig}).

As for the dynamical gauge fields, we categorize them as follows:

\begin{itemize}

\item[$i$)] the {\it generalized vierbeins} (denoted by {\Large $\bullet$}'s in
Figure \ref{wfig}):

\be
\o_{a}(s-1,s-1,\th\,)\ ,
\qquad s\se 2,3,4,...\ , \la{gvb}
\ee

which are real,

\item[$ii$)] the {\it generalized gravitini} (denoted by {\Large
$\circ$}'s in Figure \ref{wfig}):

\be
\o_{a}(s-\ft32,s-\ft12,\th\,)\ ,
\qquad s\se \ft32,\ft52,\ft72,...\ ,\la{ggr}
\ee

and their hermitian conjugates, and

\item[$iii)$] the two spin $s=1$ $SO(8)$ gauge fields (denoted
by $\star$ in Figure \ref{wfig}
):

\be
\o_{ij}(0,0)\mbox{  and   } \o_{\mi{i}16}(0,0)\ .\la{so8}
\ee

\end{itemize}

We also differentiate between dynamical and auxiliary gauge symmetries.
A {\it dynamical gauge symmetry} by definition has a nontrivial action
on a dynamical gauge field while an {\it auxiliary gauge symmetry} does
not act on any of the dynamical gauge fields. The dynamical gauge
symmetries therefore constitute the local symmetry algebra of the
dynamical equations of motion\footnote{
%
%
%%%%%% footnote %%%%%
By definition the commutator algebra of dynamical gauge transformations
closes on the dynamical gauge fields.}.
The auxiliary gauge symmetries generate algebraic shifts in the Lorentz
irreps of the auxiliary gauge fields that are not determined in terms
of the dynamical gauge fields by solving the constraints. The auxiliary
gauge symmetries can therefore be fixed uniquely by fixing a gauge
where the undetermined irreps are set equal to zero such that the
auxiliary gauge fields are given uniquely in terms of the dynamical
gauge fields.

The auxiliary gauge symmetries have parameters

\be
\vare(m,n,\th\,)\ ,\qquad |m-n|\geq 4\la{auxgauge}
\ee

and hermitian conjugates. As for the dynamical gauge symmetries, in
analogy with \eqs{gvb}{so8}, we separate them into:

\begin{itemize}

\item[$i$)] the {\it generalized Lorentz transformations}
with parameters

\be
\vare(s-2,s,\th\,)\ ,\qquad s=2,3,...\ ,\la{genlor}
\ee

and their hermitian conjugates,

\item[$ii$)] the {\it generalized reparametrizations}
with parameters

\be
\vare(s-1,s-1,\th\,)\ ,\qquad s=1,2,...\ ,\la{genrep}
\ee

which are real,

\item[$iii$)] the {\it local fermionic transformations} with parameters

\be
\vare(s-\ft52,s+\ft12,\th\,)\ ,\qquad s=\ft52,\ft72,...\ ,\la{genft}
\ee

and their hermitian conjugates,

\item[$iv$)] the {\it generalized local supersymmetries}
with parameters

\be
\vare(s-\ft32,s-\ft12,\th\,)\ ,\qquad s=\ft32,\ft52,...\ ,\la{gensusy}
\ee

and their hermitian conjugates, and

\item[$v$)] the two $SO(8)$ gauge symmetries with parameters
$\vare_{ij}(0,0)$ and $\vare_{\mi{i}16}(0,0)$.

The local fermionic transformations \eq{genft} are the fermionic
analogs of the generalized Lorentz transformations \eq{genlor}. The
role of all these symmetries in arriving at the correct number of
degrees of freedom will be analyzed in detail in section 6.

\end{itemize}

%%%%%%%%%%%%%%%%%%%%%%%%%%%%%%%%%%%%%%%%%%%%%%%%%%%%%%%%%%%%%%%%%%%%%%

\subsection{The Quasi-adjoint Representation}

%%%%%%%%%%%%%%%%%%%%%%%%%%%%%%%%%%%%%%%%%%%%%%%%%%%%%%%%%%%%%%%%%%%%%%

As explained at the end of section 2, the construction of a unitary
gauge theory based on \alg\ requires the inclusion of the spin $s\leq
\ft12$ sector shown in Table 1. To this end, one introduces a Grassmann even,
zero-form master field $\f(Y,\th)$ \cite{v10,v9,v8,v7,v6} in the
following infinite dimensional representation of \alg:

\be
\t\,(\f\,)\se\bar{\pi}\,(\f\,)\ ,\quad\quad
(\f\,)^{\dagger}\se\pi\,(\f\,)\star\C\ ,\la{Adef}
\ee

where $\C$ is the $SO(8)$ chirality operator given in \eq{chop} and $\p$ and
$\bp$ are involutions of the classical algebra product \eq{ca} defined by

\be
\begin{array}{rclrclrcl}
\pi\,(y_{\a})&=&-y_{\a}&\quad \pi\,(\yb_{\ad})&=&\yb_{\ad}&\quad
\pi\,(\th^{i})&=&\th^{i}\w2
\bar{\pi}\,(y_{\a})&=&y_{\a} &\quad\bar{\pi}\,(\yb_{\ad})&=& -\yb_{\ad}&
\quad\bar{\pi}\,
(\th^{i})&=& \th^{i}\ .
\end{array}
\la{defp}
\ee

In addition, it is useful to define the classical algebra involution

\be
\pi_{\th}(y_{\a})\se y_{\a} \ ,\qquad \pi_{\th}(\yb_{\ad})\se\yb_{\ad}\ ,\qquad
\pi_{\th}(\th^{i})\se-\th^{i}\ .\la{pth}
\ee

The maps in \eqs{defp}{pth} are also involutions of the $\star$ algebra:

\be
\pi(F\star G\,)\se \pi(F\,)\star \pi(G\,)\ ,\qquad \mbox{idem}\
\bp,\ \pi_{\th}\ ,\la{pinv}
\ee

where $F$ and $G$ are arbitrary elements of ${\cal A}$.

The representation of \alg\ on the master field $\f$ is given by the
gauge transformation

\be
\d_{\vare}\, \f\se \vare\star\f -\f \star\bar{\pi}\,(\vare\,)\ ,\la{cg}\ ,
\ee

where $\vare$ is an \alg-valued function. The covariant derivative

\be
D_{\o}\f \se d\f-\o\star \f + \f \star\bar{\pi}\,(\o\,)\ .\la{delc}
\ee

transforms as

\be
\d_{\vare}\,D_{\o}\f \se \vare\star D_{\o}\f -D_{\o}\f \star\bar{\pi}\,(\vare\,)
\la{cd}
\ee

Notice that $\f$ does not quite transform in the adjoint representation
due to the presence of the $\bp$-operation in \eq{cg}. For this reason
we shall refer to the \alg\ representation carried by $\f$ as the {\it
quasi-adjoint} representation\footnote{
%
%%%%%%%%%%% footnote
The involutions $\pi$ and $\bp$ are not $SO(3,2)$ invariant as opposed
to the anti-involution $\t$. Hence, unlike $\o$, we cannot express $\f$
in terms of a Majorana spinor of $SO(3,2)$. As a consequence, the full
theory does not possess the external $SO(3,2)$ symmetry of \alg; see
the discussion at the end of section 4.3.}.
The closure of \alg\ on $\f$ follows from \eq{pinv}:

\be
[\,\d_{\vare_{1}}\,,\,\d_{\vare_{2}}\,]\f\se
\d_{[\,\e_{2}\,,\,\vare_{1}\,]_{\star}}\,\f\ .
\ee

Let us emphasize that the main reason for the introduction of $\C$,
$\pi$ and $\bp$ in the definition \eq{Adef} of the quasi-adjoint
representation is to ensure that its spin $s\leq \ft12$ sector
correctly produces the spin $s \leq \ft12$ states that arise from the
two-singleton states tabulated in Table 1 and that must be included in
the theory in order to satisfy the unitarity requirement discussed at
the end of section 2. The reality condition in \eq{Adef}, which
involves $\C$ in a crucial way, is engineered to be consistent with the
$\tau\bp$ invariance condition imposed on $\f$. This reality condition
is necessary to obtain the correct field content\footnote{
%%%
%%%%%% footnote %%%%
We note that the reality condition stated in \eq{Adef}
differs from the reality condition used in \cite{v10}.}.
%%%%%%%%%%%%%%%%%%%%%%%%%%%%%%%%%%%%%%%%%%%%%%%%%%%%%%%%

The general solution to the condition \eq{Adef} is

\be
\f \se C+\pi\,(C^{\dagger})\star\C\ ,\la{Csol}
\ee

where the field $C$ has the expansion

\bea
C(Y,\th) \se \sum_{k=0}^{\infty} &{}&
\!\!\!\!\!\!\!\!\!\!\!\!\left(
\quad\;\sum_{ \ba{c}m-n=4k\\m,n\geq0\ea}\left( C(m,n)
+\ft1{4!}C_{i_{1}\cdots i_{4}}(m,n)\th^{i_{1}\cdots i_{4}}
+\ft1{8!}C_{i_{1}\cdots i_{8}}(m,n)\th^{i_{1}\cdots
i_{8}}\right)\right.\nn\\
&+&\sum_{\ba{c}m-n=4k+1\\m,n\geq0\ea}\left( \ft1{3!}C_{ijk}(m,n)\th^{ijk}
+\ft1{7!}C_{i_{1}\cdots i_{7}}(m,n)\th^{i_{1}\cdots i_{7}}\right)\nn\\
&+&\sum_{\ba{c}m-n=4k+2\\m,n\geq0\ea}\left(\ft1{2!} C_{ij}(m,n)\th^{ij}
+\ft1{6!}C_{i_{1}\cdots i_{6}}(m,n)\th^{i_{1}\cdots i_{6}} \right)\nn\\
&+&\left. \sum_{\ba{c}m-n=4k+3\\m,n\geq0\ea}\left( C_{i}(m,n)\th^{i}
+\ft1{5!}C_{i_{1}\cdots i_{5}}(m,n)\th^{i_{1}\cdots i_{5}}\right)\right)
\ ,\la{A}
\eea

where the bosonic fields are in the $1$, $28$ and $35_++35_-$
representations of $SO(8)$ and the fermions are in the $8$ and $56$
representations and the spin of $\f_{\mi i1k}(m,n)$ is given by
$s=\ft{m+n}{2}$.

We do not impose any reality condition on $C$. By construction $C$
contains all solutions to $\t(C)=\bp(C)$ that have number of $y$ spinor
variables greater than or equal to the number of $\yb$ spinor variables.
Using $\t(C^{\dagger})=\p(C)$ and $\p\bp\p_{\th}(C)=C$ (which follows
from $\t^2(C)=C$) we find that the hermitian conjugate of $C$ obeys

\be
\t\,(C^{\dagger}\star\C\,)\se\t\,(\C\,)\star\t\,(C^{\dagger})\se
\C\star\pi\,(C^{\dagger})\se\pi\,\pi_{\th}(C^{\dagger})\star\C\se
\bp\,(C^{\dagger}\star\C\,)\ ,\la{tcd}
\ee

Hence, if $f$ and $g$ are two ${\cal A}$ involutions that preserve
numbers of $y$ and $\yb$ and that commute with $\bar{\pi}$ and $\t$,
then $\f=f(C)+g(C^{\dagger}\star\C\,)$ obeys
$\t(\f\,)=\bar{\pi}(\f\,)$. In order to solve the hermicity condition
in \eq{A} we can then take $f$ to be the identity map and $g=\pi$:

\be
\left(C+\pi\,(C^{\dagger})\star\C\,\right)^{\dagger}\se
\C\star\,\bp\,(C)+C^{\dagger}\se
\left(\bp\,\pi_{\th}(C)+C^{\dagger}\star\C\,\right)\star\C\se
\pi\,\left(C+\pi\,(C)\star\C\,\right)\star\C\ .
\ee

Upon substituting the expansion \eq{A} into \eq{Csol} and equating
$\th$ components we find that

\bea
\f_{i_{1}\cdots i_{k}}(m,n)&=& \left\{\begin{array}{ll} C_{\mi
i1k}(m,n)&m>n\w2
\fft{(-1)^m}{(8-k)!}\e_{\mi i1k\mi j1{8-k}}C^{\star\,\mi j1{8-k}}(m,n)
& m<n\end{array}\right.\ ,\la{mn}\w2
\f(n,n)&=&  C(n,n)+\ft1{8!}(-1)^{n}\,\e^{i_{1}\cdots i_{8}}
C^{\star}_{i_{1}\cdots i_{8}}(n,n)\ ,\nn\w2
\f_{ijkl}(n,n)&=&C_{ijkl}(n,n)+\ft1{4!}(-1)^{n}\,\e_{ijklpqrs}C^{*~pqrs}(n,n)
\ ,\nn\w2
\f_{\mi{i}18}(n,n)&=&
C_{\mi{i}18}(n,n)+(-1)^{n}\e_{\mi{i}18}C^{\star}(n,n)
\ ,\la{nn}
\eea

where the allowed values of $m$, $n$ and $k$ in \eq{mn} are given by \eq{A}. Thus
there is an overcounting of degrees of freedom in the $m=n$ sector of
$C$ which is eliminated when $C$ and its hermitian conjugate are added to
give $\f$.

As we shall see in section 6.3, the field $\f(0,0)$ is the $1+1$ real
scalar of the level $L_1$ multiplet given in Table 1, and
$\f_{ijkl}(0,0)$ are the $35_{\smpl}\!+\!35_{\smm}$ real scalars of the
level $L_0$, $N=8$ supergravity multiplet \cite{cj1}. Notice that these
scalars obey the reality conditions

\bea
\f^{\star}(n,n)&=& \ft{1}{8!}(-1)^{n}\,\e^{i_{1}\cdots i_{8}}
\f_{i_{1}\cdots i_{8}}(n,n)\ ,\nn\w2
\f^{*}_{ijkl}(n,n)&=&\ft{1}{4!}(-1)^{n}\,\e_{ijklmnpq}\f^{mnpq}(n,n)\
.\la{sr}
\eea

It is gratifying to see that for $n=0$ the second equation yields the $SU(8)$
invariant reality condition on the $70$ scalars of the $N=8$
supergravity \cite{cj1,cj2,dn1,dn2}.

The left-handed fermions $C_{\a}^{i_1\cdots i_7}$ and their right-handed
hermitian conjugates constitute the spin $s=\ft12$ content of the $N=8$
supergravity multiplet, while the left-handed fermions $C_{\a}^{ijk}$ and
their right-handed hermitian conjugates precisely match the spin
$s=\ft12$ content of the level $k=1$ multiplet in Table 1.

The conditions \eq{Adef} imply that the quasi-adjoint representation
must contain an infinite dimensional spin $s\geq 1$ sector. Thus, the
inclusion of a finite number of fields with spin $s \leq \ft12$ in the
theory requires the inclusion of an infinite number of auxiliary higher
spin fields. As will be shown in section 6.1, the fields
$\f_{\mi{i}1k}(2s+n,n)$ ($s=0,\ft12,1,...$,
$n=1,2,...$) are related to the chiral components
$\f_{\mi{i}1k}(2s,0)$. For $s=0,\ft12$ the chiral components are the
physical fields of spin $s=0,\ft12$. For $s\geq 1$, the $SO(8)$ content
of the chiral component $\f_{\mi i1k}(2s,0)$ matches precisely the
generalized Weyl tensor $\x_{\mi i1k}(2s,0)$ defined in
\eq{decomp}. As will be shown in section 5.3 the
linearized field equations actually give

\bea
&&\f_{\mi i1k}(2s,0)\se \x_{\mi i1k}(2s,0)\ ,\nn\w2
&&\f_{\mi i1k}(0,2s)\se \y_{\mi i1k}(0,2s)\ ,\qquad s=1,\ft32,2,\ft52,...
\la{wey}
\eea

We shall thus refer to $\f$ as the {\it Weyl zero-form}.

%%%%%%%%%%%%%%%%%%%%%%%%%%%%%%%%%%%%%%%%%%%%%%%%%%%%%%%%%%%%%%%%%%%%%%%%%

\section{The Higher Spin Field Equations}

%%%%%%%%%%%%%%%%%%%%%%%%%%%%%%%%%%%%%%%%%%%%%%%%%%%%%%%%%%%%%%%%%%%%%%%%%

\subsection{General Discussion}

Nonlinear higher spin interactions were first constructed in
\cite{v4,v6} using the formalism of free differential algebras (FDA)
which aims at obtaining gauge invariant curvature constraints

\be
d\o\se f^\o(\o,\f)\ ,\qquad d\f\se f^\f(\o,\f)\ ,\la{fda}
\ee

where the functions $f^\o$ and $f^\f$ are to be determined, order by
order in $\f$, from the {\it integrability condition} $d^2=0$ and the
boundary condition that \eq{fda} should reduce to the trivial
constraints $R(\o)=0$ and $D_{\o}\f=0$ in the lowest order. The
structure of the gauge transformations is determined by the functions
$f^\o$ and $f^\f$ and thus reduce to \eq{dw} and \eq{cg} in the lowest
order. The diffeomorphism invariance of these equations is realized as
gauge transformations with parameter $\e=i_\rho \o$, where
$\d x^\mu=\rho^mu$.

Once the explicit form of the constraints \eq{fda} has been found, the
integrability guarantees that the first equation yields the one-form
$\o$ in terms of $\f$ up to gauge transformations. From the second
equation, one then obtains the zero-form $\f$ in terms of an
{\it initial condition} $\f|_{p} :=  \f_{p}$ where $p$ is
a fixed spacetime point. This type of initial value problem, however,
is rather untractable. Instead it is more convenient to first eliminate
all the auxiliary fields through the algebraic equations contained in
\eq{fda}, thereby obtaining a closed set of field equations involving
only the spin $s\leq \ft12$ fields and the dynamical gauge fields, and
then specify the initial data and boundary conditions for these
dynamical fields.

\bigskip
%%%%%%%%%%%%%%%%%%%%%%%%%%%%%%%%%%%%%%%%%%%%%%%%%%%%%%%%%%%%%%%%%%%%%%
\centerline{{\it Extended Free Differential Algebra}}
%%%%%%%%%%%%%%%%%%%%%%%%%%%%%%%%%%%%%%%%%%%%%%%%%%%%%%%%%%%%%%%%%%%%%%
\bigskip

As already mentioned, prior to deriving the dynamical field equations
from \eq{fda}, one first has to find the functions $f^\o$ and $f^A$.
This deformation problem turns out to be rather cumbersome in practice.
However, there exists an elegant formalism, developed by Vasiliev
\cite{v7,v8,v9,v10,v11}, to facilitate the deformation procedure.

The basic idea is to generate an order by order expansion in
$\f$ of $f^\o$ and $f^\f$ by solving an auxiliary constraint. This
constraint is formulated by means of an extended FDA with base manifold
taken to be the product of ordinary spacetime with a complex space of
an auxiliary spinor variable $Z_{\una}=(z_{\a},-\zb^{\ad})$.

The extended FDA is of the form

\be
\hat{d}A\se \hat{f}^A(A,\Phi)\ ,\qquad
\hat{d}\Phi\se \hat{f}^{\Phi}(A,\Phi)\ ,\la{extfda}
\ee

where $\Phi(x,Z;Y,\th)$ is an extended Weyl zero-form, $A(x,Z;Y,\th)$
is an extended connection one-form

\be
A\se W+V\se dx^\m W_\m+dz^\a V_\a - d\zb^{\ad}
\bar{V}_{\ad}\ ,
\la{calc}
\ee

and $\hat{d}$ is the $(x,Z)$ space exterior derivative

\be
\hat{d}\;\; := \;\; dx^\m\del_\m+dz^\a\del_\a
+d\zb^{\ad}\bar{\del}_{\ad}\;\; := \;\; d+\del+\bar{\del}
\;\; := \;\; d+d_{Z}
\ .\la{totder}
\ee

$\hat{f}^A$ and $\hat{f}^{\Phi}$ are given functions of
$A$ and $\Phi$ defined such that the integrability condition

\be
\hat{d}^{2}\se 0\la{qsquare}
\ee

is obeyed. In fact, the extended FDA describes a constraint on a
Yang-Mills curvature based on an enlarged gauge algebra \alge\ that is
a $Z$ dependent deformation of \alg\ that reduces to \alg\ when $Z=0$.
One also has

\be
\zeval{W(x,Z;Y,\th)}\se \o(Y,\th)\ ,\qquad \zeval{\Phi(x,Z;Y,\th)}
\se \f(Y,\th)\ .
\la{wphi}
\ee

Before we give the details of the extended gauge theory let us first
comment on the crucial features of the extension.\eq{qsquare} implies
that \eq{extfda} can be solved for $A$ and $\Phi$ in terms of the
initial condition $\Phi|_{(x,Z)=(p,0)}=\f_{p}(Y,\th)$ up to an extended
gauge transformation. This shows that \eq{extfda} is equivalent to a
FDA of the form \eq{fda}. The corresponding functions $f^\o$ and $f^\f$
are obtained by first solving for the $Z$ dependence of $W$, $V$ and
$\Phi$ from the components of \eq{extfda} that carries at least one $Z$
space index. Since $W$ and $\Phi$ are $Z$ space zero-forms and $V$ is a
$Z$ space one-form, this requires the initial data \eq{wphi}. (More precisely,
the extended gauge invariance can be used to fix the gauge
$\zeval{V_{\a}}=\zeval{\bar{V}_{\ad}}=0$ with unbroken \alg\ gauge
symmetry). Denoting the solutions for $W$ and $\f$ by $W[\o,\f]$ and
$\Phi[\f]$ and the spacetime components of $\hat{f}^A$ and
$\hat{f}^{\Phi}$ by $f^W$ and $f^{\Phi}$, respectively, we have

\be
f^{\o}(\o,\f)\se \zeval{f^{W}(W[\o,\f],\Phi[\f])}\ ,\qquad
f^\f(\o,\f)\se \zeval{f^{\Phi}(W[\o,\f],\Phi[\f])}\ .\la{fsol}
\ee

The fact that \eq{fsol} leads to a FDA of the form
\eq{fda} with a nontrivial $\f$-expansion relies  on the fact
that the $Z$ space is symplectic which implies the existence of a
$\star$ product of the auxiliary spinor variables. Using this $\star$
product in the definition of $\hat{f}^A$ and $\hat{f}^{\Phi}$ one finds

\be
f^{\o}(\o,\f)\se f^{W}(\o,\f)+\dots\ ,\qquad f^\f(\o,\f)\se
f^{\Phi}(\o,\f)+\dots\ ,\la{leading}
\ee

where $f^{W}(\o,\f)$ and $f^{\Phi}(\o,\f)$ result from the leading,
classical terms in the $\star$ product of the auxiliary spinor
variables, while the $\dots$ represent an expansion in
$\f$ coming from the higher order contractions of the
auxiliary spinor variables. The latter involve derivatives of
$W[\o,\f]$ and $\Phi[\f]$ with respect to
$z_{\a}$ and $\zb_{\ad}$ which when evaluated at $Z=0$ yield
nonlinear expressions in $\f$. Hence the obtained FDA of the form
\eq{fda} represents a nontrivial deformation of $R(\o)=0$ and
$D_{\o}\f=0$ provided we set

\be
f^{W}(\o,\f)\se \o\star \o\ ,\qquad f^\f(\o,\f)\se \o\star \f-\f\star
\bp(\o)\ .
\ee

This corresponds to

\bea
\hat{f}^A&=& A\star A+i\,dz^2{\cal V}(\Phi)
+i\,d\zb^2({\cal V}(\Phi))^{\dagger}\ ,\nn\w2
\hat{f}^{\Phi}&=& A \star \Phi-\Phi\star \bp(A)\ ,
\la{hatf}
\eea

where ${\cal V}$ is some function consistent with the extended
integrability condition \eq{qsquare}. Secondly, the fact that $Z$ space
is isomorphic to the space of the commuting spinors $y_{\a}$ and
$\yb_{\ad}$ allows one to define nontrivial $\star$ product
contractions between the auxiliary spinor variables $z^\a$ and
$z^{\ad}$, and the internal spinor variables $y_{\a}$ and $\yb_{\ad}$. This
allows one to construct a special function $\k(z,y)$ that projects onto
anti-chiral (i.e. $y$ independent) components, i.e.
$\zeval{\Phi\star\k}=\f|_{y=0} + \dots $. Using $\k$ in the definition of
${\cal V}$ then leads to spacetime constraints of the form \eq{wey}.

%%%%%%%%%%%%%%%%%%%%%%%%%%%%%%%%%%%%%%%%%%%%%%%%%%%%%%%%%%%%%%%%%%%%%%%%%

\subsection{Extension of The Higher Spin Superalgebra}

%%%%%%%%%%%%%%%%%%%%%%%%%%%%%%%%%%%%%%%%%%%%%%%%%%%%%%%%%%%%%%%%%%%%%%%%%

The extended associative algebra \aext\ is by definition obtained from
${\cal A}$ by first extending the set of generators $(y,\yb,\th)$ of ${\cal
A}$ with the auxiliary, commuting spinors $z_{\a}$ and $\zb_{\ad} :=
(z_{\a})^{\dagger}$ and defining the associative, manifestly
$SO(3,2)$ invariant $\star$ product on \aext\  \cite{v9,v10}

\bea
F(Z,Y)~\star~ G(Z,Y)&=& \int
F(Z+U,Y+U)~G(Z-V,Y+V)~\exp\,i
\left(u_{\a}v^{\,\a}+\ub_{\ad}\vb^{\,\ad}\right)
\ ,\nn\w2
[\,z_{\a}\,,\,\th^{i}\,]_{\star}&=&[\,\zb_{\ad}\,,\,\th^{i}\,]_{\star}\se 0\ ,
\la{yzstar}
\eea

where $Z_{\una} := (z_{\a},-\zb^{\ad})$ is a purely imaginary Majorana
spinor of $SO(3,2)$\footnote{
%
%%%%%% footnote %%%%
Our conventions differ from \cite{v10} where the reality condition
$\zb_{\ad}= -z_{\a}^{\dagger}$ is used.}
and the normalization is such that $1\star F=F$. Notice that the
hermitian conjugation acts as an anti-involution of \aext. As a
particular case of \eq{yzstar} we find \eq{mso32} and

\bea
z_{\a}\star z_{\b}&=& z_{\a}z_{\b}-i\,\e_{\a\b}\ ,\qquad
\zb_{\ad}\star \zb_{\bd}\se \zb_{\ad}\zb_{\bd}-i\,\e_{\ad\bd}\nn\w2
y_{\a}\star z_{\b}&=& y_{\a}z_{\b}-i\,\e_{\a\b}\ ,\qquad
\yb_{\ad}\star \zb_{\bd}\se \yb_{\ad}\zb_{\bd}+
i\,\e_{\ad\bd}\ ,\nn\w2
z_{\a}\star y_{\b}&=& z_{\a}y_{\b}+i\,\e_{\a\b}\ ,\qquad
\zb_{\ad}\star \yb_{\bd}\se \zb_{\ad}\yb_{\bd}-i\,\e_{\ad\bd}
\ ,\la{zz}
\eea

which can be written on the manifestly $SO(3,2)$ covariant form

\be
Z_{\una}\star Z_{\unb}\se Z_{\una}Z_{\unb}-iC_{\underline{\a\b}}\
,\qquad Z_{\una}\star Y_{\unb}\se
Z_{\una}Y_{\unb}+iC_{\underline{\a\b}}\ ,\qquad Y_{\una}\star
Z_{\unb}\se Y_{\una}Z_{\unb}-iC_{\underline{\a\b}}\ .
\la{myzso32}
\ee

This leads to contraction rules analogous to \eq{yc} with the only
difference that there is an additional factor $(-1)$ for each
contraction of type $z\star z$, $\zb\star \zb$, $y\star z$ and
$\zb\star \yb$. See Appendix D for further details.

By definition the basis elements $dz^\a$ and $d\zb^{\ad}$ of one-forms
in $Z$-space, which form a purely imaginary, Majorana spinor
$dZ^{\una}=(dz^{\a},d\zb_{\ad})$ of $SO(3,2)$,  obey

\be
dZ^{\una}\star F\se F\star dZ^{\una}\se dZ^{\una}F\ ,
\quad\quad
(dz^{\a})^{\dagger}\se d\zb^{\ad}\ ,
\la{dza}
\ee

where $F$ is an arbitrary element of \aext. Therefore, if we set

\be
S_{0}\se dZ^{\una}Z_{\una}\se dz^{\a}z_{\a}+d\zb^{\ad}\zb_{\ad} \se
(S_0)^\dagger \ ,
\la{szeroA}
\ee

then from the contraction rules \eq{appe} it follows that the exterior
derivative $d_Z$ in $Z$ space can be generate by the inner, adjoint
action of $S_{0}$:

\be
S_0 \star F_p - (-1)^p F_p \star S_0 \se -2id_Z F_p\ ,\qquad d_Z\; :=
\; dz^{\a}\pd{z}{\a}+d\zb^{\,\ad}\pd{\zb}{\ad}\ ,\la{sdel}
\ee

where $F_p$ is an \alge\ valued $(x,Z)$ space form of total degree $p$
and we have defined $dz^{\a}\wedge dx^{\m}\se-dx^{\m}\wedge dz^{\a}$.
Note that \eq{sdel} has the correct hermicity properties and that the
associativity of the $\star$ product imply the Leibniz' rule

\be
d_Z(A_{p}\star B_{q})\se d_Z A_{p}\star B_{q}+(-1)^p\,A_{p}\star d_Z
B_{q}\ .\la{delleibniz}
\ee

The maps $\t$, $\pi$ and $\bar{\pi}$ defined in \eq{deft} and \eq{defp}
are extended to \aext\ by setting

\bea
\pi\,(z_{\a})&=&-z_{\a}\ ,\qquad \bp\,(z_\a)\se z_\a\ ,\qquad\ \ \,
\t\,(z_\a)\se -i\,z_{\a}\ ,\nn\w2
\pi\,(\zb_{\ad})&=& \zb_{\ad}\ ,\qquad\ \ \, \bp\,(\zb_{\ad})\se
-\zb_{\ad}\ ,\qquad
\t\,(\zb_{\ad})\se-i\,\zb_{\ad}\ ,
\la{pz}
\eea

and declaring these maps to be involutions of the extended classical
algebra. We also define the action of these maps on the basis elements
$dz^{\a}$ and $d\zb^{\ad}$ of one-forms in $Z$-space by

\be
d_Z(\pi(F(Z,Y,\th))\se\pi(d_Z(F(Z,Y,\th))\ ,\qquad \mbox{idem}\
\bp\ \mbox{and}\ \t\ .\la{delpi}
\ee

We next define the special \aext\ element $\k(z,y)$ as

\bea
\k(z,y)& := &\exp({i\,z_{\,\a}y^{\,\a}})
\se\t\,(\k(z,y))\ ,\nn\w2
 \kb(\zb,\yb) & := & (\k(z,y))^{\dagger}\se
\exp (-i\,\zb_{\,\ad}\yb^{\,\ad}) \se\t\,(\kb(\zb,\yb))\ .\la{kappa}
\eea

Using \eq{yzstar} we find

\bea
\k\star F(z,\zb;y,\yb;\th)&=& \k\,F(y,\zb;z,\yb;\th)\ ,\qquad\ \,\quad
\kb\star F(z,\zb;y,\yb;\th)\se \kb\,F(z,-\yb;y,-\zb;\th)\ ,\nn\w2
F(z,\zb;y,\yb;\th)\star\k &=& \k\,F(-y,\zb;-z,\yb;\th)\ ,\qquad
F(z,\zb;y,\yb;\th)\star \kb\se \kb\,F(z,\yb;y,\zb;\th)\ ,\la{kstarf}
\eea

which in turn implies that the involutions $\p$ and $\bp$ have inner actions in
\aext\ given by

\bea
\pi\,(F\,)&=& \k\star F\star \k\ ,\qquad
\quad \bar{\pi}\,(F\,)\se\kb\star F\star\kb\ ,\nn\w2
\k\star\k&=&1\ ,\qquad\qquad\ \ \qquad \kb\star\kb\se 1\ ,\la{kfk}
\eea

where $F$ is an arbitrary element in \aext.

Now the enlargement \alge\ of \alg\ is defined by the extensions of
\eq{pdef} and \eq{lie}. Thus an element $\hat{P}$ in \alge\ is  Grassmann
even and it furthermore obeys

\be
\t\,(\hat{P})\se -\hat{P}\ ,\qquad (\hat{P})^{\dagger}\se -\hat{P}\ .
\la{defalge}
\ee

The extension of the finite dimensional subalgebra $OSp(8|4)$ of \alg\
generated by quadratic elements is given by $OSp(8|4)\times Sp(4,R)$
where the extra $Sp(4,R)$ factor is generated by the elements
$z_{\a}z_{\b}$, $z_{\a}\zb_{\bd}$ and $\zb_{\ad}\zb_{\bd}$. This
$Sp(4,R)$ factor commutes with $OSp(8|4)$ and generates a fictitious
$Sp(4,R)$ gauge symmetry which does not arise in the spacetime FDA
\eq{fda} where $Z$ is set equal to zero.

\bigskip
%%%%%%%%%%%%%%%%%%%%%%%%%%%%%%%%%%%%%%%%%%%%%%%%%%%%%%%%%%%
\centerline{\it The Extended Field Content}
%%%%%%%%%%%%%%%%%%%%%%%%%%%%%%%%%%%%%%%%%%%%%%%%%%%%%%%%%%%
\bigskip

The extended \alge-valued spacetime connection one-form
$W(Z,Y,\th)=dx^{\,\m}\,W_{\m}$ and Weyl zero-form $\Phi(Z,Y,\th)$
are defined by

\bea
&& \t\,(W\,)\se-W\ ,\quad\quad W^{\,\dagger}\se -W\ ,
\la{wdef}\w5
&& \t\,(\Phi\,)\se\bar{\pi}\,(\Phi\,)\ ,\quad\quad
\Phi^{\dagger}\se\pi\,(\Phi\,)\star\C\ ,
\la{phi}
\eea

such that the initial conditions defined in \eq{wphi} indeed obey
\eq{w} and \eq{Adef}. Note that $W$ and $\Phi$ are Grassmann even.
The extended curvature $2$-form and covariant exterior derivative

\bea
{\cal R}&=& dW-W\star W\ ,\nn\w2
{\cal D}\Phi&=& d\Phi - W\star\Phi + \Phi\star \bar{\pi}\,(W\,)\ ,\la{rdf}
\eea

transform covariantly under the extended gauge transformation

\bea
\d_{\hvare}\, W&=&d\hvare -[\,W\,,\,\hvare\,]_{\star}\ ,\nn\w2
\d_{\hvare}\, \Phi&=&\hvare \star \Phi - \Phi\star \bar{\pi}\,(\hvare\,)\ ,
\la{gt}
\eea

where the transformation parameter $\hvare$ is an \alge\ valued function
and the variations obey the algebra $[\d_{\hvare_1}\d_{\hvare_2}]=
\d_{[\hvare_2,\hvare_1]_{\star}}$.

It is important to notice that when evaluating \eqs{rdf}{gt} at
$Z=0$ there are contributions from the quadratic terms that involve
$\star$ contractions of terms in $W$ and $\Phi$ which are higher order
in $z$ and $\zb$ (in particular there are nontrivial cross terms coming
from $z$-$y$ and $\zb$-$\yb$ contractions):

\bea
\zeval{{\cal R}}\se R(\o)+\dots&,&\qquad
\zeval{{\cal D}\Phi}\se D_{\o}\f+\dots\ ,\nn\w2
\zeval{\d_{\hvare}W}\se \d_{\vare}\o+\dots&,&\qquad
\zeval{\d_{\hvare}\Phi}\se \d_{\vare}\f+\dots\ ,\la{zeval}
\eea

where $\vare=\hvare|_{Z=0}$ is an \alg-valued gauge parameter and the
$\dots$ represent the contributions from the higher order Taylor
coefficients in the $(z,\zb)$ expansions of $W$ and $\Phi$. Thus, if
$W$ and $\Phi$ were given in terms of the initial data \eq{wphi}, then
setting ${\cal R}$ and ${\cal D}\Phi$ equal to zero would yield a
nonlinear FDA of the form \eq{fda}.

We proceed by defining the $Z$ space the Grassmann even connection
one-form $V$ with components $V_{\una}=(V_\a,{\bar V}^{\ad})$
introduced in \eq{calc} by

\be
\t\,(V\,)\se-V\ ,\quad\quad V^{\dagger}\se -V\ .
\la{defv}
\ee

We also define a $Z$ space ``covariant derivative''

\be
S \se S_0 +2iV\ , \quad\quad \t\,(S\,)\se-S\ ,
\quad\quad S^{\dagger}\se S\ ,
\la{sconn}
\ee

where $S_0$ is the $Z$ space one-form defined in \eq{szeroA}. We take
$S$ to transform  in the adjoint representation \alge, that is

\be
\d_{\hvare} S\se \hvare\star S-S\star \hvare\ .
\la{deltas}
\ee

From \eq{sdel} it then follows that $V$ indeed transforms as a $Z$
space connection one-form

\be
\d_{\hvare}V\se d_Z \hvare - [\,V\,,\,\hvare\,]_{\star}\ .\la{xt}
\ee

whose curvature is related to $S\star S$ as follows

\be
d_Z V-V\star V\se \ft14 S\star S\ .
\la{xcurv}
\ee

From $\t(S_{\una})=-i\,S_{\una}$ it follows that
$\t^{2}(S_{\una})\se \p\bp\p_{\th}(S_{\una})\se -S_{\una}$. Combining with
\eq{kfk},  we find

\be
\k\C\star S\star \k\C\se -\kb\star S\star \kb\ ,\qquad
\k\C\star V\star \k\C\se -\kb\star V\star \kb\ .\la{ksk}
\ee

%%%%%%%%%%%%%%%%%%%%%%%%%%%%%%%%%%%%%%%%%%%%%%%%%%%%%%%%%%%%%%%%%%%%%%

\subsection{The Equations of Motion in $(x,Z)$ Space}

%%%%%%%%%%%%%%%%%%%%%%%%%%%%%%%%%%%%%%%%%%%%%%%%%%%%%%%%%%%%%%%%%%%%%%

The integrable equations of motion in $(x,Z)$ space of the higher spin
field theory are \cite{v10}
\footnote{
%
%%%%%%% footnote %%%%%%%%%
In the original formalism \cite{v9,v10} the analog of \eq{fe5} is
written by using $\k k$ and ${\bar\k}{\bar k}$, where $k,\bar k$ are
Kleinian operator \cite{fv2}, instead of our $\k \C$ and $\bar\k$,
respectively. The addition of $k$ and its hermitian conjugate
$\bar{k}$ to the set of generators of the associative algebra,
however, would give rise to new dynamical gauge fields which are
unwanted in attempting to reproduce Table 1. The use of an $SO(2N)$
chirality operators in higher spin algebras have been discussed in
\cite{kv2}.}
%%%%%%%%%%%%%%%%%%%%%%%%%%%%%%%%%%%%%%%%%%%%%%%%%%%%%

\bea
dW&=&W\star W\ ,\la{fe1}\w2 d\Phi&=&W\star \Phi-\Phi\star \bp\,(W\,)\ ,
\la{fe2}\w2
dS&=&W\star S-S\star W\ ,\la{fe3}\w2 S\star
\Phi&=&\Phi\star \bp\,(S\,)\ ,\la{fe4}\w2 S\star
S&=&i\,dz^{2}\,(1+\Phi\star \k\,\C\,)+i\, d\bar{z}^{2}\,(1+\Phi\star
\bar{\k}\,)\ .\la{fe5}
\eea

where $dz^{2}=dz^{\a}\wedge dz_{\a}$ and $d\zb^{2}=d\zb^{\ad}\wedge
d\zb_{\ad}= (dz^2)^{\dagger}$. We shall show the integrability in
section 4.4. Apart from the integrability the crucial properties of
these equations are that they preserve the representation properties
\eqs{wdef}{phi} and \eq{sconn} and that they are invariant under
the internal gauge transformations

\bea
\d W&=&d\hvare -[\,W\,,\,\hvare\,]_{\star}\ ,\nn\w2
\d \Phi&=&\hvare \star \Phi - \Phi\star \bar{\pi}\,(\hvare\,)\ ,\nn\w2
\d S&=& [\,\hvare\,,\,S\,]_{\star}\ ,\la{gauge}
\eea

where $\hvare$ is an arbitrary \alge-valued gauge parameter. The
equations are also manifestly invariant under spacetime
diffeomorphisms, since they are formulated using only spacetime
differential forms. In fact, the general coordinate transformation $\d
x^\m=\r^\m$ is incorporated into the $\hvare$-transformations by

\be
\hvare(\r)\se i_{\r}W\ .
\la{diff}
\ee

The full set of field equations \eqs{fe1}{fe5} can be derived from
\eq{fe5} and either \eq{fe2} or \eq{fe3}. To begin with \eq{fe1} is the
integrability condition for \eq{fe2} and for \eq{fe3}. Next \eq{fe4}
follows from \eq{fe5} by exploiting the associativity property $(S\star
S)\star S=S\star(S\star S)$ and by making use of \eq{kfk} and

\bea
dz^{2} \bp\,(S)\star \k\C&=& dz^{2} \k\C\star S\ ,\nn\w2
d\zb^{2} \bp\,(S)\star \kb&=& d\zb^{2}\kb \star S\ ,\la{twof}
\eea

which in turn follow from $dz^{\a}\wedge dz^{\b}\wedge dz^{\c}=0$ and
\eq{ksk}. Finally, \eq{fe3} follows from \eq{fe2} (or vice versa)
by combining \eq{fe2} with the covariant spacetime derivative of \eq{fe5}.

\bigskip
%%%%%%%%%%%%%%%%%%%%%%%%%%%%%%%%%%%%%%%%%%%%%%%%%%%%%%%%%%
\centerline{\it An\ Interaction\ Ambiguity}
%%%%%%%%%%%%%%%%%%%%%%%%%%%%%%%%%%%%%%%%%%%%%%%%%%%%%%%%%%
\bigskip

The extra factors of $\k\C$ and $\kb$ have been inserted on the right
hand side of \eq{fe5} to ensure $\t$-invariance and gauge invariance.
As already mentioned, these factors also play a crucial role in
obtaining the appropriate set of constraints on the \alg\ curvature
$R(\o)$. For example, the linearized contribution to $R(\o)$ involves
$\zeval{\Phi\star\k}=\f(0,\yb)$ which leads to a constraint of the form
\eq{wey}.

These requirements do not, however, fix the right side of \eq{fe5}
uniquely. In order to describe the interaction ambiguity we start from
the identity

\be
S\star S\se d\bar{Z} \left(1+\vf+i\C^A \vf_A\right) dZ\ ,\la{so32ss}
\ee

where $\C_A\,(A=0,1,2,3,5)$ are the $SO(3,2)$ $\C$-matrices,
$\vf^{\dagger}=\t(\vf)=\vf$ is an $SO(3,2)$ scalar and
$(\vf_{A})^{\dagger}=\t(\vf_{A})=\vf_{A}$ is an $SO(3,2)$ vector.
Eq. \eq{so32ss} is manifestly invariant under both ``internal" gauge
transformations

\be
\d_{\hvare}\,S\se [\,\hvare\,,\,S\,]_{\star}\ ,\qquad
\d_{\hvare}\,\vf\se [\,\hvare\,,\,\vf\,]_{\star}\ ,\qquad
\d_{\hvare}\,\vf_A\se [\,\hvare\,,\,\vf_A\,]_{\star}\la{vfgt}
\ee

and ``external" $SO(3,2)$ transformations that rotate
$SO(3,2)$ spinors and vectors (including $dZ^{\una}$). The external
transformations leave the $\star$ product \eq{yzstar} invariant and
they cannot be incorporated into the group of internal gauge
transformations since the external transformations rotate $dZ^{\una}$
while the internal transformations leave $dZ^{\una}$ invariant (as can
be seen from \eq{dza}).

The interaction ambiguity amounts to the degrees of freedom associated
with the choice of an $SO(3,1)$ invariant constraint expressing $\vf$
and $\vf^A$ as functions of the $SO(3,1)$ invariant, quasi-adjoint
master field $\Phi$.

For example, \eq{fe5} is the result of the $SO(3,1)$ invariant
constraint

\bea
&&\vf\se {\rm Re}\,\Phi' \ ,\qquad
\vf^5\se -{\rm Im}\,\Phi' \ ,\nn\w2
&&\vf^a\se 0\ ,\quad a=0,1,2,3\ ,\la{so31c}
\eea

where $\Phi'$ obeys

\be
\t(\Phi')\se\Phi'\ ,\qquad\Phi'^{\dagger}\se K\star \Phi'\ ,\qquad
\d_{\hvare}\,\Phi'\se [\,\hvare\,,\,\Phi'\,]_{\star}\ ,\la{phip}
\ee

and $K := \k\kb\C$, obeying $K^2=1$ and $\t(K)=K^\dagger=K$. The
reality condition on $\Phi'$ implies that
${\rm Re}\Phi'=P_+\star\Phi'$ and ${\rm Im}\Phi'=-i\,P_-\star\Phi'$,
where $P_{\pm}=\ft12 (1\pm K)$ are projectors onto the eigenvalues
$\pm1$ of $K$. This shows that the real and imaginary parts of $\Phi'$
each contain half the number of degrees of freedom of $\Phi'$. Also
notice that $K$ commutes with $W$, $\hvare$ and $\Phi'$ and
anticommutes with $S$. Thus the result of the $SO(3,1)$ invariant
constraint \eq{so31c}, the associativity and \eq{fe3} is the following
set of manifestly $SO(3,1)$ invariant equations

\bea
d\Phi'&=&W\star \Phi'-\Phi'\star W\ ,\nn\w2 S\star \Phi'&=&\Phi'\star
S\ ,\nn\w2 S\star S&=&d\bar{Z}\left(1+ {\rm Re}\Phi' + i\,
\C^{5}{\rm Im}\Phi' \right)dZ\ .
\la{so32}
\eea

All quantities in this equation are manifestly $SO(3,2)$ invariant
except the $\C^5$ term. As a consequence of breaking the manifest
external invariance from $SO(3,2)$ down to $SO(3,1)$, the
representation of the internal $OSp(8|4)$ on the fields will only be
manifestly $SO(3,1)$ invariant. The situation is analogous to the one
in $N=8$ AdS supergravity theory \cite{bf}; both equations of motion
and supersymmetry transformation rules contain explicit $\C^5$ matrices
but in such combinations that the closure of the internal
$OSp(8|4)$ algebra is not violated.

Now, eqs. \eq{so32} coincide with \eq{fe2}, \eq{fe4} and \eq{fe5}
provided that we set

\be
\Phi'\se \Phi\star\k\C\ .
\la{phipid}
\ee

The interaction ambiguity amounts to the fact this relation can be
replaced by the more general one

\be
\Phi'\se {\cal V}(\Phi\star \k\C)\ ,\la{calv}
\ee

where ${\cal V}$ is a regular, complex $\star$ function.

%%%%%%%%%%%%%%%%%%%%%%%%%%%%%%%%%%%%%%%%%%%%%%%%%%%%%%%%%%%%%%%%%%%%%%%
\subsection{Integrability of the Higher Spin Field Equations}
%%%%%%%%%%%%%%%%%%%%%%%%%%%%%%%%%%%%%%%%%%%%%%%%%%%%%%%%%%%%%%%%%%%%%%%

The integrability of the higher spin field equations \eqs{fe1}{fe5} in
$(x,Z)$ space can be made manifest by writing them in terms of
the $(x,Z)$ connection one-form $A=W+V$ (see \eq{calc}) and using the
total exterior derivative $\hat{d}=d+d_Z$ (see \eq{totder}). One then
finds that \eqs{fe1}{fe5} can be cast into the form

\bea
F &=& \ft{i}4 \,\left( dz^2\,\Phi\star \k\C + d\zb^{2}\,\Phi\star
\kb \right)\ ,\la{totr}
\w2
\hat{d} \Phi &=& A \star \Phi-\Phi\star \bp(A)\ ,\la{qphi}
\eea

where the total curvature two-form in $(x,Z)$ space and its Bianchi
identity are given by

\bea
&& F \se \hat{d}A-A\star A\ ,\nn\w2
&&\hat{d} F\se [\,A\,,\,F\,]_{\star}\ .\la{totbi}
\eea

Notice that \eq{qphi} follows from inserting \eq{totr} into \eq{totbi}
and using \eq{twof}. The gauge symmetry of \eqs{totr}{qphi} is given by

\be
\d A\se \hat{d}\hvare -[\,A\,,\hvare\,]_{\star}\ ,\qquad
\d \Phi\se \hvare\star \Phi-\Phi\star\bp(\hvare)\ .\la{totgauge}
\ee

The curvature constraint \eq{totr} can be written in components as

\bea
F_{\m\n}&=&F_{\a\m}\se F_{\ad\m}\se F_{\a\ad}\se 0\ ,\nn\w2
F_{\a\b}&=&-\ft{i}2 \e_{\a\b}\Phi\star \k\C\ ,\nn\w2
F_{\ad\bd}&=&-\ft{i}2 \e_{\ad\bd}\Phi\star \kb\ ,\la{cc}
\eea

In order to discuss the equivalence of \eqs{totr}{qphi} and
\eqs{fe1}{fe5}, we introduce the tri-grading $(q,r,s)$ of $(q+r+s)$-forms in
$(x,Z)$ space where $q$ refers to the form degree in $x$-space and the
bi-grading $(r,s)$ refers to the form degree in $(z,\bar{z})$-space
regarded as a complex space. Thus the $(2,0,0)$ components of \eq{totr}
yield
\eq{fe1}. Using
\eq{sdel} and \eq{sconn} we find that the $(1,1,0)$ and
$(1,0,1)$-components yield

\be
d_Z W + dV\se W\star V+V\star W\ ,\la{aw}
\ee

which is equivalent to \eq{fe3}. The $(0,2,0)$, $(0,1,1)$ and $(0,0,2)$
components yield

\be
d_Z V-V\star V\se\ft{i}{4}\,(dz^2\,\Phi\star \k\C+d\zb^{2}\,\Phi\star \kb)
\la{dxx}
\ee

which is equivalent \eq{fe5} using \eq{xcurv}. From the
$(1,2,0)$-components of the Bianchi identity \eq{totbi} we read off
\eq{fe2}. Finally, the $(0,2,1)$ and $(0,1,2)$-components of \eq{totbi}
are equivalent to the equation

\be
d_Z \Phi \se V\star \Phi-\Phi\star \bp(V)\la{delphi}
\ee

which yields \eq{fe4}.

As already discussed in section 4.1, the integrability of the higher
spin field equations can be used to solve for the $Z$ dependence thus
obtaining an FDA of the form \eq{fda} from which the dynamical
spacetime equations follow upon the eliminating auxiliary fields.
Another possibility is to begin by solving for the $x$ dependence of
$W$, $\Phi$ and $S$ from \eqs{fe1}{fe3} in terms of $\Phi|_{p}$ and
$S|_{p}$ (where $p$ is a fixed point in spacetime). In fact, the solution
obtained is pure gauge such that the $x$ dependence away from
$p$ is determined by an \alge\ gauge transformation. Thus the space of gauge
inequivalent solutions to the full set of equations \eqs{fe1}{fe5} is
equivalent to the space of gauge inequivalent solutions of the ``$Z$
space equations" obtained by inserting the pure gauge solution for $W$,
$\Phi$ and $S$ into the two remaining field equations \eqs{fe4}{fe5}:

\bea
S_{p}\star S_{p}&=& i\,dz^2\,(1+\Phi_{p}\star \k\,\C\,)
+i\,d\zb^2\,(1+\Phi_{p}\star \kb\,)\ ,\nn\w2 S_{p}\star \Phi_{p}&=&
\Phi_{p}\star\bp\,(S_{p})\ .
\la{algeq}
\eea

These equations are invariant under spacetime independent \alge\ gauge
transformations which are local in $Z$ space and could be a promising
starting point for obtaining other classical solutions of the theory
than the AdS vacuum solution, such as solutions with nontrivial higher
spin background fields or solutions that partially break the global
\alg\ symmetry of the AdS vacuum.

%%%%%%%%%%%%%%%%%%%%%%%%%%%%%%%%%%%%%%%%%%%%%%%%%%%%%%%%%%%%%%%%%%%%%%%%%%

\section{Expansion Around The Anti de Sitter Vacuum}

%%%%%%%%%%%%%%%%%%%%%%%%%%%%%%%%%%%%%%%%%%%%%%%%%%%%%%%%%%%%%%%%%%%%%%%%%%

\subsection{The Anti de Sitter Vacuum Solution}

%%%%%%%%%%%%%%%%%%%%%%%%%%%%%%%%%%%%%%%%%%%%%%%%%%%%%%%%%%%%%%%%%%%%%%%%%%

The field equations \eqs{fe1}{fe5} constitute an internally consistent
set of equations, but it remains to establish the physical relevance of
the equations and to make contact with $N=8$ supergravity theory. As
already discussed in section 4.1 this requires the solving of the $Z$
space equations \eqs{fe3}{fe5} and the elimination of the auxiliary
fields from the algebraic equations contained in
\eqs{fe1}{fe2}. In the following two sections we shall verify that this
yields the correct free field higher spin dynamics in the AdS vacuum, and
in section 7 we shall make contact with the linearized $N=8$ supergravity model.
Having established the physical consistency at the linearized level, one
then has an interacting higher spin field theory based on the equations
\eqs{fe1}{fe5} which is tractable in classical perturbation theory.

The $AdS_{4}$ geometry can be identified as the vacuum solution given by
\cite{v10}

\bea
\Phi_{0}(Z,Y,\th)&=&0\ ,\nn\w2
S_{0}(Z,Y,\th)&=&dz^{\a}z_{\a}+d\zb^{\,\ad}\zb_{\ad}\ ,\nn\w2
W_{0}(Z,Y,\th)&=&\fft{1}{4i}\left[~\o_{0\,\a\b}y^{\a}y^{\b}+
\bar{\o}_{0\,\ad\bd}\yb^{\,\ad}y^{\bd}+
2e_{0\,\a\bd}y^{\a}\yb^{\,\bd}\right]\;\; := \;\;\O_{0}(Y)\ ,\la{vac}
\eea

where the one-forms $\o_{0}$, $\bar{\o}_{0}$ and $e_{0}$ are the vacuum Lorentz
connection and vierbein of anti-de Sitter spacetime:

\bea
d\o_{0\,\a\b}&=&\o_{0\,\a\c}\wedge\o_{0\,\b}{}^{\c}+
e_{0\,\a\dd}\wedge e_{0\,\b}{}^{\dd}\ ,
\nn\w2
d\bar{\o}_{0\,\ad\bd}&=&\bar{\o}_{0\,\ad\cd}\wedge\bar{\o}_{0\,\bd}{}^{\cd}+
e_{0\,\d\ad}\wedge e_{0}{}^{\d}{}_{\bd}\ ,
\nn\w2
de_{0\,\a\bd}&=&\o_{0\,\a\c}\wedge e_{0}{}^{\c}{}_{\bd}
+\bar{\o}_{0\,\bd\dd}\wedge e_{0\,\a}{}^{\dd}\ .\la{ads}
\eea

To prove that \eq{vac} is a solution of the higher spin equations
\eqs{fe1}{fe5} one first observes that \eq{fe2} and \eq{fe4} are
trivially satisfied while \eq{fe1} reduces to \eq{ads}. Using \eq{sdel}
one easily verifies \eq{fe3} and \eq{fe5}.

The flat, $SO(3,2)$-valued connection one-form $\O_{0}$ defines a
AdS-covariant derivative which mixes irreducible tensors of the Lorentz
$SO(3,1)$ subgroup of $SO(3,2)$ of the same spin. It splits into the
curved, $SO(3,1)$-valued connection one-form

\be
\o_{0}\se\fft{1}{4i}\left(~\o_{0}^{\a\b}y_{\a}y_{\b}+
\bar{\o}_{0}^{\ad\bd}\yb_{\ad}y_{\bd}\,\right)\la{dl}
\ee

which defines a Lorentz covariant derivative $D_{a}$ which acts
irreducibly on Lorentz tensors, and the vierbein

\be
e_{0,\a\ad}\se -\ft12 \l~ (\s_{a})_{\a\ad}~e^{a}_{0}\ , \qquad
e^{a}_{0}\se dx^{\m}\,e_{0,\m}{}^a\ ,\la{bgvb}
\ee

where the mass parameter $\l$ is given in terms of the $AdS$ radius $a$
by

\be
\l\se {1\over  a}\ .\la{adsr}
\ee

In the following we shall need $D_{[a}D_{b]}$ acting on a Lorentz tensor
$T_{a...\a...\ad...}$:

\bea
D_{[a}D_{b]}T_{c...\c...\cd...}&=&\ft12 r_{ab,c}{}^{d}T_{d...\c...\cd...}
+~\cdots~+\ft12 r_{ab,\c}{}^{\d}T_{c...\d...\cd...}+~\cdots\nn\w2
&&+~\ft12 r_{ab,\cd}{}^{\dd}T_{c...\c...\dd...}+~\cdots\quad\ ,
\eea

where the $SO(3,1)$-valued Riemann curvature

\bea
r_{ab,cd}&=&-\l^{2}(\y_{ac}\y_{bd}-\y_{ad}\y_{bc})\ ,\nn\w2
r_{ab,\c\d}&=&\ft14 r_{ab,cd} (\s^{cd})_{\a\b}\ ,\nn\w2
r_{ab,\cd\dd}&=&\ft14 r_{ab,cd} (\sb^{cd})_{\ad\bd}\ .\la{rie}
\eea

We shall temporarily set $\l=1$ and return to the relation between $\l$ and
the dimensionful coupling of the theory in section 7.

\bigskip
%%%%%%%%%%%%%%%%%%%%%%%%%%%%%%%%%%%%%%%%%%%%%%%%%%%%%%%%%%%%%%%%%
\centerline{\it Generalized Killing Symmetries and
Admissibility Criterion for \alg}
%%%%%%%%%%%%%%%%%%%%%%%%%%%%%%%%%%%%%%%%%%%%%%%%%%%%%%%%%%%%%%%%
\bigskip

Let us emphasize that the AdS vacuum solution \eq{vac} exhibits the
full \alg\ symmetry and not just $OSp(8|4)$ symmetry. More explicitly,
the solution \eq{vac} is invariant under gauge transformations with
parameters $\hvare_{0}$ obeying the following generalized Killing
equations

\bea
&&d\hvare_{0} -[\,\O_{0}\,,\,\hvare_{0}\,]_{\star}\se 0\ ,\nn\w2
&&d_Z \hvare_{0}\se 0\la{bginv}
\eea

whose solution space forms a superalgebra isomorphic to \alg. The last
fact follows from the flatness of the $SO(3,2)$ connection $\O_{0}$.

To construct the Killing parameters explicitly we first introduce the
four commuting AdS Killing spinors $\y^{r}_{\a}(x)$ ($r=1,...,4$) and
their hermitian conjugates $\ybar^{r}_{\ad}$ obeying the AdS covariant
Killing spinor equations

\be
D\y^{r}_{\a}\se e_{0\,\a\bd}\ybar^{r\,\bd}
\ ,\qquad
D\ybar^{r}_{\ad}\se
e_{0\,\b\ad}\y^{r\,\b}\
\ .\la{ksp}\w2
\ee

We then define the hermitian, Grassmann even, commuting elements

\be
\y^{r}\se\y_{\a}^r y^{\a}+\ybar_{\ad}^r \yb^{\ad}\ ,\qquad
(\y^r)^{\dagger}\se\y^r\ , \qquad  r=1,...,4\ .
\la{yr}
\ee

From \eq{ksp} it follows that the $\y^r$ obey \eq{bginv}. By
making use of Leibniz rule it is easy to verify that

\be
\y^{\mi{r}1m\mi{i}1n}\;\; := \;\;
\y^{(r_1}\star\cdots\star\y^{r_m)}\th^{\mi{i}1n}\ ,\qquad m=1,2,...\ ,\quad
k=1,...,8\ ,\la{ymk}
\ee

also obey \eq{bginv}. Moreover, by construction

\bea
\t(\y^{\mi{r}1m\mi{i}1n}) &=& i^{m+n}\,\y^{\mi{r}1m\mi{i}1n}
\ ,\nn\w2
\left(\y^{\mi{r}1m\mi{i}1n}\right)^{\dagger} &=&
(-1)^{\ft{n(n-1)}{2}}\,\y^{\mi{r}1m\mi{i}1k}\ .
\eea

Thus, if $\l_{\mi{r}1m\mi{i}1n} := \l_{(\mi{r}1m)[\mi{i}1n]}$ are
constant Grassmann even constant coefficients obeying

\be
\left(\l_{\mi{r}1m\mi{i}1n}\right)^{\dagger}= (-1)^{\ft{n(n-1)}{2}}
\l_{\mi{r}1m\mi{i}1n}\ ,
\ee

then

\be
\vare_{0}(\l)\se i\,\l_{\mi{r}1m\mi{i}1n}\,\y^{\mi{r}1m\mi{i}1n}
\la{v0}
\ee

is a Killing parameter if $m+n=2$ mod $4$. For given $m$ and $n$, the
real dimension of the space of elements of the form \eq{v0} is
${m+3\choose m}{8\choose n}$. Recalling \eq{nbnf} we see that for
$m+n=4k+2$ ($k=0,1,...$) the space of elements of the form
\eq{v0} is isomorphic to the $k$'th level $L_{k}$ of \alg\ defined in
\eq{lk}.

Thus the construction exhausts $V_{0}$. In particular the
finite dimensional subalgebra of $V_{0}$ spanned by the $28$ global
$SO(8)$ parameters $\vare_{0}(\l_{ij})$, the $32$ unbroken AdS
supersymmetries $\vare_0(\l_{ir})$ and the $10$ Killing vectors
$\vare_0(\l_{rs})$ is isomorphic to the superalgebra $OSp(8|4)$.

%%%%%%%%%%%%%%%%%%%%%%%%%%%%%%%%%%%%%%%%%%%%%%%%%%%%%%%%%%%%%%%%%%%%%%%%%%

\subsection{Perturbative Expansion}

%%%%%%%%%%%%%%%%%%%%%%%%%%%%%%%%%%%%%%%%%%%%%%%%%%%%%%%%%%%%%%%%%%%%%%%%%%

We then proceed by an order by order analysis of \eqs{fe1}{fe5} by
expanding them in powers of $\Phi$ which is considered to be a
perturbation around the vacuum solution \cite{v10}:

\bea
\Phi &=& \Phi_{1}+\Phi_{2}+\cdots\ ,\nn\w2
S&=& S_{0}+S_{1}+S_{2}\cdots\ ,\nn\w2
W&=& W_{0}+W_{1}+W_{2}\cdots\ ,
\la{pert}
\eea

We saw that the constraint equations \eq{fe4} and \eq{fe5} yield
\eq{delphi} and \eq{dxx}, respectively, as a consequence of the
formula \eq{sdel}. Using the expansion \eq{pert}, we can express
\eq{delphi} and \eq{dxx} as follows:

\bea
d_Z \Phi_{n}&=&\ft{i}{2}\sum_{j=1}^{n-1}\Big(
\Phi_{j}\star \bar{\pi}\,(S_{n-j})-S_{j}\star \Phi_{n-j}\Big)\ ,\la{delb}
\w2
d_Z S_{n}&=& -\ft12dz^{2}\,\Phi_{n}\star \k\,\C-\ft12
d\zb^{2}\,\Phi_{n}\star\bar{\k}-\ft{i}{2}\sum_{j=1}^{n-1}S_{j}\star S_{n-j}
\ ,\qquad n=1,2,...\la{dels}
\eea

This system of linear differential equations in the spinor variable $Z$
is integrable order by order in perturbation theory. Verification of
the integrability condition $d_Z^2\Phi_{n}=0$ (assuming that $\Phi_i$
and $S_i$ obey \eqs{delb}{dels} for $i<n$) is straightforward while
verification of $d_Z^{2} S_{n}=0$ requires the use of \eq{twof}. Thus,
having integrated \eqs{delb}{dels} up to the $(n-1)$'th order one
obtains the $n$'th order solution by first integrating \eq{delb} for
$\Phi_{n}$ and then \eq{dels} for $S_{n}$:

\bea
&&\Phi_{n}(Z,Y,\th)\se \f_{n}(Y,\th)\nn\w2
&& +\ft{i}2\sum_{j=1}^{n-1}\int_{0}^{1}dt\left\{z^{\a}\left(\Phi_{j}\star
\kb\star S_{\a}^{n-j}\star \kb-
S^{j}_{\a}\star \Phi_{n-j}\right)(tZ,Y,\th)\right.\nn\w2
&&-\zb^{\ad}\left.\left(\Phi_{j}\star \kb\star \bar{S}_{\ad}^{n-j}\kb
+\bar{S}_{\ad}^{j}\star \Phi_{n-j}\right)(tZ,Y,\th)\right\}\ ,
\nn\w5
&&S^{n}_{\a}(Z,Y,\th)\se \pd{z}{\a} \x_{n}(Z,Y,\th)\nn\w2
&&+\int_{0}^{1}dt\,t\left\{z_{\a}\Big(\Phi_{n}\star \k\C+\ft{i}{2}
\sum_{j=1}^{\left[{n-1\over 2}\right]}
\left[\,S_j^{\b}\,,\,S^{n-j}_{\b}\,\right]_{\star}\Big)(tZ,Y,\th)
+\zb^{\ad}\ft{i}{2}\sum_{j=1}^{n-1}\left[\,S^{j}_{\a}\,,\,
\bar{S}^{n-j}_{\ad}\,\right]_{\star}(tZ,Y,\th)\right\}\nn\w3
&&\bar{S}^{n}_{\ad}(Z,Y,\th)\se \pd{\zb}{\ad} \x_{n}(Z,Y,\th)\nn\w2
&&+\int_{0}^{1}dt\,t\left\{\zb_{\ad}\left(\Phi_{n}\star \kb+\ft{i}{2}
\sum_{j=1}^{\left[{n-1\over 2}\right]}
\left[\,\bar{S}_j^{\bd}\,,\,\bar{S}^{n-j}_{\bd}\,\right]_{\star}
\right)(tZ,Y,\th)-z^{\a}\ft{i}{2}\sum_{j=1}^{n-1}\left[\,S^{j}_{\a}\,,\,
\bar{S}^{n-j}_{\ad}\,\right]_{\star}(tZ,Y,\th)\right\}\ ,\nn\w1
\la{phisint}
\eea

where we have applied the integration formulae \eqs{appb3}{appb5} given
in Appendix D. Notice that the $\star$ products in the quadratic terms
in the right hand side must be calculated with functions depending on
$(Z,Y,\th)$ before $Z$ is replaced by $tZ$, as explained in Appendix D
in more detail. The integration of \eq{delb} introduces the initial
condition

\be
\Phi_{n}\,(Z,Y,\th)|_{_{Z=0}}\;\; := \;\; \f_{n}\,(Y,\th)\ ,
\qquad n=1,2,...\ ,\la{bc}
\ee

in the quasi-adjoint representation \eq{Adef} of \alg. The integration
of \eq{dels} introduces an exact $Z$-space one-form $d_Z \xi_{n}$,
where
$\xi_{n}$ is an arbitrary \alge-valued $0$-form. The perturbative
expansion of the gauge transformation \eq{deltas}, with parameter
$\hvare=\hvare_{1}+\hvare_{2}+\cdots$, takes the form

\be
\delta S_{n}\se 2id_Z \hvare_{n} +\sum_{j=1}^{n-1}
[\,\hvare_{j}\,,\, S_{n-j}\,]_{\star}\ ,\qquad n=1,2,...\la{deltasn}
\ee

Hence we may eliminate $\x_{n}$ by fixing the gauge

\be
\x_{n}\se 0\ ,\qquad n=1,2,...\la{physg}
\ee

The gauge symmetries that preserve \eq{physg} have $Z$-independent
parameters $\vare(Y,\th)$ corresponding to the expected \alg\ gauge
symmetry of the field equations in $x$ space. Notice that in the
perturbative expansion \eq{physg} is imposed order by order, so that at
$n$'th order the remaining gauge symmetries have \alg\ valued
parameters
$\vare_{i}(Y,\th)$ ($i=1,..,n$) and \alge\ valued parameters
$\hvare_{i}(Z,Y,\th)$ ($i\geq n+1$).

Having obtained the solutions for $\Phi$ and $S$ one proceeds by solving
for $W$ by inserting the perturbative expansions for $S$ and $W$ into
\eq{fe3}. This yields the integrable set of equations

\be
d_Z W_{n}\se -\ft{i}{2}dS_{n}+\ft{i}{2}
\sum_{j=0}^{n-1}(W_{j}\star S_{n-j}+S_{n-j}\star W_{j})\ ,\qquad
n=1,2,...\ ,\la{delw}
\ee

which allows one to obtain $W_{n}$ in terms of $\f_{i}$ ($i=1,...,n$)
and the initial conditions $\o_{i}$ ($i=1,...,n$) where

\be
W_{n}\,(Z,Y,\th)|_{_{Z=0}}\;\; := \;\;\o_{n}(Y,\th)\ ,\qquad
n=1,2,...\la{bcw}
\ee

With help of the formula \eq{appb3} given in Appendix D, we can
integrate \eq{delw} subject to the initial condition \eq{bcw}, and we
thus find

\bea
&&W_{n}(Z,Y,\th)\se \o_{n}(Y,\th)\nn\w2
&&-\ft{i}{2}\int_{0}^{1}dt\left\{z^{\a}\Big(dS^{n}_{\a}
-\sum_{j=0}^{n-1}[\,W_{j}\,,\,S^{n-j}_{\a}\,]_{\star}\Big)(tZ,Y,\th)
+\zb^{\ad}\Big(d\bar{S}^{n}_{\ad}
-\sum_{j=0}^{n-1}[\,W_{j}\,,\,\bar{S}^{n-j}_{\ad}\,]_{\star}\Big)(tZ,Y,\th)
\right\}\ .\nn\w1
&&\qquad\qquad\qquad\qquad\qquad\qquad n=1,2,...\la{wint}
\eea

Having solved \eqs{fe3}{fe5} and thus obtained expressions for $W$ and
$\Phi$ in terms of the initial conditions \eq{bc} and \eq{bcw}, one
proceeds by inserting $W$ and $\Phi$ into the remaining two equations
\eqs{fe1}{fe2}, that is, ${\cal R}(Z,Y,\th)=0$ and ${\cal
D}\Phi(Z,Y,\th)=0$. These equations only need to be evaluated at $Z=0$,
since if \eqs{fe3}{fe5} hold then $[S,{\cal R}]_{\star}=0$ and
$S\star {\cal D}\Phi+ {\cal D}\Phi\star \bar{\pi}\,(S\,)=0$ are also
satisfied. The latter equations, when
expanded around the $AdS$ vacuum, yield linear $Z$-space differential
equations which imply that ${\cal R}$ and ${\cal D}\Phi$ vanish for all
$Z$, provided that

\be
\zeval{{\cal R}}\se 0\ ,\qquad
\zeval{{\cal D}\Phi}\se 0\la{rdphi}
\ee

are satisfied.

The initial conditions $\f_{i}$ defined in \eq{bc} and $\o_{i}$ defined
in \eq{bcw} are introduced independently. Since the structure of the
perturbative solution is quite complicated, it seems that the
perturbative solution scheme would lead to proliferation of degrees of
freedom. This is misleading however, since the integrability of the
exact equations implies that $\Phi$, $S$ and $W$ depend only on the
initial condition \eq{wphi}, that is on the quantities

\bea
\f(Y,\th)& := &\zeval{\Phi\,(Z,Y,\th)}\se
\f_{1}(Y,\th)+\f_{2}(Y,\th)+\cdots
\ ,\nn\w2
\o(Y,\th)& := &\zeval{W(Z,Y,\th)}\se
\O_{0}(Y)+\o_{1}(Y,\th)+\o_{2}(Y,\th)+
\cdots\ .\la{init}
\eea

%%%%%%%%%%%%%%%%%%%%%%%%%%%%%%%%%%%%%%%%%%%%%%%%%%%%%%%%%%%%%%%%%%%%%%%%%%%

\subsection{The Linearized Curvature Constraints}

%%%%%%%%%%%%%%%%%%%%%%%%%%%%%%%%%%%%%%%%%%%%%%%%%%%%%%%%%%%%%%%%%%%%%%%%%%%

At the linearized level, the equations \eq{delb}, \eq{dels} and
\eq{delw} read

\bea
d_Z \Phi_{1}&=&0\ ,
\la{p1}\w2
d_Z S_{1}&=& -\ft12\left(
dz^{2}\Phi_{1}\star
\k\,\C+d\zb^{2}\Phi_{1}\star \bar{\k}\right)\ ,
\la{p2}\w2
d_Z W_{1}&=&-\ft{i}{2}dS_{1}+\ft{i}{2}[\,\O_{0},\,S_{1}\,]_{\star} \ .
\la{p3}
\eea

From \eq{phisint}, \eq{physg} and \eq{wint} we find

\bea
\Phi_{1}(Z,Y,\th)&=& \f(Y,\th)\ ,\nn\w2
S_{1}(Z,Y,\th)&=&\int_{0}^{1}dt\,t\,\left( dz^{\a}z_{\a}\,
\f(-tz,\bar{y},\th)\,\k(tz,y)\star\C+d\zb^{\ad}\zb_{\ad}\, \f(y,t\zb,\th)\,
\bar{\k}(t\zb,\yb)\right)\nn\w2
W_{1}(Z,Y,\th)&=&\o(Y,\th)+\O_{1}(Z,Y,\th)\ ,\nn\w2
\O_{1}(Z,Y,\th)&=&-\ft{1}{2}\int_{0}^{1}\,dt'\int_{0}^{1}\,dt\,t\left\{
\left(itt'\,\o_{0}{}^{\a\b}z_{\a}z_{\b}+e_{0}{}^{\a\bd}z_{\a}\bar{\del}_{\bd}
\right)\f(-tt'z,\yb,\th)\,\k(tt'z,y)\star \C\right.\nn\w2
&&\qquad +\left.\left(itt'\bar{\o}_{0}{}^{\ad\bd}\zb_{\ad}
\zb_{\bd}-e_{0}{}^{\a\bd}
\zb_{\bd}\del_\a\right)\f(y,tt'\zb,\th)\,\kb(tt'\zb,\yb)\right\}\ ,
\la{w1sol}
\eea

where we have made the replacements $\f_1\rightarrow \f$ and
$\o_1\rightarrow\o$ and $\del_{\a}=\ft{\del}{\del y^{\a}}$. Notice
that in verifying the $\t$ invariance of \eq{w1sol} one has to make use
of $\t(\f(z,\yb,\th)=\pi\bp(\f(z,\yb,\th))$. Next we insert
$\Phi_1$ and $W_1$ into the linearization of \eq{rdphi}. Since $\f$
has no background value the linearization of $\zeval{{\cal D}\Phi}=0$
is simply $D_{\O_{0}}\f=0$. To linearize $\zeval{{\cal R}}=0$ it is
important to notice that $\zeval{\{\O_{0},\star\O_{1}\}_{\star}}\neq
0$, even though $\zeval{\O_{0}}=\zeval{\O_{1}}=0$. Hence the linearized
higher spin equations of motion in spacetime are given by

\bea
&&R_1\se
\zeval{\{\,\O_{0}\,,\, \O_{1}\,\}_{\star}}\ ,\nn\w2
&&d\f-\O_{0}\star \f+\f\star\bp(\O_{0})\se 0\ ,
\la{fdasol1}
\eea

where the AdS covariant linearized curvature is defined by

\be
R_1\;\; := \;\; d\o-\{\,\O_{0}\,,\, \o\,\}_{\star}\la{r1}
\ee

which obeys the Bianchi identity

\be
dR_1\se [\,\O_0\,,\, R_1\,]_{\star}\ .
\la{r1bi}
\ee

We proceed by substituting the definition of $\O_{0}$ given in
\eq{vac} into the equations \eq{fdasol1} and expanding all the $\star$
products. After some algebra and making use of \eq{appe} we obtain the
 Lorentz covariant, linearized curvature constraints

\bea
R_1(y,\yb,\th)&=&-\ft{i}{4}e_{0}{}^{\a\bd}\wedge
e_{0}{}^{\c}{}_{\bd}~\del_{\a}\del_{\c}\f(y,0,\th)-
\ft{i}{4}e_{0}{}^{\b\ad}\wedge
e_{0\,\b}{}^{\cd}~\bar{\del}_{\ad}\bar{\del}_{\cd}\f(0,\yb,\th)\star\C
\ ,\nn\w2
D\f(y,\yb,\th)&=& -i\,e_{0}{}^{\a\bd}\left(y_{\a}\yb_{\bd}
-\del_\a\bar{\del}_{\bd}\right) \f(y,\yb,\th)\ .\la{dc1}
\eea

Similarly, evaluating the $\star$ products in \eq{r1} and \eq{r1bi}
gives the identities

\bea
R_1(y,\yb,\th)& :=& D\o(y,\yb,\th)-e_{0}{}^{\a\bd}
\wedge\left(y_{\a}\bar{\del}_{\bd} +
\yb_{\bd}\del_\a\right)\o(y,\yb,\th)\ ,\la{rbi11}\w2
D\,R_1(y,\yb,\th)&=&e_{0}{}^{\a\bd}\wedge(y_{\a}\bar{\del}_{\bd}+\yb_{\bd}
\del_{\a})R_1(y,\yb,\th)\ ,\la{adscurv}
\eea

where $D$ is the Lorentz covariant, exterior derivative acting on a
$p$-form $X$ as

\be
DX \se dX-\o_{0}\star X + (-1)^{p} X\star\o_{0}\ .
\ee

Substituting the expansions for $\o$ and $\f$ given in \eq{wexp},
\eq{Csol} and \eq{A} into \eq{dc1} and expanding in $y$
and $\bar{y}$ gives the following component form of the linearized
curvature constraints

\bea
&& R^1_{\a\b,\mi\c1{2s-2}}(\th\,)\se C_{\a\b\mi\c1{2s-2}}(\th\,)
\ ,\quad\quad s=1,\ft32,2,...\ ,
\la{wt2}\w4
&& R^1_{\a\b,\mi\c1k\mi{\cd}{k+1}{2s-2}}(\th)\se 0\ ,\quad\quad
\quad\quad\quad s=\ft32,2,\ft52,...\ ,\quad k=0,..,2s-3
\la{rv1}\w4
&& D_{\a\ad}C_{\mi\b1m\mi{\bd}1n}(\th\,)\se
i\,C_{\a\mi\b1m\ad\mi{\bd}1n}(\th\,)-i\,mn\,\e_{\a\b_1}\,\e_{\ad\bd_1}\,
C_{\mi\b2m\mi{\bd}2n}(\th\,)
\nn\w2
&&\qquad \qquad\qquad\qquad\qquad\qquad\qquad
\qquad\quad\quad \ \  m,n=0,1,2,...\la{dcl1}
\eea

and the hermitian conjugates of \eqs{wt2}{rv1}. Similar manipulations
of \eqs{r1}{r1bi} yield

\bea
&& R^1_{\a_1\a_2,\mi\b1m\mi{\bd}1n}(\th\,)
\se 2\, D\o_{\a_1\a_2,\mi\b1m\mi{\bd}1n}(\th\,)
\nn\w2
&& -m\,\e_{\a_1\b_1}\,
      \o_{\a_2}{}^{\cd}{}_{,\mi\b2m\cd\mi{\bd}1n}(\th\,)
   -n\,\o_{\a_1\bd_1,\a_2\mi\b1m\mi{\bd}2n}(\th\,)\ ,
\la{rbi1}\w2
&& D_{\ad}{}^{\c}R^1_{\c\a,\mi\b1m\mi{\bd}1n}-
   D_{\a}{}^{\cd}R^1_{\cd\ad,\mi\b1m\mi{\bd}1n}
\nn\w2
&& \se  m\left[R^1_{\a\b_1,\mi\b2m\ad\mi{\bd}1n}-\e_{\a\b_1}
R^1_{\ad}{}^{\cd}{}_{,\mi\b2m\cd\mi{\bd}1n}\right]
\nn\w2
&& \qquad -n\left[R^1_{\ad\bd_1,\a\mi\b1m\mi{\bd}2n}-\e_{\ad\bd_1}
R^1_{\a}{}^{\c}{}_{,\c\mi\b1m\mi{\bd}2n}\right]\ .
\la{rbi}
\eea

In writing \eqs{wt2}{rbi} we have converted the curved indices of the
forms into flat indices using the AdS vierbein \eq{bgvb} and set

\bea
&&D_{\a\ad}\;\; := \;\;(\s^a)_{\a\ad}D_{a}\ ,
\qquad\qquad\qquad\quad\ \ \
\o_{\a\bd}(Y,\th)\;\; := \;\;(\s^{a})_{\a\bd}\,\o_{a}(Y,\th)\ ,\nn\w2
&&R^1_{\a\b}(Y,\th)\;\; := \;\;\ft12(\s^{ab})_{\a\b}R^1_{ab}(Y,\th)
\ ,\qquad D\o_{\a\b}(Y,\th)\;\; := \;\; \ft12(\s^{ab})_{\a\b}\,D_{a}\o_{b}(Y,\th)\ .
\la{defns}
\eea

The linearized curvature constraints thus assume the structure
suggested by the discussion at the end of section 3; the
$SO(8)$ content of the generalized Weyl tensor $C_{\mi\c1{2s}}(\th)$
matches that of the chiral curvature $R^1_{\c_1\c_2,\mi\c3{2s}}(\th\,)$,
as can be seen by comparing the expansions \eq{wexp} and \eq{A}, and
hence the curvature constraints \eq{wt2} and \eq{dcl1} have well-defined $\th$
expansions.

\bigskip
%%%%%%%%%%%%%%%%%%%%%%%%%%%%%%%%%%%%%%%%%%%%%%%%%%%%%%%%%%%%%%%%%%
\centerline{{\it Independent Constraints}}
%%%%%%%%%%%%%%%%%%%%%%%%%%%%%%%%%%%%%%%%%%%%%%%%%%%%%%%%%%%%%%%%%%
\bigskip

Not all constraints in \eqs{wt2}{dcl1} are independent. The
relationships among them are due to the Bianchi identities \eq{rbi} and
the fact that some of the constraints are just trivial identifications
of components in $R^1$ and $\f$.

To begin with, for $s\geq \ft52$ , the Bianchi identity \eq{rbi} and
the $k=1$ and $k=2$ components of the constraint \eq{rv1} yields

\be
R^1_{\a\b,\ad\mi{\bd}1n}(\th\,)-
\e_{\a\b}R^1_{\ad}{}^{\cd}{}_{,\cd\mi{\bd}1n}(\th\,)\se0\ ,
\qquad n=2,3,...\ ,\la{co}
\ee

which implies that for $s\geq \ft52$, eq. \eq{wt2} simply identifies
$C_{\mi\a1{2s}}(\th)$ with the generalized Weyl tensor of spin $s$.
Eq. \eq{co} also implies that the $k=0$ component of \eq{rv1} is
trivial.

For $s=2$, \eq{rbi} and the $k=1$ component of \eq{rv1} yield

\be
R^1_{\ad\bd,\a\b}(\th\,)-\e_{\ad\bd}R^1_{\a}{}^{\c}{}_{,\c\b}(\th\,)\se
R^1_{\a\b,\ad\bd}(\th\,)-\e_{\a\b}R^1_{\ad}{}^{\cd}{}_{,\cd\bd}(\th\,)\ .
\la{rbistwo}
\ee

This implies that when $R^1_{ab,\a\b}(\th)$ is expressed in terms of
the vierbein by solving the $k=1$ component of \eq{rv1} for
$\o_{a,\a\b}(\th)$, then $R^1_{ab}(ab,\a\b)(\th)$ consists of $20$ real
components: $10$ components in the Weyl tensor $R_{(\a\b,\c\d)}(\th)=
(R_{(\ad\bd,\cd\dd)}(\th))^{\dagger}$, $9$ components in $R_{\a\b,\cd\dd}(\th)=
R_{\cd\dd,\a\b}(\th)$ and $1$ component in $R^{\a\b}{}_{,\a\b}(\th)=
(R^{\ad\bd}{}_{,\ad\bd}(\th))^{\dagger}$. Hence, for $s=2$, \eq{wt2} decomposes
into two independent equations one of which sets the $SO(3,1)$ singlet
equal to zero and the other identifying $C_{\a\b\c\d}(\th)$ with the
spin $s=2$ Weyl tensor. The $k=0$ component of \eq{rv1} is an
independent constraint that sets the $9$ of $SO(3,1)$ equal to zero. In
section 7 we shall show that the $9+1$ constraints contain the Ricci
tensor, and they yield the linearized Einstein's equation with
cosmological constant.

For $s=\ft32$, eq. \eq{wt2} decomposes into two independent equations
one of which is $R^1_{\a}{}^{\b}{}_{,\b}(\th)=0$ and the other
identifying $C_{\a\b\c}(\th)$ with the spin $s=\ft32$ Weyl tensor
$R^1_{(\a\b,\c)}(\th)$. Eq. \eq{rv1}, where only $k=0$ is allowed, is
the independent constraint $R^1_{\a\b,\cd}(\th)=0$. In section 7 we
shall show that the independent constraints in this sector contain the
gravitino field equation.

For $s=1$ \eq{wt2} identifies $C_{\a\b}(\th)$ with the
$SO(8)\times SO(8)$ field strength $R^1_{\a\b}(\th)$ and likewise, for
$(m,n)=(0,0)$ \eq{dcl1} identifies $C_{\a\ad}(\th)$ with the derivatives
of the scalars.

Turning to \eq{dcl1}, we note that the Bianchi identity
\eq{rbi} and the identifications in \eq{wt2} yield

\be
D_{\ad}{}^{\b}C_{\b\mi\a1l}(\th)\se 0\ ,\qquad l=2,3,...\ ,
\ee

which implies that \eq{dcl1} is simply an identification of
$C_{\a\mi\b1m\ad}(\th)$ with
$D_{\ad(\a }C_{\mi\b1m)}(\th)$ for $m\geq3$. On the other hand, for
$(m,n)=(2,0)$, $(1,0)$, $(1,1)$ it is clear that \eq{dcl1} yields
independent constraints, leading to the spin $s\leq 1$ equations of
motion. The remaining components of \eq{dcl1} (that is for $m+n\geq3$
and $m,n\geq1$) are redundant. To show this we assume that \eq{dcl1}
holds for $D_{\a\ad}C_{\mi\b1{m-1}\mi{\bd}1{n-1}}(\th)$. Then it
follows that

\bea
&&D_{\b_1}{}^{\ad}C_{\mi\b2{m+1}\ad\mi{\bd}1{n-1}}(\th\,)\se 0\ ,\nn\w2
&&D^{\a\ad}C_{\a\mi\b1{m-1}\ad\mi{\bd}1{n-1}}(\th\,)\se -i\,(m+1)(n+1)
C_{\b1{m-1}\mi{\bd}1{n-1}}(\th\,)\ ,\la{dcl3}
\eea

which in turn implies that \eq{dcl1} holds for
$D_{\a\ad}C_{\mi\b1m\mi{\bd}1n}(\th)$.

To summarize, the independent constraints are

\begin{itemize}

\item[$(i)$] the $k\geq 1$ components of \eq{rv1} for $s\geq 2$:

\bea
&&R^1_{\a\b,\mi\c1k\mi{\cd}{k+1}{2s-2}}(\th\,)\se 0\ ,\nn\w2
&&R^1_{\ad\bd,\mi\c1k\mi{\cd}{k+1}{2s-2}}(\th\,)\se 0\ ,\qquad s=2,\ft52,3,...
\ ,\qquad k=1,...,2s-3\ ,
\la{i}
\eea

\item[$(ii)$] the following components of \eq{wt2} and \eq{rv1} for $s=2$:

\bea
&&R^1_{\a\b,}{}^{\a\b}(\th\,)\se 0
\ ,\nn\w2
&&R^1_{\a\b,\cd\dd}(\th\,) \se 0\ ,
\la{iii}
\eea

\item[$(iii)$] the following components of \eq{wt2} and \eq{rv1}
for $s=\ft32$:

\bea
&&R^1_{\a\b,}{}^{\b}(\th\,)\se R^1_{\ad\b,}{}^{\b}(\th\,)\se 0
\ ,\nn\w2
&&R^1_{\a\b,\cd}(\th\,)\se R^1_{\ad\bd,\c}(\th\,)\se 0\ ,
\la{iv}
\eea

\item[$(iv)$] and the following components of \eq{dcl1} for $s=0,\ft12,1$:

\bea
D_{\a\ad}C_{\b\c}(\th\,)&=& i\,C_{\a\b\c\ad}(\th\,)\ ,\nn\w2
D_{\a\ad}C_{\b}(\th\,)&=& i\, C_{\a\b\ad}\ ,\nn\w2
D_{\a\ad}\f_{\b\bd}(\th\,)&=&i\,\f_{\a\ad}(\th\,)-i\,\e_{\a\b}\e_{\ad\bd}
\,\f(\th\,)\ ,
\la{va}
\eea

together with the identities

\be
C_{\a\b}(\th)\se R^1_{\a\b}(\th)\ ,\qquad
\f_{\a\ad}(\th)\se-i\,D_{\a\ad}\f(\th)\ .
\la{vb}
\ee

\end{itemize}

The above analysis shows the auxiliary status of the zero-forms
$\f(m,n,\th)$ with $m+n\geq 2$ and the dynamical status of the spin
$s\leq\ft12$ fields $\f(m,n,\th)$ with $m+n\leq 1$.

\bigskip
%%%%%%%%%%%%%%%%%%%%%%%%%%%%%%%%%%%%%%%%%%%%%%%%%%%%%%%%%%%%%
\centerline{{\it Symmetries of the Linearized Equations}}
%%%%%%%%%%%%%%%%%%%%%%%%%%%%%%%%%%%%%%%%%%%%%%%%%%%%%%%%%%%%%
\bigskip

Inserting the expansion \eq{pert}, and similar expansions for the gauge
parameters where the leading term is the Killing parameter, into the
expression \eq{gauge} for the gauge transformations, linearizing and
setting $Z=0$ using \eq{zeval} we find that the linearized field
equations \eqs{wt2}{dcl1} are invariant under the Killing symmetries

\be
\d_{\vare_{0}}\o\se \zeval{[\,\vare_{0}\,,\,W_{1}\,]_{\star}}
\ ,\qquad
\d_{\vare_{0}} \f\se \vare_{0}\star \f- \f\star \bp\,(\vare_{0}\,)\ ,\la{gl}
\ee

where $\vare_{0}$ obey \eq{bginv}, and the local gauge transformations

\be
\d_{\vare} \o\se d\vare-[\,\O_{0}\,,\,\vare\,]_{\star}
\ ,\qquad
\d_{\vare}\f\se 0\ ,
\la{lo}
\ee

where $\vare$ is an arbitrary \alg-valued parameter. These
transformation in components read

\bea
\d\o_{\a\ad,\mi\b1m\mi{\bd}1n}&=& D_{\a\ad}\vare_{\mi\b1m\mi{\bd}1n}
+m \,\e_{\a\b_1}\vare_{\mi\b2m\mi{\bd}1n}+n\,
\e_{\ad\bd_1}\vare_{\mi\b1m\mi{\bd}2n}\ .\nn\w2
\la{lingauge}
\eea

The linearized field equations transform into each other under \eq{gl}
but they are separately invariant under \eq{lo}. The local symmetries
\eq{lo} will be used to impose gauge conditions on the gauge field
$\o$.

%%%%%%%%%%%%%%%%%%%%%%%%%%%%%%%%%%%%%%%%%%%%%%%%%%%%%%%%%%%%%%%%%%%%%%%%%

\section{Spectral Analysis}

%%%%%%%%%%%%%%%%%%%%%%%%%%%%%%%%%%%%%%%%%%%%%%%%%%%%%%%%%%%%%%%%%%%%%%%%%

In section 6.1 we shall analyze the elimination of the auxiliary gauge
fields $\o(m,n,\th)$ with $|m-n|>1$ by solving \eq{i}. In section 6.2
we shall analyze the equations of motion in the $s\geq
\ft32$ sector of the theory. For $s=\ft32$ and $s=2$, these follow from
\eq{iii} and \eq{iv}, respectively, and for $s\geq \ft52$ they follow
from \eq{i}. Finally, in section 6.3 we shall analyze the equations of
motion in the $s\leq 1$ sector following from \eq{va}.

 For $s=\ft52,\ft72,...$, the dynamical field equations are
 ``Dirac like" first order equations obtained from some of the Lorentz irreps
of the physical fermionic curvatures $R_1(s-\ft32,s-\ft12,\th)$ ( the
ones that are not used up in solving for the auxiliary gauge fields
$\o(s-\ft52,s+\ft12,\th)$). For $s=3,4,...$ the dynamical field equations
are the ``Klein-Gordon like" second order equations obtained from some
of the Lorentz irreps of the auxiliary bosonic curvatures
$R_1(s-2,s,\th)$ (the ones that are not used up in solving for
$\o(s-3,s+1,\th)$)
\footnote{
%%%
%%%%%%%%%%%%%%%%%%%%%% footnote %%%%%%%%%%%%%%
In the action formalism the gauge invariant quadratic action leading to
the correct dynamical equations of motion fails in producing the
constraints allowing to solve for $\o(m,n,\th)$ with
$|m-n|>2$. These therefore have to imposed by hand, since it is
crucial to solve for all auxiliary fields at the linearized level in
order for higher order interactions to make sense \cite{v1}. Thus, the
description of the interacting higher spin gauge theory seems to be
more cumbersome in the action formalism than in the free differential
algebra approach where the field equations are obtained from an
integrable set of constraints.}.

%%%%%%%%%%%%%%%%%%%%%%%%%%%%%%%%%%%%%%%%%%%%%%%%%%%%%%%%%%%%%%%%%%%%%%%%

\subsection{Elimination of Auxiliary Fields}

%%%%%%%%%%%%%%%%%%%%%%%%%%%%%%%%%%%%%%%%%%%%%%%%%%%%%%%%%%%%%%%%%%%%%%%

The strategy for solving the generalized torsion constraints \eq{i} is
to first decompose the gauge fields and their derivatives and the gauge
transformations \eq{lingauge} into Lorentz irreducible tensors. An
auxiliary gauge field is then a gauge field that can be eliminated
completely in the sense that each irrep of the gauge field can either
be solved for algebraically using \eq{i} or be set equal to zero by
using the gauge transformations that take the form of Stuckelberg type
shifts.

In order to solve the constraints \eqs{i}{iv}, we first solve
\eq{i} which can be written as $R^1_{ab}(m,n)=0$,
%%%%%%%%%%%%%%  footnote
\footnote{ In this and the next section we will
suppress the $\th$-dependence of all the fields and parameters.}
%%%%%%%%%%%%%%%%%%%%%%%%%
for $|m-n|=0,1,2$ and the constraints \eqs{iii}{iv}. Using these
results, we then solve the constraints \eq{i} for the remaining values
$|m-n|\ge 3$.

\newpage

%%%%%%%%%%%%%%%%%%%%%%%%%%%%%%%%%%%%%%%%%%%%%%%%%%%%%%%%%%%%%
{\bf The $m-n=0$, $s \geq 2$ Sector:}
%%%%%%%%%%%%%%%%%%%%%%%%%%%%%%%%%%%%%%%%%%%%%%%%%%%%%%%%%%%%%
\bigskip

For convenience let us write $m=n=s-1$. From the decomposition rules
\eq{decomp}, we find that the constraint $R^{1}_{ab}(s-1,s-1)=0$ has the
schematical structure (suppressing the exact
$s$-dependent values of the coefficients)

\bea
&&\l(s+1,s-1)+\bar{\z}(s+1,s-1)\se 0\ ,
\nn\w2
&&\l(s-1,s-1)+\z(s-1,s-1)+\bar{\z}(s-1,s-1)\se 0\ ,
\nn\w2
&&\l(s-3,s-1)+\z(s-3,s-1)\se 0\ ,
\la{solv1}
\eea

where $\l$'s are the three irreps that appear in the decomposition of
the self-dual part of $D_{[a}\o_{b]}(s-1,s-1)$ and $\z$ and $\bar{\z}$
are irreps that appear in the decomposition of the generalized Lorentz
connection $\o_{\a\ad}(s-2,s)$ and its hermitian conjugate
$\o_{\a\ad}(s,s-2)$. Thus we can solve for $\y(s-1,s+1)$ in terms of $
\l(s+1,s-1)^{\dagger}$, and $\z(s-1,s-1)$ in terms of
$\l(s-1,s-1)$ and $\l(s-1,s-1)^\dagger$, and $\z(s-3,s-1)$ in terms of
$\l(s-3,s-1)$. The irrep $\z(s-3,s+1)$ for $s\geq 3$ does not appear in
\eq{solv1} and therefore it remains undetermined. Putting these results
together, we find the following relation between the generalized
Lorentz connections and generalized vierbeins:

\bea
\o_{\a\ad,\mi{\b}1{s-2}\mi{\bd}1{s}}&=&
 -\ft{s-3}{s-1}D\o_{\ad\bd_1,\a\mi{\b}1{s-2}\mi{\bd}2{s}}
+D\o_{\bd_1\bd_2,\a\mi{\b}1{s-2}\ad\mi{\bd}3s}\nn\w2
&&+\ft3{s+1}\e_{\ad\bd_1}D\o_\a{}^{\c}{}_{,\c\mi\b1{s-2}\mi{\bd}2{s}}
+\ft{s-2}{s+1}\e_{\ad\bd_1}
D\o_{\b_1}{}^{\c}{}_{,\c\a\mi\b2{s-2}\mi{\bd}2s}
\nn\w2
&&+(s-2)\e_{\a\b_1}\z_{\mi\b2{s-2}\ad\mi{\bd}1s}\ ,
\quad\quad s=2,3,4,...
\la{lcsol}
\eea

Another way of understanding the presence of the undetermined irrep
$\z(s-3,s+1)$ in \eq{lcsol} for $s\geq 3$ is to note that the gauge
symmetry with parameter $\vare(s-3,s+1)$ transforms the generalized
Lorentz connection but not the generalized vierbein; from
\eq{lingauge} and \eq{lcsol} it follows that $\d\z(s-3,s+1)=\vare(s-3,s+1)$
($s\geq 3$). Since this gauge symmetry acts by shifting $\z(s-3,s+1)$ it
can be fixed uniquely by imposing the gauge condition

\be
\z(s-3,s+1)\se 0\ ,\qquad s=3,4,...\ .
\la{solv4}
\ee

Hence the generalized Lorentz connections and the gauge symmetry with parameter
$\vare(s-3,s+1)$ are auxiliary.

There are also constraints which arise from the Bianchi identities
\eq{rbi} upon the use of the use of the vanishing curvature constraints
\eq{i}. Using the decomposition rules \eq{decomp} and using the constraints
$R^{1}_{ab}(s-1,s-1)=0$ in the Bianchi identity \eq{rbi} for $s\geq 2$, we
obtain

\bea
&&\x(s,s)\se \bar{\y}(s,s)\ ,
\nn\w2
&&\y(s-2,s-2)\se\bar{\x}(s-2,s-2)\ ,
\nn\w2
&&(s-2)\x(s-2,s)+\y(s-2,s)\se 0\ , \qquad s=2,3,...\ ,
\la{solv2}
\eea

where $(\x,\y)$ and $({\bar \x}, {\bar y})$ are the irreps that appear
in the decompositions of $R^1_{ab}(s-2,s)$ and $R^1_{ab}(s,s-2)$,
respectively. The first two equations state that $\x(s,s)$ and
$\y(s-2,s-2)$ are real.

\bigskip
%%%%%%%%%%%%%%%%%%%%%%%%%%%%%%%%%%%%%%%%%%%%%%%%%%%%%%%%%%%%%
{\bf The $|m-n|=2$, $s \geq 2$ Sector:}
%%%%%%%%%%%%%%%%%%%%%%%%%%%%%%%%%%%%%%%%%%%%%%%%%%%%%%%%%%%%%
\bigskip

For convenience, we set $m=s-2$ and $n=s$. Taking into account
\eq{solv2} in the constraint $R^{1}_{ab}(s-2,s)=0$, we find

\be
\x(s,s)\se \y (s-2,s-2)\se 0\ ,\qquad s=2,3,...\ ,
\la{solv5}
\ee

and

\bea
&&\x(s-2,s)+ \z^{(-)}(s-2,s)\se0\nn\w2
&&\x(s-4,s)\se
\y(s-2,s+2)\se 0\ ,\qquad s=3,4,...
\la{solv6}
\eea

None of the irreps in \eq{solv6}, nor any other of the remaining
vanishing curvatures components, depend on the generalized vierbein.
By examining the exact coefficients of the last equation in \eq{solv2}
and the first equation in \eq{solv6} one can show that these two
equations are linearly independent and that the latter one is
independent of the generalized vierbein. Moreover the generalized
vierbein does not appear undifferentiated in any of the remaining
curvature constraints. Hence the generalized vierbeins are dynamical
and obey the equations of motion \eq{solv5}, which can be written more
explicitly as

\be
e_{0,\a_1}{}^{\bd}\wedge R^1_{\mi\a2{s-1}\bd\mi{\ad}1{s-1}}\se 0\ ,
\qquad s=2,3,...
\la{beom}
\ee

\bigskip
%%%%%%%%%%%%%%%%%%%%%%%%%%%%%%%%%%%%%%%%%%%%%%%%%%%%%%%%%%%%%
{\bf  The $|m-n|=1$, $s \geq \ft32$ Sector}
%%%%%%%%%%%%%%%%%%%%%%%%%%%%%%%%%%%%%%%%%%%%%%%%%%%%%%%%%%%%%
\bigskip

For convenience, we set $m=s-\ft32$ and $n=s-\ft12$. The constraint
$R^1_{ab}(s-\ft32,s-\ft12)=0$ for $s\geq \ft52$, together with \eq{iv}
and the Bianchi identity \eq{rbi} imply

\bea
&&\x(s+\ft12,s-\ft12)\se  0\ ,\nn\w2
&&(s-\ft32)\z^{(-)}(s-\ft32,s-\ft52)\se 0\ ,\nn\w2
&&(s-\ft32)\x(s-\ft32,s-\ft12)+\y(s-\ft32,s-\ft12)\se 0\ ,
\qquad s=\ft32,\ft52,...
\la{solv3}
\eea

and

\bea
&&\x(s-\ft32,s-\ft12)+\y(s-\ft32,s-\ft12)\se 0\nn\w2
&&(s-\ft52)\x(s-\ft72,s-\ft12)\se\z^{(-)}(s-\ft32,s+\ft32)
\se 0\ ,\qquad s=\ft52,\ft72,...\ .
\la{solv9}
\eea

Following steps analogous to those used in the analysis of
\eqs{solv5}{solv6} we find that none of the irreps in \eq{solv9}, nor
any other of the remaining vanishing curvatures components, depend on
the generalized gravitini. The two linear combinations of
$\x(s-\ft32,s-\ft12)$ and $\y(s-\ft32,s-\ft12)$ that
appear in \eq{solv3} and \eq{solv9} have different coefficients; the
linear combination in \eq{solv9} is independent of the generalized
gravitino. Hence the generalized gravitini are dynamical and obey the
equations of motion \eq{solv3}, which can be written more explicitly as

\be
e_{0,\a_1}{}^{\bd}\wedge R^1_{\mi\a2{s-3/2}\bd\mi{\ad}1{s-1/2}}\se 0\ ,
\qquad s=\ft32,\ft52,...
\la{feom}
\ee

\bigskip
%%%%%%%%%%%%%%%%%%%%%%%%%%%%%%%%%%%%%%%%%%%%%%%%%%%%%%%%%%%%%%%%%%%%
{\bf The General Case:}
%%%%%%%%%%%%%%%%%%%%%%%%%%%%%%%%%%%%%%%%%%%%%%%%%%%%%%%%%%%%%%%%%%%%
\bigskip

Suppose that we have solved $R^1_{ab}(m+2,n-2)=0$ for the auxiliary
gauge fields $\o_{\a\ad}(m+1,n-1)$ in terms of $D_{[a}\o_{b]}(m+2,n-2)$
and fixed the auxiliary gauge symmetries with parameters
$\vare(m,n)$ for $n\geq m+ 6\geq6 $ in the bosonic sector or
$n\geq m+ 5\geq5 $ in the fermionic sector. We then turn to solving
$R^1_{ab}(m+1,n-1)=0$. In doing so, we make use of the the Bianchi identity
\eq{rbi} for $R^1_{ab}(m+2,n-2)$ which reads

\be
\x(m+3,n-1)\se \x(m+1,n-1)+\y(m+1,n-1)\se
\y(m+1,n-3)\se 0\ ,
\la{solv8}
\ee

where $\x$ and $\y$ are the irreps that appear in the decomposition of
$R^1_{ab}(m+1,n-1)$. Hence the remaining independent components
of $R^1_{ab}(m+1,n-1)=0$ are given by

\be
\x(m+1,n-1)+\y(m+1,n-1)\se\x(m-1,n-1)\se
\y(m+1,n+1)\se0 \ ,
\la{solv10}
\ee

where the two linear combinations of $\x(m+1,n-1)$ and $\y(m+1,n-1)$ in
\eq{solv8} and \eq{solv10} are independent.
Since \eq{solv10} reduces to \eq{solv1}, \eq{solv6} and \eq{solv9} we
may as well define the range of $m$ and $n$ in \eq{solv10} to be
$|m-n|>1$, and it is straight forward to obtain:

\bea
&&\o_{\a\ad,\mi\b1m\mi{\bd}1n}\se \ft1{2(m+1)}D\o_{\bd_1\bd_2,\a\mi\b1m
\ad\mi{\bd}1{n-2}}
\nn\w2
&&+\e_{\ad\bd_1}\ft{n}{2(n+1)}\left[\ft{n-1}{m+n+2}D\o_{\bd_2}
{}^{\cd}{}_{,\a\mi\b1m\cd\mi{\bd}3n}\right.
\nn\w2
&&+\ft{n+1}{(m+1)(m+n+2)}\left(
D\o_{\a}{}^{\c}{}_{,\c\mi\b1m\mi{\bd}2n}
+m\,D\o_{\b_1}{}^{\c}{}_{,\c\a\mi\b2m\mi{\bd}2n}\right)
\nn\w2
&&\left.-\ft{m}{(m+1)(m+2)}\e_{\a\b_1}D\o^{\c\d}{}_{,\c\d\mi\b2m
\mi{\bd}2n}\right]
\nn\w2
&&+m\,\e_{\a\b_1}\y_{\mi\b2m\ad\mi{\bd}1n}\ ,\qquad |m-n|>1\ ,
\la{auxsolv}
\eea

where $\y(m-1,n+1)$ is undetermined. Notice that $R^1_{ab}(1,n-1)=0$
determines $\o_{\a\ad}(0,n)$ uniquely for $n\geq 2$. From
\eq{lingauge} it follows that the transformation of $\y(m-1,n+1)$
under the gauge symmetry with parameter $\vare(m-1,n+1)$ is given by

\be
\d\y(m-1,n+1)\se
\vare(m-1,n+1)\ ,\qquad |m-n|>1\ .\la{solv11}
\ee

This gauge symmetry can thus be fixed uniquely by imposing the
algebraic gauge condition

\be
\y(m-1,n+1)\se 0\ ,\qquad |m-n|>1\ .
\la{solv12}
\ee

Hence the gauge fields in \eq{auxsolv} and the gauge symmetries used to
impose \eq{solv12} are auxiliary.

\bigskip
%%%%%%%%%%%%%%%%%%%%%%%%%%%%%%%%%%%%%%%%%%%%%%%%%%%%%%%%%%%%%%%%%%%%%%
\centerline{{\it Summary of Dynamical Fields, Gauge Symmetries and
Equations of Motion}}
%%%%%%%%%%%%%%%%%%%%%%%%%%%%%%%%%%%%%%%%%%%%%%%%%%%%%%%%%%%%%%%%%%%%%%
\bigskip

We thus conclude that the dynamical degrees of freedom of the higher
spin theory and the corresponding dynamical equations of motion and
gauge symmetries are:

\begin{itemize}

\item[$i$)] the generalized vierbeins $\o(s-1,s-1,\th)$ ($s=2,3,...$)
obeying the second order equations \eq{beom}, with invariance under the
generalized reparametrizations and generalized Lorentz transformations
given by

\bea
\d \o_{\a\ad,\mi\b1{s-1}\mi{\bd}1{s-1}}
&=& D_{\a\ad}\vare_{\mi\b1{s-1}\mi{\bd}1{s-1}}
+(s-1)\e_{\a\b_1}\vare_{\mi\b2{s-1}\ad\mi{\bd}1{s-1}}
\nn\w2
&& +(s-1)\e_{\ad\bd_1}\vare_{\a\mi\b1{s-1}\mi{\bd}2{s-1}}\ ,
\la{gengt}
\eea

\item[$ii$)] the generalized gravitini fields
$\o(s-\ft32,s-\ft12,\th)$($s=\ft32,\ft52,...$) and their hermitian
conjugates obeying the first order equations \eq{feom} with invariance
under the generalized local supersymmetries and the local fermionic
transformations given by

\bea
\d \o_{\a\ad,\mi\b1{s-3/2}\mi{\bd}1{s-1/2}}
&=&  D_{\a\ad}\vare_{\mi\b1{s-3/2}\mi{\bd}1{s-1/2}}
+(s-\ft32)\e_{\a\b_1}\vare_{\mi\b2{s-3/2}\ad\mi{\bd}1{s-1/2}}
\nn\w2
&&+(s-\ft12)\e_{\ad\bd_1}\vare_{\a\mi\b1{s-3/2}\mi{\bd}2{s-1/2}}\ ,
\la{gengtf}
\eea

\item[$iii)$] the two $SO(8)$, spin $s=1$ gauge fields $\o_{ij}(0,0)$
and $\o_{\mi{i}16}(0,0)$ transforming as

\be
\d \o_{ij} \se d \vare_{ij}\ ,\quad\quad
\d \o_{\mi{i}16} \se d \vare_{\mi{i}16}\ ,
\ee

the spin $s=\ft12$ fermions $C_{\a}^{ijk}$,
$C_{\a}^{i_1\cdots i_7}$ and their hermitian conjugates and the
scalars $\f$ and $\f_{ijkl}$ defined in \eq{nn} and obeying the reality
condition \eq{sr}. The fields with $s \leq 1$ obey equations of motion
obtainable from \eq{va} and \eq{vb}.

\end{itemize}

%%%%%%%%%%%%%%%%%%%%%%%%%%%%%%%%%%%%%%%%%%%%%%%%%%%%%%%%%%%%%%%%%%%%%%%%

\subsection{The Analysis of the Spin $s\geq\ft32$ Equations of Motion}

%%%%%%%%%%%%%%%%%%%%%%%%%%%%%%%%%%%%%%%%%%%%%%%%%%%%%%%%%%%%%%%%%%%%%%%%

\centerline{\it The Bosonic Sector}

%%%%%%%%%%%%%%%%%%%%%%%%%%%%%%%%%%%%%%%%%%%%%%%%%%%%%%%%%%%%%%%%%%%%%%%%

In the bosonic case, the linearized equations of motion for the generalized
vierbein $\o_{\a\ad}(s-1,s-1)$ ($s\geq 2$) is given by \eq{beom}. Combining
this equation with \eq{rbi11} we find

\bea
&&e_{0\,\a_1}{}^{\bd}\wedge\left[
D\o_{\mi\a2{s-1}\bd\mi{\ad}1{s-1}}+e_{\bd}{}^{\b}\wedge
\o_{\b\mi\a2{s-1}\mi{\ad}1{s-1}}
\right.\nn\w2
&&\qquad \left.+(s-1)e_{\ad_{1}}{}^{\b}\wedge
\o_{\b\mi\a2{s-1}\bd\mi{\ad}2{s-1}}
\right]\se0\ ,\qquad\quad\  s\se 2,3,4,...
\la{bose}
\eea

Combining \eq{bose} with the expression \eq{lcsol} for the generalized
Lorentz connection $\o_{\a\ad}(s-2,2)$ in terms of the generalized
vierbein and the gauge condition \eq{solv4} we find that the linearized
equations are second order in derivatives and that they are invariant
under the gauge transformations \eq{gengt}. In order to determine the
spectrum we shall use the gauge invariance to fix a gauge in which the
equations reduce to a Klein-Gordon like equation
$(D^2+M^2(s))\y(s,s)=0$ where $\y(s,s)$ is the spin $s$ irrep of the
generalized vierbein and
$M^2(s)$ a critical AdS-mass such that the equation possesses a
residual, on-shell gauge symmetry which leaves only two physical modes
in $\y(s,s)$. These two modes correspond to the ``massless"
representations of AdS group $SO(3,2)$. As for the ordinary photon
representations of the Poincar\'e algebra these massless
representations show the characteristic feature of multiplet shortening
(where the on-shell gauge modes correspond to the null-states of the
representation modules). The requirement of multiplet shortening amounts
to an $s$ dependent condition on the $SO(3,2)$ Casimir $C_2[SO(3,2)]$
which when turned into a differential operator acting on the modes of
the higher spin fields generates the critical value of the mass term
$M^2(s)$.

The gauge symmetries \eq{gengt} allow us to fix the generalized Lorentz
type gauge

\bea
&&D^{\a\ad}\o_{\a\ad,\mi\b1{s-1}\mi{\bd}1{s-1}}\se0\ ,
\la{bg1}\w2
&&\o_{\ad}{}^{\b}{}_{,\b\mi\b1{s-2}\mi{\bd}1{s-1}}\se 0\ ,\qquad
s=2,3,...
\la{bg2}
\eea

The algebraic condition \eq{bg2} eliminates all Lorentz irreps in the
generalized vierbein except the $(s,s)$ irrep. The gauge condition
\eq{bg1} and the $(s-2,s-2)$ component of \eq{bg2} fix the generalized
reparametrizations up to residual gauge transformations with parameters
obeying

\be
[D^{2}+1-s^2]\vare_{\mi\b1{s-1}\mi{\bd}1{s-1}}\se0\ ,\qquad
D^{\a\ad}\vare_{\a\mi\b1{s-2}\ad\mi{\bd}1{s-2}}\se0\ .
\la{resb}
\ee

The residual gauge transformations also involve compensating
generalized local Lorentz transformations with parameters given by

\be
\vare_{\mi\a1{s-2}\mi{\ad}1s}\se \ft1s D^{\b}{}_{(\ad_1}
\vare_{\,|\b\mi\a1{s-2}|\mi{\ad}2s)}\ ,
\ee

such that the first derivatives of the parameters of the generalized
reparametrization are only subject to the second condition in
\eq{resb}. The $(s-2,s)$ component of \eq{bg2} fixes uniquely the
generalized Lorentz gauge parameter $\vare(s-2,s)$. For $s=2$ and
$\th=0$ the gauge condition \eq{bg1} yields transversality condition
$D^{\m}\o_{\m,b}$ on the graviton field $\o_{\m,a}$, the $(0,0)$
component of \eq{bg2} yields the tracelessness condition
$\y^{ab}\o_{a,b}$ and the $(0,2)$ component of \eq{bg2} implies the
Lorentz gauge $\o_{[a,b]}=0$.

From \eq{bg1} and \eq{bg2} it follows that the derivatives of the
generalized vierbein obey

\be
D\o_{\a}{}^{\b}{}_{,\b\mi\b1{s-2}\mi{\bd}1{s-1}}\se0\ ,
\la{gfr}
\ee

which implies that $D\o_{\a\b,\mi\c1{s-1}\mi{\cd}1{s-1}}$ is symmetric
in all its undotted indices. The solution \eq{lcsol} for the generalized
Lorentz connection then simplifies to

\be
\o_{\a_1\ad_1,\mi\a2{s-1}\mi{\ad}2{s+1}}\se \ft2{s-1}
D\o_{\ad_1\ad_2,\mi\a1{s-1}\mi{\ad}3{s+1}}\ .\la{lcsolgf}
\ee

After a straightforward calculation, where one repeatedly makes use of
\eqs{bg1}{bg2}, \eq{gfr} and the expression \eq{rie} for the Lorentz
curvature, one finds

\be
\left[D^2+3-(s-1)^{2}\right]\o_{\a_1\ad_1,\mi\a2s\mi{\ad}2s}\se0\ .
\la{gvbe}
\ee

This equation describes an irreducible, massless spin
$s$ field $\o_{\a_1\ad_1,\mi\a2s\mi{\ad}2s}$ with $AdS$-energy

\be
E_{0}\se s+1\ .
\la{adse}
\ee

To verify \eq{adse} we consider the harmonic expansion on the coset
$SO(3,2)/SO(3,1)$. The positive energy representations of $SO(3,2)$ can
be characterized in the $SO(3)\times SO(2)$ basis, where $SO(3)$
is generated by the spatial rotations $M_{ij}$ ($i,j=1,2,3$) and
$SO(2)$ by the energy operator $M_{04}$. These representations are
labeled by the lowest energy $E_{0}$ and the highest eigenvalue
$s_{0}$ of $M_{12}$ when the energy is fixed to be $E_{0}$.
Following the procedure described in \cite{es1,es2}, we
Euclideanize the $AdS$ group to $SO(5)$, with irreps labeled by
highest weights $(n_1,n_2)$, and the Lorentz group to $SO(4)$,
with irreps labeled by highest weights $(j_1,j_2)$. The quadratic
Casimir eigenvalues for these groups are

\bea
C_{2}[SO(3,2)]&=&E_0(E_0-3)+s(s+1)\ ,
\w2
C_2[SO(5)]&=& n_1(n_1+3)+n_2(n_2+1)\ .
\eea

This suggests that in continuing $SO(5)$ back to $SO(3,2)$ we identify
$n_1$ with $-E_{0}$ and $n_2$ with $s_0$. Since the $SO(4)$ content of
the gauge fixed generalized vierbein $\o_{\a_1\ad_1,\mi\a2s\mi{\ad}2s}$
is $(j_1,j_2)=(s,0)$ we can expand $\o_{\a_1\ad_1,\mi\a2s\mi{\ad}2s}$ in
terms of representation functions of $SO(5)$ as follows

\be
\o_{\a_1\ad_1,\mi\a2s\mi{\ad}2s}(x)\se \sum_{n_1\geq s \geq n_2\geq 0}
\sum_{p}\o_{p}^{(n_1n_2)}
D_{\mi\a1s\mi{\ad}1s,\,p}^{(n_1n_2)}(L_{x}^{-1})\ ,
\la{he1}
\ee

where $\o_{p}^{(n_1n_2)}$ are constant expansion coefficients,
$D_{\mi\a1s\mi{\ad}1s,\,p}^{(n_1n_2)}(L_x^{-1})$, known as Wigner
functions, refer to the representation of the coset representative
$L_{x}^{-1}$ with rows labeled by $\mi\a1s$ and $\mi{\ad}1s$ and
columns by $p=1,...,dim(n_1n_2)$. The eigenvalues of the d'Alembertian
acting on the Wigner functions are computed from the formula

\be
D^{2} L_x^{-1}\se
-\l^{2}\Big(C_2[SO(5)]-C_2[SO(4)]\Big)
L_x^{-1}\ ,
\la{d2}
\ee

where $D^{2}$ is the d'Alembertian in the Euclidean metric (which have
opposite sign to the $AdS$ d'Alembertian) and the quadratic Casimir of
$SO(4)$ is given by

\be
C_2[SO(4)]\se j_1(j_1+2)+j_2^{2}\ .
\la{so4}
\ee

Thus, using that in continuing back to $AdS$ one has to let
$D^2\rightarrow -D^2$ and $n_1\rightarrow -E_{0}$, and setting the
inverse $AdS$ radius $\l=1$, we find from \eq{gvbe} that the energy
eigenvalues are to be solved from the characteristic equation

\be
E_{0}(E_0-3)-s+3-(s-1)^{2}\se 0\ ,
\ee

with the positive energy solution \eq{adse}.

To calculate the number of massless modes we notice that the gauge
transformations generated by the residual parameters obeying \eq{resb}
obey the gauge fixed equations of motion \eq{gvbe}. Hence the number of
real on-shell degrees of freedom is given by the number of components
of the spin $s$ irrep ($(s+1)^2$), minus the number of gauge conditions
\eq{bg1} linear in derivatives ($s^2$), minus the number of residual
gauge symmetries, which is equal to the number of degrees of freedom in
$\vare(s-1,s-1)$ ($s^2$) minus the number of constraints \eq{resb}
linear in derivatives ($(s-1)^2$). Thus there are

\be
(s+1)^2-s^2-[s^2-(s-1)^2]\se2\ ,
\ee

on-shell degrees of freedom with spin $s=2,3,4,...$ and energy $E_{0}=s+1$
describing massless higher spin bosons.

\bigskip
%%%%%%%%%%%%%%%%%%%%%%%%%%%%%%%%%%%%%%%%%%%%%%%%%%%%%%%%%%%%%%%%%%
\centerline{\it The  Fermionic Sector}
%%%%%%%%%%%%%%%%%%%%%%%%%%%%%%%%%%%%%%%%%%%%%%%%%%%%%%%%%%%%%%%%%%
\bigskip

In the fermionic case, we combine the equation \eq{feom} with \eq{rbi11}
to obtain the following first order equation for the generalized gravitini
fields $\o(s-\ft32,s-\ft12)$ and their hermitian conjugates
$\o(s-\ft12,s-\ft32)$:

\bea
&&e_{0\,\a_1}{}^{\bd}\wedge\left[
D\o_{\mi\a2{s-1/2}\mi{\ad}1{s-1/2}}+
e_{\bd}{}^{\b}\wedge\o_{\b\mi\a2{s-1/2}\mi{\ad}1{s-3/2}}\right.\nn\w2
&&\qquad\left. +(s-\ft32)
e_{\ad_{1}}{}^{\b}\wedge\o_{\b\mi\a2{s-1/2}\bd\mi{\ad}2{s-3/2}}
\right]\se0\ ,\qquad s\se \ft32,\ft52,\ft72,...
\la{fermi}
\eea

The local symmetries of this equation are given in
\eq{gengtf} and they allow us to fix the following gauge

\bea
D^{\a\ad}\o_{\a\ad,\mi\b1{s-3/2}\mi{\bd}1{s-1/2}}&=&0\ ,\la{fgf1}\w2
\o_{\a}{}^{\ad}{}_{,\mi\b1{s-3/2}\ad\mi{\bd}1{s-3/2}}&=&0\ ,\la{fgf2}\w2
\o^{\a}{}_{\bd_1,\a\mi\b1{s-5/2}\mi{\bd}2{s+1/2}}&=&0\ .
\la{fgf3}
\eea

Eqs. \eq{fgf1} and \eq{fgf2} fix the generalized local supersymmetries
up to residual symmetries generated by parameters obeying the Dirac
equation

\be
D_{\a}{}^{\ad}\vare_{\mi\b1{s-3/2}\ad\mi{\bd}1{s-3/2}}-
(s+\ft12)\,\vare_{\a\mi\b1{s-3/2}\mi{\bd}1{s-3/2}}\se0\ .
\la{resf}
\ee

The residual supersymmetry transformations also involve a compensating
fermionic gauge transformation with parameter $\vare(s-\ft52,s+\ft12)$
given by

\be
\vare_{\mi\a1{s-5/2}\mi{\ad}1{s+1/2}}\se \ft1{s-\ft12}
D^{\b}{}_{(\ad_1}\vare_{\,|\b\mi\a1{s-5/2}|\mi{\ad}2{s+1/2}})\ ,
\ee

The gauge condition \eq{fgf3} fixes uniquely the gauge parameters
$\vare(s-\ft52,s+\ft12)$. Together \eq{fgf2} and \eq{fgf3} eliminate all
Lorentz irreps in the generalized gravitino except the spin $s$ irrep.

The gauge choice allows us to rewrite the fermionic equation \eq{fermi}
as the generalized Dirac equation

\be
D_{\a}{}^{\ad}\o_{\b_1\ad,\mi\b2{s-1/2}\mi{\bd}1{s-1/2}}-
(s-\ft12)\o_{\a\bd_1,\mi\b1{s-1/2}\mi{\bd}2{s-1/2}}\se 0\ .
\la{de}
\ee

By combining \eq{de} with its hermitian conjugate and making repeated
use of \eqs{fgf1}{fgf3} we obtain the second order equation

\be
\left[D^2+s+\ft52-(s-\ft12)^2\right]
\o_{\a_1\ad_1,\mi\a2{s-1/2}\mi\ad2{s+1/2}} \se0\ .
\la{fer}
\ee

Applying the techniques for spectral analysis described in the bosonic
xase, we find that the Euclideanized, gauge fixed generalized gravitino
belongs to the $(j_1j_2)=(s,-\ft12)$ representation of $SO(4)$ (and its
hermitian conjugate belongs to $(j_1j_2)=(s,\ft12)$). The harmonic
expansion now involves $SO(5)$ irreps satisfying $n_1\geq s\geq n_2\geq
\ft12$. The $AdS$ energies therefore solve the characteristic equation

\be
E_0(E_0-3)+\ft94-(s-\ft12)^{2}\se0\ ,
\ee

which has the positive energy root $E_0=s+1$.

To find the number of massless modes we verify that the gauge
transformations generated by the residual parameters obeying \eq{resf}
also obey the Dirac equation \eq{de}. Hence the count of on-shell
degrees of freedom (following rules analogous to those given for the
bosonic count and taking into account the fact that the Dirac operator
in \eq{de} has half the maximum rank) shows that there are

\be
(s+\ft12)(s+\ft32)-(s-\ft12)(s+\ft12)-\Big[(s-\ft12)(s+\ft12)-
(s-\ft32)(s-\ft12)\Big]\se 2
\ee

real, on-shell degrees of freedom with spin $s=\ft32,\ft52,\ft74,...$
and energy $E_{0}=s+1$ describing massless higher spin fermions.

%%%%%%%%%%%%%%%%%%%%%%%%%%%%%%%%%%%%%%%%%%%%%%%%%%%%%%%%%%%%%%%%%%%%%%%%

\subsection{The Spin $s\leq1$ Sector}

%%%%%%%%%%%%%%%%%%%%%%%%%%%%%%%%%%%%%%%%%%%%%%%%%%%%%%%%%%%%%%%%%%%%%%%%

%%%%%%%%% spin-1

In this sector the equations of motion follows from \eq{va}. In the
spin $s=1$ sector we obtain

\be
D_{a}R^1_{bc}(\th\,)\se \ft{i}4(\s_{bc})^{\a\b}(\s_{a})^{\c\dd}
C_{\a\b\c\dd}(\th\,)
+\ft{i}4(\sb_{bc})^{\ad\bd}(\sb_a)^{\cd\d}C_{\d\ad\bd\cd}(\th\,)\ .
\ee

from the $(m,n)=(2,0)$ component of \eq{dcl1}, the identity
$C_{\a\b}(\th)=R^1_{\a\b}(\th)$ and \eq{rbi1}. Multiplying this
equation with $\y^{ab}$ and $\e^{abcd}$, and using the ``membrane"
identities

\be
(\s^{ab})^{(\a\b}(\s_{a})^{\c)\dd}\se0
\ ,\qquad
\e^{abcd}(\s_{bc})^{(\a\b}(\c_{d})^{\c)\dd}\se 0\ ,
\la{memb}
\ee

and expanding the $\th$ dependence using \eq{wexp}, one finds the spin
$1$ equations of motion and the Bianchi identities:

\bea
D^{a}R_{ab}^{1\,ij}\se0\ ,\qquad \e^{abcd}D_{b}R_{cd}^{1\,ij}\se 0\ ,\nn\w2
D^{a}R_{ab}^{1\,\mi{i}16}\se0\ ,\qquad \e^{abcd}D_{b}R_{cd}^{1\,\mi{i}16}\se 0
\ .\la{sone}
\eea

The gauge transformations $\d\o(\th)=d\e(\th)$ allow us to impose the
Lorentz gauge $D^a\o_a(\th)=0$ in which the equations of motion take
the form

\be
\left(D^2+3\right)\o_a^{ij}\se0\ ,\qquad
\left(D^2+3\right)\o_a^{i_1\cdots i_6}\se0\ .
\la{son}
\ee

The corresponding characteristic equation has the critical root $E_0=2$
and the residual gauge symmetries as usual cancel the longitudinal
on-shell mode. Hence the theory contains two massless $SO(8)$ vector
fields (gauging the $SO(8)_{+}$ and $SO(8)_-$ discussed in section
2).

%%%%%%%% spin 1/2

In the spin $s=\ft12$ sector the equations of motion are given by the
$(m,n)=(1,0)$ component of \eq{dcl1}:

\be
D_{\a\ad}C_{\b}(\th)\se i \,C_{\a\b\ad}(\th)\ .
\ee

Expanding the $\th$ dependence of the $(0,1)$ component of this equation
one finds the first order Dirac equations

\be
D^{\a}{}_{\ad}C_{\a}^{ijk}\se 0\ ,\qquad
D^{\a}{}_{\ad}C_{\a}^{i_1\cdots i_7}\se 0\ ,
\la{shalf}
\ee

giving two real, on-shell fermionic degrees freedom. Squaring these
equations gives

\be
\left(D^{2}+3\right)C_{\a}^{ijk}\se 0\ ,\qquad
\left(D^{2}+3\right)C_{\a}^{i_1\cdots i_7}\se 0\ ,
\ee

with critical energy $E_{0}=\ft32$, as expected for massless spin
$\ft12$ fermion fields.

%%%%%%%% spin 0

Finally in the scalar sector the $(m,n)=(0,0)$ and $(1,1)$ components
of \eq{dcl1} read

\be
D_{\a\ad}\f(\th)\se i\,\f_{\a\ad}(\th)\ ,\qquad
D_{\b\bd}\f_{\a\ad}(\th)\se -i\,\e_{\b\a}\e_{\bd\ad}\f(\th)+i\,
\f_{\a\b\ad\bd}(\th)\ ,\la{seco}
\ee

where $\f(\th)$ comprise the complex scalar fields introduced in
\eq{nn}. Evaluating the $D^{\a\ad}$ divergence of the first equation in
\eq{seco} using the latter equation and expanding in $\th$ yields the
scalar field equations

\be
\left(D^2+2\right)\f\se0\ ,\qquad
\left(D^2+2\right)\f^{ijkl}\se0\ ,
\la{szero}
\ee

with critical energies $E_0=1$ and $E_0=2$, as expected for massless
scalars.

%%%%%%%%%%%%%%%%%%%%%%%%%%%%%%%%%%%%%%%%%%%%%%%%%%%%%%%%%%%%%%%%%%%%%%%%%

\section{The Linearized $N=8$ AdS Supergravity}

%%%%%%%%%%%%%%%%%%%%%%%%%%%%%%%%%%%%%%%%%%%%%%%%%%%%%%%%%%%%%%%%%%%%%%%%%

We expect the linearized field equations and supersymmetry
transformations of the level $k=0$ multiplet of Table \ref{table} to
agree with those of gauged $N=8$ supergravity. Let us derive the exact
correspondence and relate the coupling constants of the higher spin
theory to the gravitational coupling constant $\k$ and the $SO(8)$
gauge coupling $g$ of the $N=8$ theory.

The level $k=0$, spin $s\leq\ft12$ equations are given by \eq{sone},
\eq{shalf} and \eq{szero}, while the gauge invariant spin $s=\ft32$
equation is given by \eq{fermi}. In the spin $s=2$ sector the
linearized equation of motion follow from the spin $s=2$ component of
\eq{i} and the constraints listed in \eq{iii}, i.e. the $9+1$ real
components

\be
R^1_{\a\b,\ad\bd}\se R^1_{\ad\bd,\a\b}\se 0\ ,\qquad R^1_{\a\b,\a\b}\se
R^1_{\ad\bd,\ad\bd}\se0\ .
\la{91cpt}
\ee

From the discussion following \eq{rbistwo} we recall that the torsion
constraint given by the spin $s=2$ component of \eq{i} together with
the Bianchi identity \eq{rbi} imply the reality of the quantities in
\eq{91cpt} as well as the vanishing of the $6$ real components in

\be
R^1_{\a_1}{}^{\b}{}_{,\a_2\b}\se R^1_{\ad_1}{}^{\bd}{}_{,\ad_2\bd}\se
0\ .
\la{6cpt}
\ee

In order to compare the spin $s=2$ equation to the $N=8$ theory it is
convenient to rewrite it as the linearization of Einstein's vacuum
equation with a cosmological constant (including higher orders to
the spin $s=2$ field equation will of course lead to more complicated
terms in the right side of the Einstein equation). To this end we define the
Ricci tensor $r_{ab}$ in the usual way as

\be
r_{ab}\se e_{a}{}^{\m}e_{c}{}^{\n}r_{\m\n,b}{}^{c}\ ,
\la{ricci}
\ee

where $r_{\m\n\,ab}$ is the $SO(3,1)$-valued Riemann
tensor and
$e_{a\m}$ denotes the inverse of the vierbein $e_{\m a}$ defined
in \eq{appb}. Linearizing \eq{ricci} around the AdS vacuum \eq{rie}, we
find

\bea
r_{ab}&=& \L\y_{ab}+r^1_{ab}\ ,\qquad \L\se -3\l^2\ ,\w2
r^1_{ab}&=& r^1_{ac,b}{}^{c}+2\,\o_{a,b}+\y_{ab}\,\o^{c}{}_{,c}\ ,
\la{linricci}
\eea

where $\l$ is the inverse AdS radius defined in \eq{adsr}, and the
linearized Riemann curvature

\be
r^1_{\m\n,ab}\se 2D_{[\m}\o_{\n],ab}\ , \la{linrie}
\ee

where $D_{\m}$ now denotes the background Lorentz covariant derivative.
To obtain the Einstein equation from the constraints it is more
convenient to treat the Lorentz connection as an independent field
(rather than substituting the solution for the Lorentz connection
obtained from the torsion constraint into the Riemann tensor). The
Ricci tensor \eq{ricci} then contains $16$ real components constrained
by the now independent components in \eq{91cpt} and \eq{6cpt}. From
\eq{rbi1} it follows that the linearized, AdS covariant curvature
$R^1_{ab,cd}$ and the linearized Riemann tensor $r^1_{ab,cd}$ are
related by

\bea
R^1_{\a_1\a_2,\b_1\b_2} &=&\ r^1_{\a_1\a_2,\b_1\b_2}
-2\,\e_{\a_1\b_1}\o_{\a_2}{}^{\cd}{} _{\b_2\cd}\ ,
\nn\w2
R^1_{\a_1\a_2,\bd_1\bd_2}&=& r^1_{\a_1\a_2,\bd_1\bd_2}
-2\,\o_{\a_1\bd_1,\a_2\bd_2}\ ,
\la{adsrie}
\eea

and hermitian conjugates, where

\bea
r^1_{\a\b,\c\d}&=&\ft18(\s^{ab})_{\a\b}(\s^{cd})_{\c\d}r_{ab,cd}\ ,\nn\w2
r^1_{\a\b,\cd\dd}&=&\ft18(\s^{ab})_{\a\b}(\sb^{cd})_{\cd\dd}r_{ab,cd}\ ,\nn\w2
\o_{\a\ad,\b\bd}&=&-\ft12(\s^{a})_{\a\ad}(\s^b)_{\b\bd}\o_{a,b}\ ,\la{norms}
\eea

as follows from \eq{defns} and \eq{appb}. The constraints
\eq{91cpt} and \eq{6cpt} then yield

\bea
&&r^1_{c\{a,b\}}{}^{c}-2\o_{\{a,b\}}\se0\ ,\nn\w2
&&r^1_{ab,}{}^{ab}+6\o^{a}{}_{,a}\se0\ ,\nn\w2
&&r^1_{c[a,b]}{}^{c}-2\o_{[a,b]}\se0\ ,\la{ee1}
\eea

where $\{ab\}$ denotes the traceless symmetric part. In deriving
\eq{ee1} one has to make use of the self-duality properties \eq{app6}
and notice that the pairs of indices $ab$ and $cd$ of the Riemann
tensor in the right side of in the two first equations in
\eq{norms} are projected onto (anti-)selfdual components. Eqs. \eq{ee1}
then follow by adding up selfdual and anti-selfdual components of the
constraints \eq{91cpt} and \eq{6cpt}. Combining \eq{linricci} and
\eq{ee1} yields $r^1_{ab}=0$, that is, in the linearized approximation
the equations of motion for the spin $s=2$ vierbein can be written as
the Einstein's equation with cosmological constant.

In summary, the linearized equations of motion of the level $k=0$ multiplet
are given by

\bea
\mbox{spin $s=2$}      &:&\qquad r_{ab}\se -3\l^2\,\y_{ab}\ ,\nn\w2
\mbox{spin $s=\ft32$}  &:&\qquad(\s^{abc})_{\a}{}^{\ad}\left(
D_{b}\,\o_{c,\ad}^{\,i}-\ft12\l\,(\s_b)_{\ad}{}^{\b}\o_{c,\b}^{\,i}\right)\se0
\ ,\nn\w2
\mbox{spin $s=1$}      &:&\qquad D^{a}D_{[a}\o_{b\,]}^{\,ij}\se0\ ,\nn\w2
\mbox{spin $s=\ft12$}  &:&\qquad D_{\a}{}^{\ad}C_{\ad}^{ijk}\se0\ ,\nn\w2
\mbox{spin $s=0$}      &:&\qquad \left(D^2+2\l^2\right)\f^{ijk\,l}\se0\ ,
\la{sgm}
\eea

where we have reintroduced the inverse $AdS$ radius $\l$ also in the
spin $s\leq \ft32$ equation.

The linearized, gauged $N=8$ supergravity model is described by the
quadratic action \cite{dn1}

\bea
e^{-1}{\cal L}_2&=& {1\over 2} R  -{\k^{2}\over 8g^{2}}
F_{\m\n,IJ}F^{\m\n,IJ}
-{1\over 96} \del_{\m}\phi_{ijkl}\del^{\m}\phi^{ijkl}
+{g^{2}\over 24\k^{2}}\, \phi_{ijkl}\phi^{ijkl} +{6g^2\over\k^2}\nn
\w2
&&
+\left(\ft{i}{2}
\bar{\psi}_{L\m}^{i}\c^{\m\n\r}D_{\n}\psi_{L\r i} +
{ig\over\sqrt{2}\k}\bar{\psi}^{i}_{R\m}\c^{\m\n}\psi_{L\n,i}
-{1\over 2} {\bar\chi}^{ijk}_L\c^{\m}D_{\m}\chi_{Lijk}\ +h.c.\right) \ ,
\la{lagr}
\eea

where $i,j,..=1,...,8$ are $SU(8)$ indices and $I,J,..=1,...,8$ are
$SO(8)$ indices, $\phi_{ijkl}$ are the $35+35$ scalars obeying \eq{sr}
and the fermions are Weyl. The complex conjugation changes chirality,
and consequently both chiralities occur for the gravitini as well as the
spin 1/2 fields. Thus, the theory is vector like. In writing \eq{lagr}
we have assigned (energy) dimension $\ft12$ to all the fermions,
dimension $0$ to the scalars and dimension $1$ to the vector fields.
Thus the Lagrangian ${\cal L}_2$ has dimension $2$.

We find that \eq{sgm} is in perfect agreement with \eq{lagr}
provided that we make the identifications

\bea
\o_\m{}^{\a\ad} \ \ &\rightarrow& \ \ e_\m{}^a\ ,\qquad\
\o_\m^{ij} \ \ \rightarrow \ \  A_\m^{IJ}\ ,
\nn\w2
\o_{\m\a}^i \ \ &\rightarrow& \ \  \psi_{L\m}^i\ ,\qquad
C_\a^{ijk} \ \ \rightarrow \ \  \chi_L^{ijk}\ ,
\eea

and identify the following important relation between the Newton's
constant $\k$, the $SO(8)$ gauge coupling $g$ and the inverse AdS
radius:

\be
{g^2\over \k^2} \se \ft12\l^2\ .
\la{gk}
\ee

The free parameters of the higher spin theory are therefore the gauge
coupling $g$ and the inverse $AdS$ radius $\l$. The gauge coupling is
introduced into the full set of higher spin equations \eqs{fe1}{fe5}
and
\eq{gauge} by replacing $W\rightarrow g\, W$. These equations are
consistent with the assignment of dimension $0$ to the master fields $W$
and $\Phi$. Using the dimensionful coupling $\l$ one then defines
component fields $\tilde{\o}_{\m}(m,n;\th)$ and $\tilde{C}(m,n;\th)$
with canonical mass dimensions as follows \cite{fv1}

\bea
\tilde{\o}_{\m}(m,n;\th\,)&=& \left\{\begin{array}{ll}
\l^{-1+\ft{|m-n|}{2}}\o_{\m}(m,n;\th\,)\ ,&(m,n)\neq(0,0)\ ,\\ &\\
\o_{\m}(0 ,0,\th\,)\ ,& (m,n)=(0,0)\end{array}\right.\nn\w2
\tilde{C}(m,n;\th\,)&=& \l^{\ft{m+n}{2}} C(m,n;\th\,) \ ,\qquad m,n=0,1,...
\la{dim}
\eea

Since the mass dimension of $\o_{\m}(m,n,\th)$ is equal to $1$ this
means that the generalized vierbein $\tilde{\o}_{a}(m,m;\th\,)$ has mass
dimension $0$, the generalized gravitino $\tilde{\o}_{a}(m-1,m;\th\,)$
has dimension $\ft12$ and the generalized Lorentz connection
$\tilde{\o}_{a}(m-2,m;\th\,)$ and the $SO(8)$ vector fields
$\tilde{\o}_{\m}(0,0,\th\,)$ has dimension $1$. The generalized Weyl
tensor $\tilde{C}(m+2,0;\th)$ ($m\geq 1$) has dimension $\ft{m+2}{2}$,
which is the same as the dimension of the pure curvature component
$\tilde{R}_{\a\b}(m,0;\th)$. Finally the fermions $\tilde{C}_{\a}^{ijk}$
and $\tilde{C}_{\a}^{i_1\cdots i_7}$ hve dimension $\ft12$ and the
scalars $\tilde{\f}$ and $\tilde{\f}_{ijkl}$ have dimension $0$. Thus,
in particular, \eq{dim} yields the correct canonical dimensions of
fields of the $N=8$ supergravity multiplet.

%%%%%%%%%%%%%%%%%%%%%%%%%%%%%%%%%%%%%%%%%%%%%%%%%%%%%%%%%%%%%%%%%%%%%%%%%%%

\section{Discussion}

%%%%%%%%%%%%%%%%%%%%%%%%%%%%%%%%%%%%%%%%%%%%%%%%%%%%%%%%%%%%%%%%%%%%%%%%%%%

The results of this paper suggest that the $D=4, N=8$ AdS supergravity
can be embedded into a higher spin gauge theory. The fully nonlinear
equations are consistent but we have shown the embedding at the
linearized level. The important next step in this program is to study
the interactions, starting with the quadratic fields in the equations of
motion. Various aspects of such interactions have been studied before
but they have not been compared to those of $N=8$ AdS supergravity.
Given the facts that:

\begin{itemize}
\item[(a)]  the full higher spin equations of motion are consistent,
\item[(b)]  they yield the correct $N=8$ AdS supergravity equations
            at the linearized level,
\item[(c)] the interactions in the $N=8$ AdS supergravity seem to be
           unique due to the highest possible supersymmetry in the theory,
\end{itemize}

one may expect that the higher spin equations at hand already contain
the full fledged $N=8$ AdS supergravity, along with the sector for the
spin $s\geq\ft52$ fields. If so, then one would also expect to uncover
the $E_7/SU(8)$ coset structure of Cremmer-Julia \cite{cj1,cj2} which
plays an important role in the description of both the $N=8$ Poincar\'e
\cite{cj1,cj2} as well as the AdS $N=8$ supergravity \cite{dn1,dn2}.

Relevant to the problem of finding the hidden symmetries in the theory
is the question of how unique is the higher spin AdS supergravity. This
issue has already been addressed in section 4.3. The full significance
of an interaction ambiguity discussed in that section is not clear to
us at present. It may as well play a role in the search for the hidden
$E_7$ symmetry. Recalling the uniqueness of the $N=8$ AdS supergravity,
we expect that there should be no ambiguity in the interactions
provided that we insist on the consistent truncation of the theory to
the pure
$N=8$ AdS supergravity. A careful comparison of the first interaction
terms in the higher spin theory and the $N=8$ AdS supergravity is
required to settle this question.

The $N=8$ supersingleton propagating at the boundary of $AdS$ spacetime
serves as a spectrum generating representation for the massless higher
spin theory propagating in the bulk of the $AdS$ spacetime. It is
tempting to believe that the singletons could play a more fundamental
role in the derivation of the effective bulk action from the
bulk/boundary duality prescription of \cite{jm,ew,gkp}. Notice that
both bulk theory and boundary theory has the same dimensionful coupling
$\l$(the inverse $AdS$ radius). Since the (massless) spectra of the bulk and
the boundary theories agree, the essential test of the bulk/boundary
duality is therefore whether it is possible to represent the \alg\
symmetry algebra as charges of conserved currents in the $N=8$
supersingleton theory. In that case we would expect that the
nonlinearities of the bulk theory would be reproduced by the
interactions between the composite singleton states (the dimensionless
gauge coupling $g$ would the be introduced in the boundary theory by a
rescaling of the composite states describing the gauge fields). While
this program by and large remains to be realized, the construction of
higher spin currents has been recently investigated \cite{anselmi}, at
least for low lying spins, in the context of $AdS_5$. Higher spin
supercurrents have also been constructed \cite{fz}, also in the context
of $AdS_5$.

Since the boundary theory involves massless as well as massive composite
states, bulk/boundary duality would also yield higher spin bulk
interactions including both massless and massive ``matter" sectors
(allowing a consistent truncation to the massless sector). Inclusion of
massive sectors could generate mechanisms for spontaneous breaking
\cite{v9,bsst1} of the \alg\ symmetry in which the massive multiplets
are ``eaten" by the massless gauge multiplets. This point is of great
physical interest since we do not presently know how to fit massless
higher spins into an $M$-theoretical framework.

A desirable formal consequence of bulk/boundary duality would be to
ultimately reduce the rather cumbersome calculations implied by the
perturbative expansion around the AdS vacuum outlined in section 5.2 to
calculations entirely within the free boundary supersingleton field
theory. An interesting related issue is whether it is possible to
accommodate the auxiliary spinor variables in the boundary theory and
derive the higher spin $(x,Z)$ space field equations \eqs{fe1}{fe5}
using background field methods.

The bulk/boundary duality of the type discussed above may also exist for
the $4D$ doubletons (vector multiplet) propagating at the boundary of
$AdS_5$ and the $6D$ doubletons (tensor multiplet) propagating at the
boundary of $AdS_7$, respectively \cite{g3}. The possibility of
constructing higher spin interactions in $AdS$ spaces of dimension
larger that four has been investigated in \cite{v2}. A more tractable
example is the $3D$ higher spin theory described in \cite{v11}. In this
case, the study of bulk/boundary duality is expected to be simpler
because the $2D$ boundary theory is a more tractable conformal field
theory. Another reason for the tractability of the $3D$ case is that it
may be possible to construct an action for a theory based on the $3D$
higher spin AdS superalgebra, combining the elements of the work of
\cite{blen} that involve a Chern-Simons action and the work described in
\cite{v11}.

It would, of course, be desirable to find an action which yields the
consistent, fully nonlinear equations of motion of Vasiliev that we have
studied here. Indeed, $R\wedge R$ type actions have been considered
before \cite{fv0,fv1,v2,v4}, but one drawback of these actions is that
the spin $s \leq 1/2$ sector of the theory does not fit in a natural and
geometrical way into the part of the action that describes the fields
with spin $s \geq 3/2$. On the other hand, the fact that there is a
gauge master field $W$ and a matter master field $\Phi$ in the Vasiliev
formalism studied here is suggestive of a $D=5$ origin in which the
matter master field may emerge as the fifth component of the gauge
master field upon dimensional reduction to $D=4$. Furthermore, it is
encouraging that there exists the possibility of a Chern-Simons type
Lagrangian of the form $tr(R\wedge R\wedge W)$ in five dimensions (the
definition of traces of higher spin algebras is explained in
\cite{v11}).

The construction of higher spin superalgebras and free field equations
of motion for higher spin fields becomes rapidly very complicated in
higher than four dimensions \cite{v2} and understandably not much
progress has been made in this front for sometime. However, the recent
exciting developments in M-theory provide ample motivation for exploring
the $D=11$ origin of the AdS higher spin supergravity studied here.
Starting from the $D=11$ supergravity theory alone, Kaluza-Klein
compactification gives rise to massless and massive fields of maximum
spin two. The corners of M-theory where it can be treated
perturbatively, on the other hand, can give rise to infinite towers of
higher spin fields, but all of these fields are massive for spin $s >2$.
Therefore, one is led to speculate about the existence of either a new
corner of M-theory, or supermembrane theory, which may give rise to the
higher spin AdS supergravity in $D=4$ in a certain limit, or a new kind
of $D=11$ limit which modifies the well known $D=11$ supergravity theory
in a profound way. The first scenario is in line with the previous
studies on the supermembrane-supersingleton connection
\cite{duff1,bd1,nst,bsst1,bsst2,bst0}. The latter scenario is motivated
by a recent construction of the singleton/doubleton representations of
a candidate $D=11$ AdS supergroup in \cite{g5}, which turn out to be of
rather unusual kind. It is not known yet if an action, or equations of
motion, can be written down to describe these representations in the
ten dimensional boundary of $AdS_{11}$, but the possibility is
certainly tantalizing and we expect that it would be highly relevant to
the massless higher spin theory in the M-theory framework. In this
context, it is interesting to note that the representations of $SO(9)$
group as the little group classifying the massless degrees of freedom
of an eleven dimensional supergravity has been studied recently
\cite{ramond} with interesting results that hint at the possibility of
higher spin massless fields.

%%%%%%%%%%%%%%%%%%%%%%%%%%%%%%%%%%%%%%%%%%%%%%%%%%%%%%%%%%%%%%%%%%%%%%%%%%

\noindent {\bf Acknowledgements}

We thank M. Duff, J. Maldacena and P.K. Townsend for useful
discussions. We are grateful to E. Witten for his comments on the first
version of this paper which led us to add several clarifying remarks.

%%%%%%%%%%%%%%%%%%%%%%%%%%%%%%%%%%%%%%%%%%%%%%%%%%%%%%%%%%%%%%%%%%%%%%%%%%

\pagebreak

%%%%%%%%%%%%%%%%%%%%%%%%%%%%%%%%%%%%%%%%%%%%%%%%%%%%%%%%%%%%%%%%%%%%%%%%%%

\appendix

%%%%%%%%%%%%%%%%%%%%%%%%%%%%%%%%%%%%%%%%%%%%%%%%%%%%%%%%%%%%%%%%%%%%%%%%%%

\section{Spinor Conventions}

%%%%%%%%%%%%%%%%%%%%%%%%%%%%%%%%%%%%%%%%%%%%%%%%%%%%%%%%%%%%%%%%%%%%%%%%%%

For $SO(3,1)$ we take $\y_{ab}=\mbox{diag}(-+++)$ and work with
two-component Weyl spinors
\bea
y^{\a}&=& \e^{\a\b}y_{\b}\ ,\qquad y_{\a}\se y^{\b}\e_{\b\a}\ ,\nn\w2
\yb_{\ad}&=&  (y_{\a})^{\dagger}\ , \qquad \yb^{\ad}\se
(y^{\a})^{\dagger}\ ,\nn\w2
\yb^{\ad}&=&\e_{\ad\bd}\yb_{\ad}\ ,\qquad
\yb_{\ad}\se \yb^{\bd}\e^{\bd\ad}\ ,
\la{app1}
\eea

where the charge conjugation matrix
$\e_{\a\b}=\e^{\a\b}=\e_{\ad\bd}=\e^{\ad\bd}$ obeys
$\e_{\a\b}\e^{\a\c}=\d_{\b}^{\c}$. Using the Pauli matrices $\s^{1,2,3}$
we define the van der Waerden symbols $(\s^a)_{\a\bd}$ ($a=0,1,2,3$)
\bea
(\s^{a})_{\a\bd}& := &(1,\,\s^{1}\,,\s^{2},\,\s^{3}\,)\ ,\nn\w2
(\sb^{a})^{\ad\b}& := &(1,\,-\s^{1}\,,-\s^{2},\,-\s^{3}\,)
\se\e^{\ad\cd}\e^{\b\d}(\s^{a})_{\d\cd}\ ,
\la{app2}
\eea

with the hermicity properties
\be
\left((\s^{a})_{\a\bd}\right)^{\dagger}\se (\sb^{a})_{\ad\b}\se (\s^a)_{\b\ad}\ ,
\qquad
\left((\s^{a})^{\a\bd}\right)^{\dagger}\se(\sb^{a})^{\ad\b}\se (\s^a)^{\b\ad}\ .
\la{app3}
\ee

The van der Waerden symbols obey the completeness relations
\bea
(\s^{a})_{\a\ad}(\sb_{a})^{\bd\b}&=&-2\,\d_{\a}^{\b}\,\d_{\ad}^{\bd}\ ,
\nn\w2
(\s^{a})_{\a}{}^{\ad}(\sb^{b})_{\ad}{}^{\b}&=&\y^{ab}\d_{\a}^{\b}\
+\ (\s^{ab})_{\a}{}^{\b}
\ ,\nn\w2
(\sb^{a})_{\ad}{}^{\a}(\s^{b})_{\a}{}^{\bd}&=&\y^{ab}\d^{\bd}_{\ad}\
+\ (\sb^{ab})_{\ad}{}^{\bd}
\ ,\la{app4}
\eea

where $(\s^{ab})_{\a\b}=-(\s^{ba})_{\a\b}=(\s^{ab})_{\b\a}$ and
$(\sb^{ab})_{\ad\bd}=-(\sb^{ba})_{\ad\bd}=(\sb^{ab})_{\bd\ad}$. These
quantities obey the decomposition rules
\bea
(\s^{ab})_{\a}{}^{\c}(\s^{c})_{\c\,\bd}&=&
i\e^{abcd}(\s_{d})_{\a\bd}\ +\ 2\,i\, (\s^{[a})_{\a\bd}\,\,\y^{b]c}\ ,
\nn\w2
(\s^{ab})_{\a}{}^{\c}(\s^{cd})_{\c}{}^{\b}&=&\left(i\,\e^{abcd}+2\y^{b[c}\y^{d\,]a}
\right)\d_{\a}^{\b}\ +\ 4(\s^{[a|[d})_{\a}{}^{\b}\,\,\y^{c]|b\,]}
\ ,\la{app5}
\eea

and they have the duality properties
\be
\ft12 \e_{abcd}(\s^{cd})_{\a\b}\se i(\s_{ab})_{\a\b}\ ,\qquad
\ft12 \e_{abcd}(\sb^{cd})_{\ad\bd}\se -i(\sb_{ab})_{\ad\bd}\ ,\la{app6}
\ee

where the tensorial $\e$-symbol is defined by $\e^{0123}=-\e_{0123}=1$.

For $SO(3,2)$ we take $\y_{AB}=\mbox{diag}(-+++-)$ ($A,B=0,1,2,3,5$)
and work with four-component, Majorana spinors $\Psi_{\una}$ and
$\C$-matrices
\be
(\C^A)_{\una}{}^{\unb}\se \mx{\{}{ll}{i(\c^a\c^5)_{\una}{}^{\unb}&\mbox{for }
A=a\w2 i(\c^5)_{\una}{}^{\unb}&\mbox{for }A=5}{.}\ ,
\la{app9}
\ee

where $\c^a$ are symmetric $SO(3,1)$ $\c$-matrices and
$\c^5=i\c^0\c^1\c^2\c^3$ (such that $(\c^5)^2=1$). In the Dirac
representation we choose
\bea
C_{\underline{\a\b}}\se
\mx{(}{ll}{\,\e_{\a\b}&0\\0& \,\e^{\ad\bd}}{)}\ ,
\nn\w1
(\c^{a})_{\una}{}^{\unb}\se \mx{(}{ll}{0&(\s^{a})_{\a\bd}\\
-(\sb^{a})^{\ad\b}&0}{)}\ ,\qquad
(\c^5)_{\una}{}^{\unb}\se\mx{(}{ll}{-\d^{\b}_{\a}&0\\0&\d^{\ad}_{\bd}}
{)}\ .
\la{app10}
\eea

An $SO(3,1)$ Weyl spinor $\psi_{\a}$ and its hermitian conjugate
$(\psi_{\a})^{\dagger}=\bar{\psi}_{\ad}$ can be used to represent a
real Majorana spinor $\Psi^{(+)}_{\una}$ as well as a purely imaginary
Majorana spinor $\Psi^{(-)}_{\una}$:
\be
\Psi^{(\pm)}_{\una}\se \mx{(}{c}{\psi_{\a}\\ \pm\bar{\psi}^{\ad}}{)}
\ ,\qquad
\overline{\Psi}^{(\pm)}\;\; := \;\; (\Psi^{(\pm)})^{\dagger}\c^{0}\se
\pm(\Psi^{(\pm)})^{T}C\ .
\la{app11}
\ee

%%%%%%%%%%%%%%%%%%%%%%%%%%%%%%%%%%%%%%%%%%%%%%%%%%%%%%%%%%%%%%%%%%%%%%%%%%

\section{The $OSp(8|4)$ Subalgebra of \alg}

%%%%%%%%%%%%%%%%%%%%%%%%%%%%%%%%%%%%%%%%%%%%%%%%%%%%%%%%%%%%%%%%%%%%%%%%%%

In the expansion \eq{p}, we find the quadratic, homogeneous polynomial
\be
P_{2}(Y,\th)\;\; := \;\;\fft1{4i}\left(
\l^{ij}\th_{ij}+2~\e_{i\,\a}\,y^{\a}\th^{i}
+2~\bar{\e}_{i\,\ad}\,\yb^{\ad}\th^{i}
+ \ell_{\a\b}\,y^{\a}y^{\b}+2~a_{\a\bd}\,y^{\a}\yb^{\bd}
+\ell_{\ad\bd}\,\yb^{\ad}\yb^{\bd}\right)\ ,\la{osp}
\ee

where $\l^{ij}, \e_{i\a}, \ell_{\a\b}, a_{\a\bd}$ are the parameters for
$SO(8)$, supersymmetry, Lorentz transformations and translations
respectively. The corresponding generators
\bea
Q_{\a i}&=&\ft12y_{\a}\th_{i}\ ,\qquad Q_{\ad i}\se \ft12\yb_{\ad}\th_{i}\ ,
\qquad\quad T^{ij}\se\ft{i}2\th^{ij}\nn\w2
M_{\a\b}&=&\ft12y_{\a}y_{\b}\ ,\qquad \ P_{\a\bd}\se
\ft12y_{\a}\yb_{\bd}\ ,\qquad M_{\ad\bd}\se\ft12\yb_{\ad}\yb_{\bd}\ ,\la{ospgen}
\eea

obey the $D=4$, $N=8$ anti de Sitter superalgebra $OSp(8|4)$
\bea
\{\,Q_{\a i}\,,\,Q_{\b j}\,\}_{\star}&=&\d_{ij}\, M_{\a\b}+ \e_{\a\b}\,T_{ij}
\ ,\qquad \quad
\{\,Q_{\a i}\,,\,Q_{\bd j}\,\}_{\star}\se\d_{ij}\, P_{\a \bd}\ ,\nn\w2
{}[\,T_{ij}\,,\,Q_{\a k}\,]_{\star}&=& i\,\d_{jk}\,Q_{\a i}-
i\,\d_{ik}\,Q_{\a j}\ ,\qquad\quad\ [\,T_{ij}\,,\,T_{kl}\,]_{\star}\se
i\,\d_{jk}\,T_{il}+\mbox{3 terms}\ ,\nn\w2
{}[\,M_{\a\b}\,,\,Q_{\c i}\,]_{\star}&=& i\,\e_{\a\c}\,Q_{\b i}+
i\,\e_{\b\c}\,Q_{\a i}
\ ,\qquad\ \
{}[\,P_{\a\bd}\,,\,Q_{\c i}]_{\star}\se i\,\e_{\a\c}\,Q_{\bd i}\ ,\nn\w2
{}[\,M_{\a\b}\,,\,M_{\c\d}\,]_{\star}&=&i\,\e_{\a\c}\,M_{\b\d}+\mbox{3
terms}\ ,\qquad [\,M_{\a\b},\,,P_{\c\dd}\,]_{\star}\se i\,\e_{\a\c}\,P_{\b\dd}+
i\,\e_{\b\c}\,P_{\a\dd}\ ,\nn\w2
{}[\,P_{\a\bd}\,,\,P_{\c\dd}\,]_{\star}&=&i\,\e_{\bd\dd}\,M_{\a\c}+
i\,\e_{\a\c}\,M_{\bd\dd}
\ ,\la{ospalg}
\eea

and hermitian conjugates. The following change of basis for the
$SO(3,2)$ subalgebra
\be
M_{ab}\se
\ft14(\s_{ab})^{\a\b}M_{\a\b}+\ft14(\sb_{ab})^{\ad\bd}M_{\ad\bd}
\ ,\qquad M_{a4}\se \ft12(\s_a)^{\a\bd}M_{\a\bd}\ ,\qquad a,b=0,1,...,3
\ ,
\ee

yields
\be
{}[\,M_{AB}\,,\,M_{CD}\,]_{\star}\se -i\y_{BC}M_{AD}+\mbox{3 terms}\
,\qquad A,B=0,1,...,4\ .
\ee

If we let $\O$ to be the $SO(3,2)$ connection one-form given
in the two $SO(3,2)$ bases by
\be
\O\se\ft1{4i}\left(\o_{\a\b}y^{\a}y^{\b}+\bar{\o}_{\ad\bd}\yb^{\ad}\yb^{\bd}
+2\o_{\a\ad}y^{\a}\yb^{\ad}\right)\se\ft1{2i}\o_{AB}M^{AB}\ ,\la{adsconn}
\ee

then we find the following relation between the components gauge fields
\be
\o_{\a\b}\se\ft14(\s^{ab})_{\a\b}\o_{ab}\ ,\qquad
\o_{\ad\bd}\se\ft14(\sb^{ab})_{\ad\bd}\o_{ab}\ ,\qquad
\o_{\a\ad}\se -\ft12(\s^{a})_{\a\ad}\o_{a}\ ,\la{appb}
\ee

where $\o_{a} := \o_{a4}$. One may notice that
\be
y_{\mi\a1{4k+2-n}}\th_{\mi{i}1n}\se
M_{(\a_1\a_2}\star\cdots \star M_{\a_{4k+1-2m}\a_{4k+2-2m}}\star
Q_{\a_{4k+3-2m}[i_1}\star\cdots \star Q_{\a_{4k+2-m})i_m]}\ .\la{env}
\ee

Hence, considered as a vector space, \alg\ can be identified with the
subspace of the $OSp(8|4)$ enveloping algebra which is spanned by odd,
fully (anti-)symmetrized functions. However, when considered as
algebras, \alg\ and the $OSp(8|4)$ enveloping algebra differ from each
other. Actually, the $OSp(8|4)$ relations \eq{ospalg} in combination
with \eq{env} do not suffice to determine the commutator of two general
elements in \alg. Moreover, a representation of \alg\ does not need to
represent the relation \eq{env}, as is the case, for instance , with
the tensor product representation \eq{tprule}.

%%%%%%%%%%%%%%%%%%%%%%%%%%%%%%%%%%%%%%%%%%%%%%%%%%%%%%%%%%%%%%%%%%%%%%%%%%

\section{Oscillator Realization of \alg}

%%%%%%%%%%%%%%%%%%%%%%%%%%%%%%%%%%%%%%%%%%%%%%%%%%%%%%%%%%%%%%%%%%%%%%%%%%

One way to obtain unitary representations of \alg\ is consider tensor
products of the Fock space ${\mit\Phi}$ obtained by acting on a ground
state $\ket{0}$ with a pair of bosonic creation operators $\htad_p$
($p=1,2$) and fermionic creation operators $\htp_A$ ($A=1,...,4$)
obeying \cite{kv1,kv2}
\bea
[\,\hta_p\,,\,\htad_q\,]&=&\d_{pq}\ ,\qquad p,q=1,2\ ,\nn\w2
\{\,\htp_{A}\,,\,\htpd_{B}\,\}&=& \d_{AB}\ ,\qquad A,B=1,2,3,4\nn\w2
\hta_p\ket{0}&=&\htp_A\ket{0}\se 0\ .
\eea

The representation on ${\mit\Phi}$ of the element $F$ of ${\cal A}$
given in \eq{F} is given by
\be
\hat{F}\se F(\hty,\htyb;\htt)\ ,
\ee

where $\hty_\a$, $\htyb_\ad$ and $\htt^i$ are given by
\be
\hta_1\se \ft12(\hty_1+i\htyb_{\dot{2}})\ ,\qquad
\hta_2\se \ft12(\htyb_{\dot{1}}+i\hty_2)\ ,\qquad
\htp_A\se \ft12(\htt_{2A-1}+i\htt_{2A})\ .
\ee

Notice that the operator $\hat{F}$ is fully (anti)symmetrized, or Weyl
ordered. The $\star$ product of elements in ${\cal A}$ given in
\eqs{yc}{thc} is then represented in ${\mit\Phi}$ by the ordinary operator
product:
\be
\widehat{(F\star G)}\se\hat{F}\hat{G}\ .
\ee

The resulting unitary representation of \alg\ acts reducible on
${\mit\Phi}$, since the elements of \alg\ are even polynomials, and as a
result ${\mit\Phi}$ actually splits into two UIR's of \alg, namely
\be
{\mit\Phi}\se {\mit\Phi}_{e}\oplus {\mit\Phi}_{o}\ ,
\la{eo}
\ee

where the states in ${\mit\Phi}_{e}$ (${\mit\Phi}_{o}$) are made up by
acting on $\ket{0}$ with an even (odd) total number of fermionic and
bosonic oscillators. The two spaces ${\mit\Phi}_{e}$ and
${\mit\Phi}_{o}$ remain irreducible under the $OSp(8|4)$ subalgebra
\eq{ospalg} of \alg\ and, using the notation introduced in \eq{dirac},
they are labeled as follows
\be
{\mit\Phi}_{e}\se [D(\ft12,0)\otimes 8_s]\oplus[D(1,\ft12)\otimes
8_c]\ ,\qquad
{\mit\Phi}_{o}\se [D(\ft12,0)\otimes 8_c]\oplus[D(1,\ft12)\otimes
8_s]\ ,
\ee

where $8_s$ ($8_c$) are the two $8$-dimensional subspaces of the
$16$-dimensional Fock space of the fermionic oscillators, obtained by
acting with an even (odd) number of fermionic creation operators on the
vacuum. The representation of the element $P$ in \alg\  on the tensor product
${\mit\Phi}\otimes{\mit\Phi}$ is defined by
\be
\hat{P}(\ket{u}\otimes\ket{v})\;\; := \;\; (\hat{P}\ket{u})\otimes
\ket{v}+(-1)^{uP}\ket{u}\otimes (\hat{P}\ket{v})\ ,
\la{tprule}
\ee

where $u$ and $P$ in the exponents denote Grassmann parities
(the vacuum is taken to be even). Eq. \eq{tprule} implies that
the tensor product is {\it not} a representation of the $\star$ algebra
${\cal A}$. The tensor product form a reducible, unitary representation
of \alg, containing the invariant subspaces
$\left[{\mit\Phi}_{\l}\otimes{\mit\Phi}_{\l'}\right]_{S,A}$, where
$\l,\l'=e,o$ (see \eq{eo}), obtained by symmetrization ($S$) and
anti-symmetrization ($A$) of the tensor product according to the rule
\bea
\Big[\ket{u}\otimes\ket{v}\Big]_{S}\se \ket{u}\otimes\ket{v}
+(-1)^{uv}\ket{v}\otimes\ket{u}\ ,\nn\w2
\Big[\ket{u}\otimes\ket{v}\Big]_{A}\se \ket{u}\otimes\ket{v}
-(-1)^{uv}\ket{v}\otimes\ket{u}\ .
\eea

The result in Table 1 for the $SO(3,2)\times SO(8)$ content of
$\left[{\mit\Phi}_{e}\otimes{\mit\Phi}_{e}\right]_{S}$ is derived using
the following decomposition rules for the $SO(3,2)$ content
\bea
\left[D(\ft12,0)\otimes D(\ft12,0)\right]_{S}&=& D(1,0)\oplus
D(3,2)\oplus D(5,4)\oplus D(7,6)\oplus\cdots\ ,\nn\w2
\left[D(\ft12,0)\otimes D(\ft12,0)\right]_{A}&=&
D(2,1)\oplus D(4,3)\oplus D(6,5)\oplus\cdots\ ,\nn\w2
\left[D(1,\ft12)\otimes D(1,\ft12)\right]_{S}&=&
D(2,1)\oplus D(4,3)\oplus D(6,5)\oplus\cdots\ ,\nn\w2
\left[D(1,\ft12)\otimes D(1,\ft12)\right]_{A}&=&
D(2,0)\oplus D(3,2)\oplus D(5,4)\oplus D(7,6) \oplus \cdots\ ,\nn\w2
\left[D(\ft12,0)\otimes D(1,\ft12)\right]_{S,A}&=&
D(\ft32,\ft12)\oplus D(\ft52,\ft32)\oplus D(\ft72,\ft52)\oplus \cdots\ ,
\la{diractp}
\eea

and the $SO(8)$ content
\bea
8_s\otimes 8_s&=& 1_S+28_A+35^{+}_S\ ,\nn\w2
8_s\otimes 8_c&=& 8_c\otimes 8_s\se 8_v+56\nn\w2
8_c\otimes 8_c&=& 1_A+28_S+35^{-}_A\ ,
\la{so8tp}
\eea

where the odd Grassmann parity of the states in $8_c$ has been taken
into account in the last equation.

%%%%%%%%%%%%%%%%%%%%%%%%%%%%%%%%%%%%%%%%%%%%%%%%%%%%%%%%%%%%%%%%%%%%%%%%%%

\section{Symplectic Differentiation and Integration Formula}

%%%%%%%%%%%%%%%%%%%%%%%%%%%%%%%%%%%%%%%%%%%%%%%%%%%%%%%%%%%%%%%%%%%%%%%%%%

Using \eq{yzstar}, one finds the following contraction rules
\bea
y_{\a}\star F(Z,Y)&=&y_{\a} F(Z,Y)+
\left[-i\pd{z}\a+i\pd{y}\a\right]F(Z,Y)\ ,\nn\w2
z_{\a}\star F(Z,Y)&=& z_{\a} F(Z,Y)+
\left[-i\pd{z}\a+i\pd{y}\a\right]F(Z,Y)\ ,\nn\w2
F(Z,Y)\star y_{\a}&=&y_{\a} F(Z,Y)+
\left[-i\pd{z}\a-i\pd{y}\a\right]F(Z,Y)\ ,\nn\w2
 F(Z,Y)\star z_{\a}&=& z_{\a} F(Z,Y)+
\left[i\pd{z}\a+i\pd{y}\a\right]F(Z,Y)\ ,\nn\w5
\yb_{\ad}\star F(Z,Y)&=&\yb_{\ad} F(Z,Y)+
\left[i\pd{\zb}{\ad}+i\pd{\yb}{\ad}\right]F(Z,Y)\ ,\nn\w2
\zb_{\ad}\star F(Z,Y)&=& \zb_{\ad} F(Z,Y)+
\left[-i\pd{\zb}{\ad}-i\pd{\yb}{\ad}\right]F(Z,Y)\ ,\nn\w2
F(Z,Y)\star \yb_{\ad}&=&\yb_{\ad} F(Z,Y)+
\left[i\pd{\zb}{\ad}-i\pd{\yb}{\ad}\right]F(Z,Y)\ ,\nn\w2
 F(Z,Y)\star \zb_{\ad}&=& \zb_{\ad} F(Z,Y)+
\left[i\pd{\zb}{\ad}-i\pd{\yb}{\ad}\right]F(Z,Y)\ .\la{appe}
\eea

where $F(Z,Y)$ is an arbitrary function.
The linear differential equations in $Z$-space of the type
\bea
&&\del f\se g \se dz^{\a}g_{\a}+d\zb^{\ad} g_{\ad}\ ,
\la{appb1}\w2
&&\del (dz^{\a}f_{\a}+d\zb^{\ad}f_{\ad})\se h\se
\ft12 dz^2 h+\ft12d\zb^{2} \tilde{h}
+dz^{\a}\wedge d\zb^{\ad} h_{\a\ad}\ ,\la{appb2}
\eea

encountered in the perturbative expansion of the higher spin equations
around the anti de Sitter vacuum, have the solutions \cite{v10}
\bea
f(z,\zb)&=& f(0,0)+\int_{0}^{1}dt\left[\,z^{\a} g_{\a}(tz,t\zb)
+\zb^{\ad}g_{\ad}(tz,t\zb)\right]\ ,
\la{appb3}\w2
f_{\a}(z,\zb)&=& \pd{z}{\a}k(z,\zb)-\int_{0}^{1}dt\,t\left[
\,z_{\a}\,h(tz,t\zb)+\zb^{\ad}h_{\a\ad}(tz,t\zb)\right]\ ,
\la{appb4}\w2
f_{\ad}(z,\zb)&=& \pd{\zb}{\ad}k(z,\zb)-\int_{0}^{1}dt\,t\left[
\,\zb_{\ad}\,\tilde{h}(tz,t\zb)-z^{\a
}h_{\a\ad}(tz,t\zb)\right]\ ,
\la{appb5}
\eea

where $f(0)$ is an arbitrary constant and $k(z,\zb)$ an arbitrary
function. In proving these formula one makes use of
\be
t\fft{d}{dt}h(tz)\se z^{\a}\pd{z}{\a}h(zt)\ .\la{chrule}
\ee

Therefore, to apply them correctly to \eqs{p1}{p3}, one must expand the
$\star$ products in the right hand sides of \eqs{p1}{p3} {\it before}
one replaces $z$ and $\zb$ by $tz$ and $t\zb$ as indicated in
\eqs{appb3}{appb5}. This is so because the contractions alter the
functional dependence on $z$ and $\zb$ such that if one would replace
$z$ and $\zb$ by $tz$ and $t\zb$ before one expands the $\star$ products
then \eq{chrule} is no longer valid. As a matter of fact, if we let $A
:= A(tz,t\zb,y,\yb)$ and $B := B(tz,t\zb,y,\yb)$ then we have
\be
t\fft{d}{dt}(A\star B)\se \left[z^{\a}\pd{z}{\a}+\zb^{\ad}\pd{\zb}{\ad}\right]
(A\star B)-2i\,\e^{\a\b}\pd{z}{\a}A\star \pd{z}{\b}B
-2i\,\e^{\ad\bd}\pd{\zb}{\ad}A\star \pd{\zb}{\bd}B\ .\la{appe2}
\ee

\pagebreak

%%%%%%%%%%%%%%%%%%%%%%%%%%%%%%%%%%%%%%%%%%%%%%%%%%%%%%%%%%%%%%%%%%%%%%%%%%

%%%%%%%%%%%%%%%%%%%%%%%%%%%%%%%%%%%%%%%%%%%%%%%%%%%%%%%%%%%%%%%%%%%%%%%%%%%

\ed